\documentclass[prodmode,acmtopc]{acmsmall} 

\usepackage[ruled]{algorithm2e}
\usepackage{dblfloatfix}

\SetAlFnt{\small}
\SetAlCapFnt{\small}
\SetAlCapNameFnt{\small}
\SetAlCapHSkip{0pt}
\IncMargin{-\parindent}
\newcommand{\bq}{\begin{equation}}
\newcommand{\eq}{\end{equation}}

\newcommand{\flop}{\mbox{flop}}
\newcommand{\flops}{\mbox{flops}}

\newcommand{\GBS}{\mbox{GB/s}}

\newcommand{\LUPS}{\mbox{LUP/s}}

\newcommand{\GLUPS}{\mbox{GLUP/s}}
\newcommand{\GHZ}{\mbox{GHz}}

\newcommand{\LUP}{\mbox{LUP}}
\newcommand{\LUPs}{\mbox{LUPs}}

\newcommand{\bytes}{\mbox{bytes}}
\newcommand{\byte}{\mbox{byte}}

\newcommand{\MB}{\mbox{MB}}

\newcommand{\MiB}{\mbox{MiB}}

\newcommand{\cycles}{\mbox{cy}}

\newcommand{\eos}{~.}

\newcommand{\ecmm}{ECM model}

\definecolor{orange}{rgb}{1,0.5,0}

\newcommand{\construction}[1]{}

\newcommand{\olsep}{\|}
\newcommand{\nolsep}{|}
\newcommand{\ecmspace}{\,}
\newcommand{\ecm}[6]{\mbox{$\left\{{#1}\ecmspace\olsep\ecmspace {#2}\ecmspace\nolsep\ecmspace {#3}\ecmspace\nolsep\ecmspace {#4}\ecmspace\nolsep\ecmspace {#5}\right\}\ecmspace{#6}$}}
\newcommand{\epsep}{\rceil}
\newcommand{\ecmp}[5]{\mbox{$\left\{{#1}\ecmspace\epsep\ecmspace {#2}\ecmspace\epsep\ecmspace {#3}\ecmspace\epsep\ecmspace {#4}\right\}\ecmspace{#5}$}}

\newcommand{\etal}{\textit{et al.}}

\usepackage[lofdepth=1,lotdepth=1]{subfig}
\usepackage{float}
\graphicspath{{./figures/}}

\usepackage[acronym]{glossaries}
\newacronym{rapl}{RAPL}{Running Average Power Level}
\newacronym{cl}{CL}{Cache Line}
\newacronym{ecm}{ECM}{Execution-Cache-Memory}
\newacronym{fma}{FMA}{Fused Multiply-Add}
\newacronym{fit}{FIT}{Finite Integration Technique}
\newacronym{thiim}{THIIM}{Time Harmonic Inverse Iteration Method}
\newacronym{fdfd}{FDFD}{Finite-Difference Frequency Domain}
\newacronym{pv}{PV}{Photovoltaic}
\newacronym{fed}{FED}{Fixed-Execution to Data}
\newacronym{tg}{TG}{Thread Group}
\newacronym{cfl}{CFL}{Courant-Friedrichs-Lewy}
\newacronym{llc}{LLC}{Last-Level Cache}
\newacronym{ptx}{PTX}{Parallel Thread Execution}
\newacronym{cta}{CTA}{Cooperative thread array}
\newacronym{fpu}{FPU}{Floating-Point Unit}
\newacronym{smx}{SMX}{Streaming Multiprocessor}
\newacronym{gpu}{GPU}{Graphics Processing Unit}
\newacronym{lifo}{LIFO}{Last In First Out}
\newacronym{lup}{LUP}{Lattice-site Update}
\newacronym{mic}{MIC}{Many Integrated Cores}
\newacronym{uma}{UMA}{Uniform Memory Access}
\newacronym{numa}{NUMA}{Non-Uniform Memory Access}
\newacronym{knc}{KNC}{Knight's Corner}
\newacronym{tgs}{TGS}{Thread Group Size}
\newacronym{fifo}{FIFO}{First In First Out}
\newacronym{hpc}{HPC}{High Performance Computing}
\newacronym{pde}{PDE}{Partial Differential Equation}
\newacronym{mwd}{MWD}{Multi-threaded Wavefront Diamond blocking}
\newacronym{swd}{1WD}{Single-threaded Wavefront Diamond blocking}
\newacronym{cpu}{CPU}{Central Processing Unit}
\newacronym{mpi}{MPI}{Message Passing Interface}
\newacronym{tlb}{TLB}{Translation Lookaside Buffer}
\newacronym{dsl}{DSL}{Domain Specific Language}
\newacronym{simd}{SIMD}{Single Instruction Multiple Data}
\newacronym{avx}{AVX}{Advanced Vector Extensions}
\newacronym{dp}{DP}{Double Precision}
\newacronym{sp}{SP}{Single Precision}
\newacronym{toc}{ToC}{Table of Contents}
\newacronym{los}{LoS}{List of Symbols}
\newacronym{loa}{LoA}{List of Abbreviations}
\newacronym{phd}{PhD}{Doctoral}
\newacronym{MS}{MS}{Masters}
\newacronym{Ms}{MS}{Microsoft}
\newacronym{CD}{CD}{Compact Disc}
\newacronym{kaust}{KAUST}{King Abdullah University of Science and Technology}

\usepackage{listings}
\lstdefinestyle{lstc} {
  showstringspaces=false,
  frame=lines,
  escapechar=\%,
  escapebegin=\color{black}\ttfamily\bfseries,
  language=C,
  basicstyle=\footnotesize\ttfamily,
  keywordstyle=\ttfamily,
  commentstyle=\ttfamily,
  aboveskip=0.2cm,belowskip=0.3cm,
  numberstyle=\tiny,
  numbers=none,
  numbersep=-2.5pt
}
\lstdefinestyle{lst_alg}  {
  frame=lines,
  language=C,
  basicstyle=\footnotesize,
  numbers=left,
}

\newcommand*{\sfwidth}{4.3cm} 

\acmVolume{0}
\acmNumber{0}
\acmArticle{0}
\acmYear{0}
\acmMonth{0}

\setcopyright{none}

\doi{0000001.0000001}

\issn{1234-56789}

\begin{document}

\markboth{T. Malas et al.}{Multi-dimensional intra-tile parallelization for memory-starved stencil computations}

\title{Multi-dimensional intra-tile parallelization for memory-starved stencil computations}
\author{TAREQ M. MALAS
\affil{Extreme Computing Research Center (ECRC).
King Abdullah University of Science and Technology}
GEORG HAGER
\affil{Erlangen Regional Computing Center (RRZE).
Friedrich-Alexander University of Erlangen-Nuremberg}
HATEM LTAIEF
\affil{ECRC.
King Abdullah University of Science and Technology}
DAVID E. KEYES
\affil{ECRC.
King Abdullah University of Science and Technology}}

\begin{abstract}
  Optimizing the performance of stencil algorithms has been the
  subject of intense research over the last two decades. Since many
  stencil schemes have low arithmetic intensity, most optimizations
  focus on increasing the temporal data access locality, thus reducing
  the data traffic through the main memory interface with the
  ultimate goal of decoupling from this bottleneck. There are,
  however, only few approaches that explicitly leverage the shared
  cache feature of modern multicore chips. If every thread works on
  its private, separate cache block, the available cache space can
  become too small, and sufficient temporal locality may not be
  achieved.

  We propose a flexible multi-dimensional intra-tile parallelization
  method for stencil algorithms on multicore CPUs with a shared
  outer-level cache. This method leads to a significant reduction in
  the required cache space without adverse effects from hardware
  prefetching or TLB shortage.  Our \emph{Girih} framework includes an
  auto-tuner to select optimal parameter configurations on the target
  hardware. We conduct performance experiments on two contemporary
  Intel processors and compare with the state-of-the-art stencil
  frameworks PLUTO and Pochoir, using four corner-case stencil schemes
  and a wide range of problem sizes. \emph{Girih} shows substantial
  performance advantages and best arithmetic intensity at almost
  all problem sizes, especially on low-intensity stencils with
  variable coefficients. We study in detail the performance behavior
  at varying grid size using phenomenological performance modeling.
  Our analysis
  of energy consumption reveals that our method can save energy by
  reduced DRAM bandwidth usage even at marginal performance gain.
  It is thus well suited for future architectures that will be strongly
  challenged by the cost of data movement, be it in terms of performance
  or energy consumption.
\end{abstract}

%
%


%
%



\acmformat{Tareq M. Malas, Georg Hager, Hatem Ltaief, and David E. Keyes, 2015. Multi-dimensional intra-tile parallelization for memory-starved stencil computations.}

\begin{bottomstuff}

\end{bottomstuff}

\maketitle

\section{Introduction}
\label{sec:intro}

Regular stencil computations arise as kernels in structured grid finite-difference, finite-volume, and finite-element discretizations of partial differential equation conservation laws and constitute the principal innermost kernel in many temporally explicit schemes for such problems.
They also arise as a co-principal innermost kernel of Krylov solvers for temporally implicit schemes on regular grids.
In~\cite{asanovic2006landscape}, they constitute the fifth of the ``seven dwarfs,'' the classes floating point kernels that receive the greatest attention in high performance computing.


A decisive property of stencil operators is their local spatial extent, which derives from the truncation error order of the finite discretization scheme.
In contemporary applications that drive fast stencil evaluations, such as seismic imaging, eighth order is the most common in industrial applications of which we are aware.
This contrasts with the second-order schemes for which the greatest amount of performance-oriented research has been done to date.


Another property of stencil operators with fundamental impact on achievable efficiency is the spatial dimension.
Our contribution concentrates on the most common case of three dimensions, though our illustrations of concepts often retreat into one or two dimensions, so that space and time fronts can be visualized on planar figures.
In principle, each spatial dimension may be treated in the same or in a different way with respect to partitioning and participation in wavefronts, as we show in our work and the literature.


We contribute a multi-dimensional intra-tile parallelization scheme using multi-core wavefront diamond blocking for optimizing practically relevant stencil algorithms.
The results demonstrate a substantial reduction in cache block size and memory bandwidth requirements using four corner-case stencil types that represent a full range of practically important stencil computations.
In contrast to many temporal blocking approaches in the literature, our approach efficiently utilizes the shared cache between threads of modern processors.
It also provides a controllable tradeoff between memory bandwidth per thread and frequency of synchronization to alleviate the bottleneck at the \gls{cpu} or memory interface, as needed when applying a stencil to a particular architecture.

Our scheme provides cache block sharing along the leading dimension (the dimension of most rapid index advance in a Cartesian ordering) that results in better utilization of loaded \gls{tlb} pages and hardware prefetching to the shared cache level.
We introduce a novel \gls{fed} wavefront parallelization technique that reduces the data movement in the cache hierarchy of the processor by using tiling hyperplanes that are parallel to the time dimension.
Our approach achieves hierarchical cache blocking by using large tiles in the shared cache level and fitting subsets of the tiles in the private caches of the threads, providing cache blocks that span multiple cache domains.
Our implementation uses an efficient runtime system to dynamically schedule tiles to thread groups.
We also develop an efficient fine-grained synchronization scheme to coordinate the work of the thread groups and avoid race conditions.
Finally, coupled with auto-tuning, our cache block sharing algorithm provides a rich set of run-time configurable options that allow architecture-friendly data access patterns for various setups.

We make several contributions in addition to \cite{malas_sisc2015} and \cite{wellein5254211}. 
We generalize our approach to multi-dimensional intra-tile parallelism and show efficient implementation for intra-tile parallelism in different tiling techniques in different dimensions.
The testbed framework, named \emph{Girih}, is described in details.
The auto-tuning approach is generalized to include all the tunable parameters.
Finally, we introduce a fixed-execution to data multi-core wavefront approach that allows each thread to update local grid data to its private caches, while the wavefront traverse the grid of the solution domain.

Using single-thread per cache block, as in \cite{Strzodka:2011:CAT}\cite{Bondhugula2008} is not sufficient to decouple from main memory in several situations, even with very efficient temporal blocking approaches.
Larger cache block sizes than available cache memory are required to reduce the main memory bandwidth requirements.
Cache block sharing can be used to alleviates the memory bandwidth pressure by providing sufficient data reuse through larger shared cache blocks.
Total cache block requirement reduction factor is correlated with the number of threads updating each cache block.
For example, when two threads share each cache block, the total cache block size requirement can be reduced nearly by a factor of two.

%

\section{Background}

\subsection{Cache blocking}

Typical scientific applications use larger grid size, on the order of GiBs, than a processor's cache memory, which is on the order of MiBs.
In the ``na\"{i}ve'' approach, the grid cells are updated in lexicographic order.
Each grid cell update involves loading the neighbor grid cells to perform the stencil computation.
This neighbor access results in loading the same data multiple times from main memory in each iteration, with no data reuse across iterations.
Consequently, low \flops/\byte\ ratio is observed in memory-bound stencil computations with the ``na\"{i}ve'' approach.~\cite{datta09}

Temporal blocking techniques allow more in-cache data reuse to reduce the memory bandwidth pressure.
These techniques  reorder the updates to perform several time step updates to the grid points while in the cache memory.
Wavefront blocking and diamond tiling are important temporal blocking techniques, so we describe them in more details.

\subsubsection{Wavefront temporal blocking}\label{sec:wavefront}

\begin{figure}[tbp]
    \centering
    \subfloat[Grid traversal in chronological order using the na\"{i}ve scheme. Each grid point is updated once per full sweep in the domain.]{
    \centering
        \includegraphics[clip=true, trim = 0.0cm 0.0cm 0.0cm 0.0cm ,width=3.5cm]{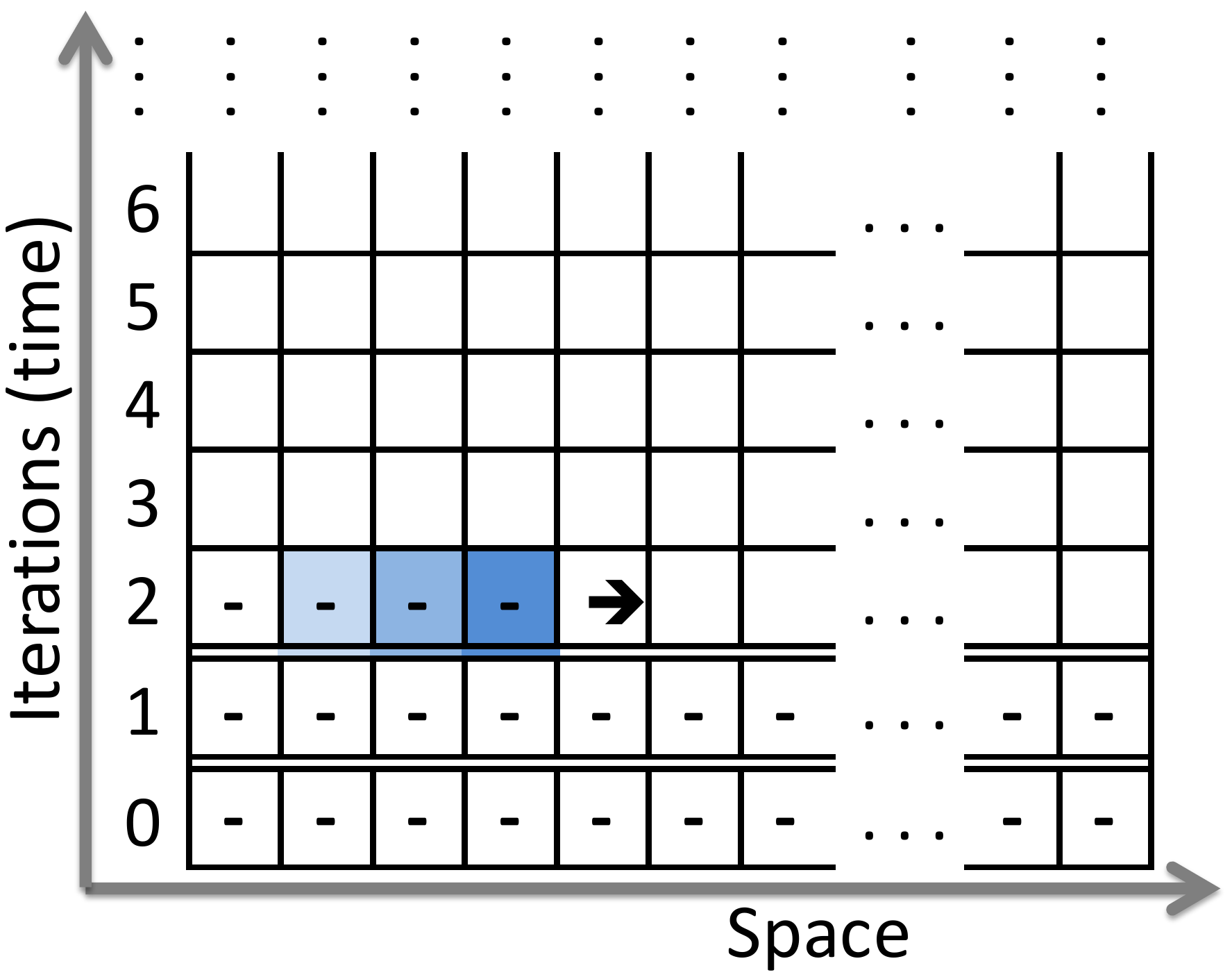}
    \label{fig:naive_3pt_1d}
    }
     \quad
    \subfloat[Grid traversal using single-thread wavefront blocking. The red arrows indicate the data dependency across iterations, which controls the slope of the wavefront.]{
    \centering
        \includegraphics[clip=true, trim = 0.0cm 0.0cm 0.0cm 0.0cm ,width=3.5cm]{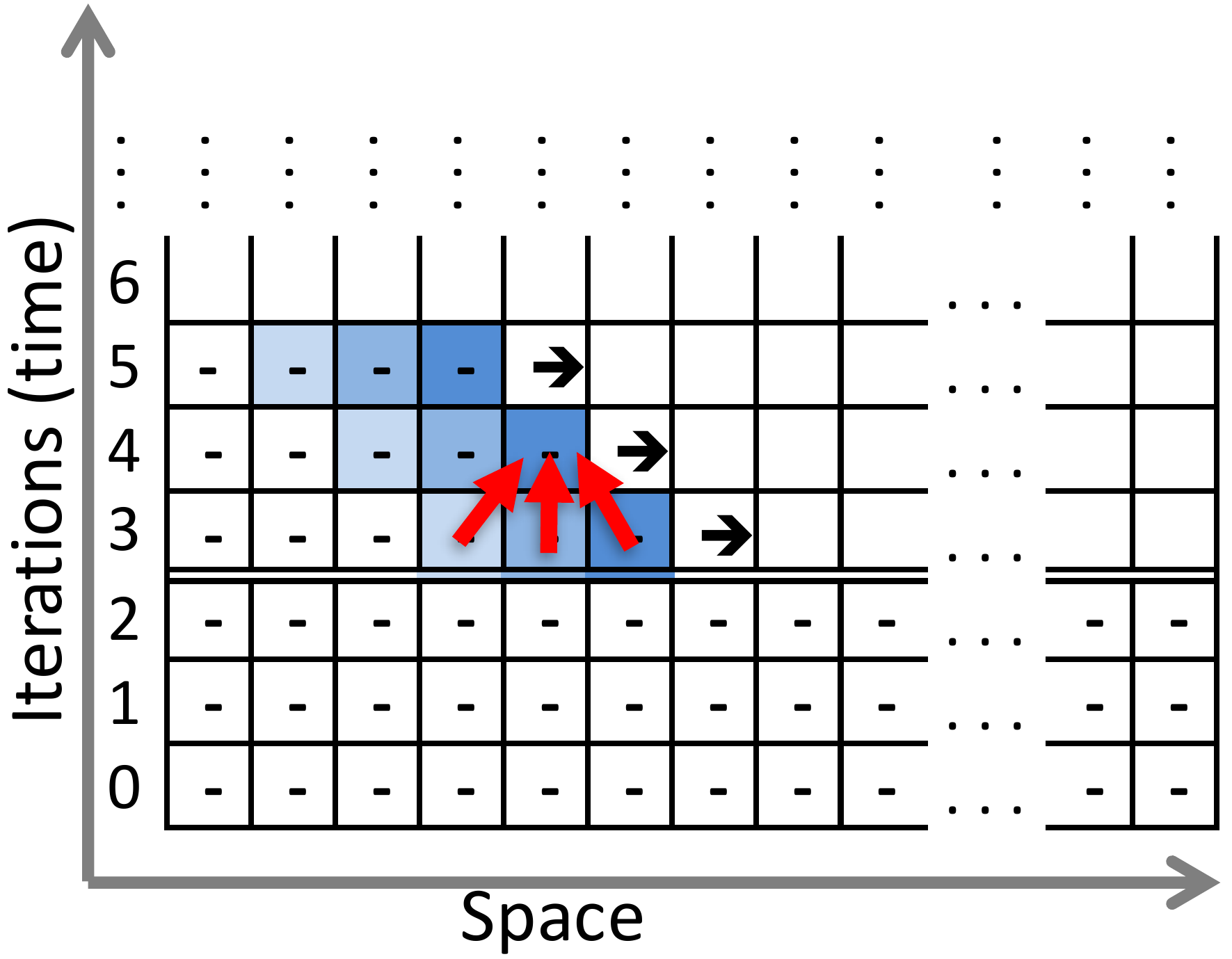}
    \label{fig:1_core_wavefront_3pt_1d}
    }
    \quad
    \subfloat[Multi-thread wavefront traversal in space-time blocks.
    The symbols/colors in the cells represent the update of different threads.
    Each thread updates one time step of the wavefront tile in this example.
    Extra spacing between the threads (i.e., steeper tile slope) allows concurrent update of the wavefront tile.]{
    \centering
        \includegraphics[clip=true, trim = 0.0cm 0.0cm 0.00cm 0.0cm ,width=5.0cm]{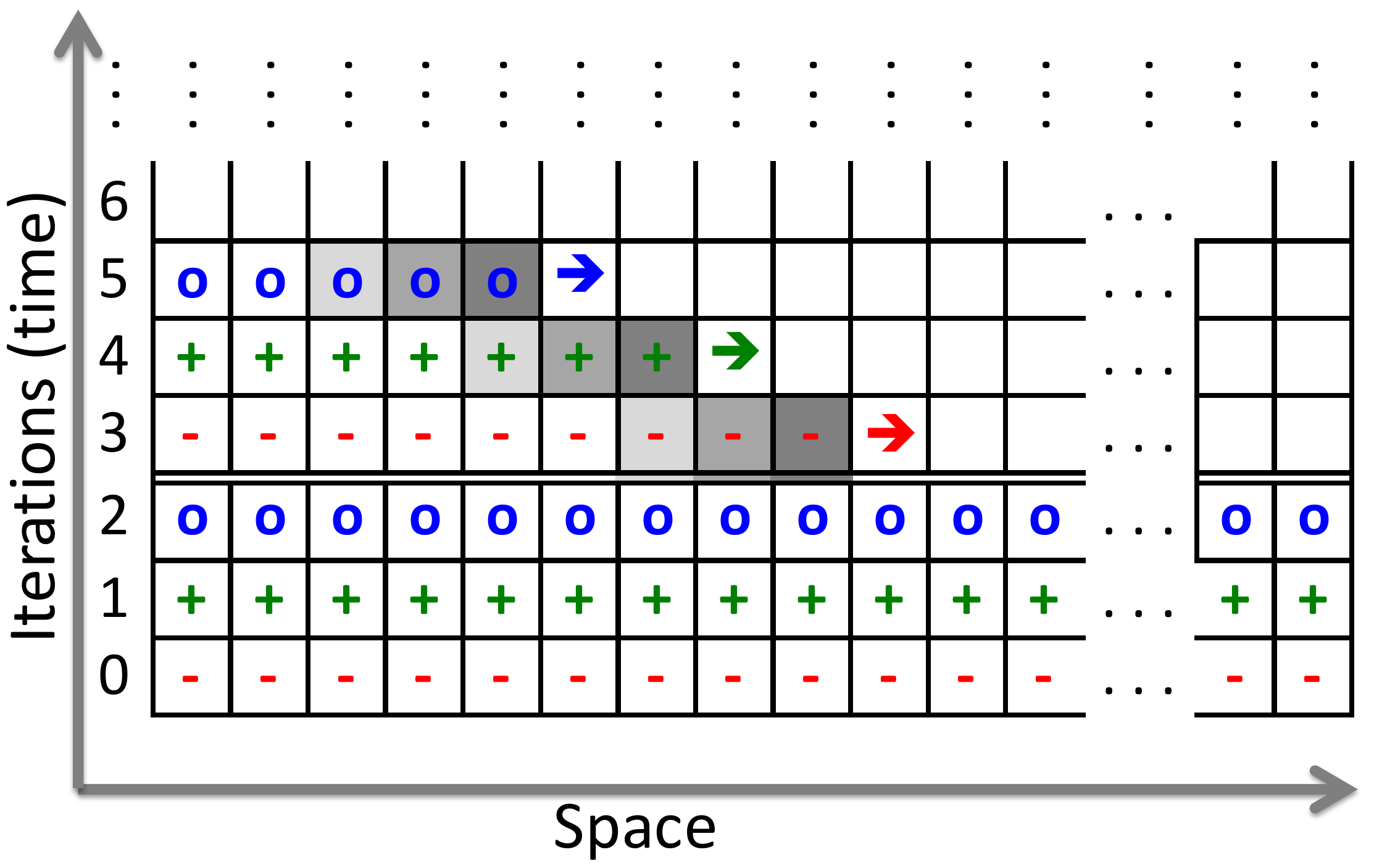}
    \label{fig:multi_core_wavefront_3pt_1d}
    }
    \caption{Na\"{i}ve and wavefront approaches to update one-dimension stencil computation. Darker boxes (in blue and gray) indicate more recent updates, cf.~\protect\cite{malas_sisc2015}. }
\end{figure}

Different tiling techniques can be used in each dimension in the grid, so we describe the tiling techniques using a 3-point stencil in one dimensional grid, using the C language syntax:\\
\verb.for(t=0; t<T; t++){.\\
\verb.  for(x=1; x<Nx-1; x++){.\\
\verb.    A[(t+1)%2][x] = W1 * A[t%2][x] + W2 * (A[t%2][x-1] + A[t%2][x+1]).\\
\verb.}}.\\

The na\"{i}ve update scheme of the 3-point stencil is shown in Fig~\ref{fig:naive_3pt_1d}.
The colored cells represent the most recent updates, where the darkest is most recent.

Wavefront temporal blocking was first introduced in~\cite{Lamport:1974}.
The update order in this technique maximizes the reuse of most recently visited grid cells.
The basic idea is shown in Fig.~\ref{fig:1_core_wavefront_3pt_1d}.
The space-time tile is traversed from left to right (in the directions of the arrows).
The red arrows show the data dependency across time steps for each grid cell, which is important to get correct results.
The wavefront slope increases as the spatial order of the stencil increases, as the data dependency of the stencil operator extends in space.
The wavefront blocking works only if the wavefront tile fits in the cache memory.
Using larger blocks in time increases the data reuse and the cache block size of the tile.

The wavefront tile can be updated using single thread per wavefront \cite{Strzodka:2011:CAT,nguyen18993} or multiple threads  per wavefront \cite{wellein5254211}.
Single-thread wavefronts require dedicated cache block per thread in the cache memory, without utilizing any shared cache hardware feature in the processor.
On the other hand, explicitly multi-core aware wavefront tiling utilizes the shared cache feature among the threads of the processor.
The multi-core aware wavefront pipelines the data among a thread group that shares a cache memory level.
This cache sharing reduces the cache memory size requirements and allows the use of larger cache blocks.
As a result, the memory bandwidth pressure can be reduced compared to the single-thread wavefront.
Fig.~\ref{fig:multi_core_wavefront_3pt_1d} shows the multi-thread wavefront variant proposed in \cite{wellein5254211}, where each thread's update is assigned different symbol/color.
Each thread is assigned to one or more consecutive time steps of the wavefront tile.
To enable concurrent update in the wavefront tile, the slope of the wavefront is increased to add spacing between the threads.
For example, the additional cell spacing in Fig.~\ref{fig:multi_core_wavefront_3pt_1d} allows the three threads to update the wavefront tile concurrently.

Combining the wavefront blocking scheme with other tiling techniques is essential in three-dimensional grids.
Otherwise, using only the wavefront scheme in multiple dimensions would require much larger cache block size than typical cache sizes, when a reasonable grid size is used.
Wavefront tiling is commonly combined with tiling approaches such as trapezoidal, parallelogram, or diamond tiling.
These tiling approaches limit the tile size in other dimensions, such that the wavefront tile size fits in the desired cache memory.
Since diamond tiling is proven to provide maximum the data reuse of a loaded spatial cache block~\cite{orozco2009ICPP}, we describe it in greater detail in the next section.

Wavefront blocking is ideal when the block size in time is fixed by a different factor, for example, by the tiling approach in a different spatial dimension.
By pipelining the data to the cache memory, the wavefront method loads each element once to the cache memory and performs the maximum time updates before evicting the results from the cache memory.
On the other hand, loading data blocks separately and updating their elements by other tiling techniques would not allow maximum data reuse at the edges of the loaded block due to the data dependencies.

\subsubsection{Diamond tiling} \label{sec:diamond_tiling}
\begin{figure}[tbp]
\centering
\includegraphics[width=10cm]{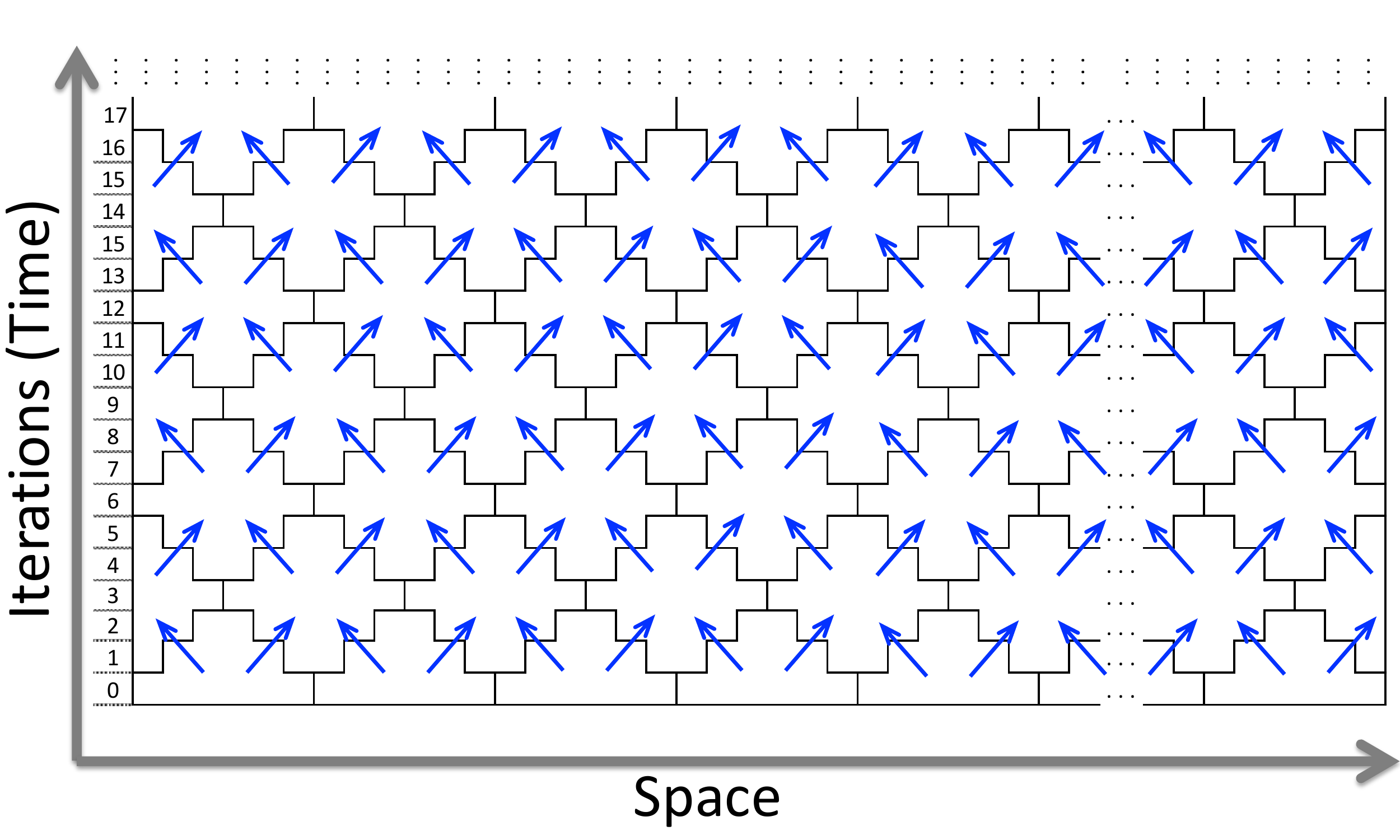}
\caption{Diamond tiling structure for the 3-point stencil in one-dimensional grid. The blue arrows show the data dependency of the diamond tiles, cf.~\protect\cite{malas_sisc2015}.}
\label{fig:diamond_tiling_1d}
\end{figure}

Diamond tiling is an important temporal blocking technique in the literatures.
The basic idea of diamond tiling is shown in Figure~\ref{fig:diamond_tiling_1d}.
Diamond tiles have dependency over the two tiles below, as indicated by the blue arrows.
The tile slope rely on the stencil semi-bandwidth, where $S=\pm 1/R$.
Diamond tiling has several advantages: it allows maximum data reuse~\cite{orozco2009ICPP}, allows more concurrency in the tiles update, and has a unified tile shape, making the implementation simple.

\subsection{Analytic and phenomenological performance modeling}\label{sec:models}

Performance optimization of numerical applications is often carried
out without knowing whether the resulting code is ``good enough.''
Analytic performance models can answer the question what ``good
enough'' actually means, in the sense that they predict the optimal
performance of an algorithm and/or an implementation in view of the
available resources. The Roof\/line model is a well-known example,
whose principles date back into the 1980s
\cite{Kung:1986:MRB:17407.17362,Callahan88,hockney89,schoenauer00} and
which has received revived interest in the context of cache-based
multi-core processor architectures in recent
years~\cite{roofline:2009}. It predicts the performance of
``steady-state'' loops or loop nests on a CPU, assuming that one of
two possible bottlenecks apply: either the runtime is limited by the
execution of instructions (execution bottleneck) or by the required
data transfers through the memory hierarchy (data bottleneck),
whichever takes longer. ``Steady state'' in this context means that
start-up and wind-down effects are ignored, and that the CPU executes
the same mix of instructions with the same data requirements for a
long time. The assumptions behind the Roof\/line model are fulfilled
for many algorithms in computational science that are either very
data-bound or very compute-bound, i.e., whenever the data transfer and
execution times differ strongly. Unfortunately, stencil algorithms
with temporal blocking optimizations do not fall in this category. 

The \ecmm~\cite{th09,CPE:CPE3180,stengel14} is an extension of
the Roof\/line model. In contrast to the latter, it
does not assume perfect overlap between in-core and data transfer times.
Instead, it assumes that the in-core execution time, i.e., the time required
to execute a number of loop iterations with data coming from the L1 cache,
is composed of an overlapping  and a non-overlapping part. The
non-overlapping part is serialized with all transfer times between
adjacent memory hierarchy levels down to where the data originally
resided. We briefly review the model in the following;
details can be found in~\cite{stengel14}.

The in-core execution time $T_\mathrm{core}$ is set by the execution
unit (pipeline) taking the largest number of cycles to execute the
instructions for a given number of loop iterations. The
non-overlapping part of $T_\mathrm{core}$, called $T_\mathrm{nOL}$,
consists of all cycles in which a LOAD instruction retires.  As all
data transfers are in units of cache lines (CLs), we usually consider
one cache line's ``worth of work.''  With a double-precision stencil
code, one unit of work is $8$~iterations.

The time for all data transfers required to execute the work
unit is the ``transfer time.'' Latency is neglected altogether
(although it can be brought into the model as a correction factor,
see~\cite{HFEHW15}), so
the cost for one CL transfer can be determined by the
maximum bandwidth (cache lines per cycle) between adjacent memory hierarchy
levels. 
E.g., on the Intel Haswell architecture, one CL transfer takes one cycle
between the L1 and L2 caches and two cycles between L2 and L3.
Moving a 64-\byte\ CL between memory and L3 takes
$64\,\bytes\cdot f/b_\mathrm S$ cycles. Here $f$ is the
clock frequency of the CPU and $b_\mathrm S$ is the fully saturated
memory bandwidth. 

If $T_\mathrm{data}$
is the transfer time,
$T_\mathrm{OL}$ is the overlapping part of the core execution,
and $T_\mathrm{nOL}$ is the non-overlapping part, then
\bq
T_\mathrm{core} =  \max\left(T_\mathrm{nOL},T_\mathrm{OL}\right)\quad\mbox{and}\quad
T_\mathrm{ECM} =  \max(T_\mathrm{nOL}+T_\mathrm{data},T_\mathrm{OL})\label{eq:T}\eos
\eq
$T_\mathrm{data}$ is the sum of the data transfer times through
the memory hierarchy. If, e.g., the data is in the L3 cache,
$T_\mathrm{data}=T_\mathrm{L1L2}+T_\mathrm{L2L3}$. 
%
We use a shorthand notation for the cycle times in the model for
executing a unit of work:
\ecm{T_\mathrm{OL}}{T_\mathrm{nOL}}{T_\mathrm{L1L2}}{T_\mathrm{L2L3}}{T_\mathrm{L3Mem}}{}. 
Adding up the contributions from $T_\mathrm{data}$ and $T_\mathrm{nOL}$
and applying (\ref{eq:T}) one can predict the cycles for executing the loop
with data from any given memory level.
E.g., if the model is \ecm{4}{4}{2}{4}{9}{\cycles}, the prediction
for L3 cache will be $\max\left(4,4+2+4\right)\,\cycles=10\,\cycles$.
For predictions we use  ``$\epsep$''
as the separator to present the information in a similar way as in the model.
In the example above this would be
$T_\mathrm{ECM}=\ecmp{4}{6}{10}{19}{\cycles}$. It is easily possible
to get from time to performance (work divided by time)
by calculating the fraction  $P=W/T_\mathrm{ECM}$, where $W$ is the
work.
If $T_\mathrm{ECM}$ is given in clock cycles but the unit of work
is a \LUP\ and the performance metric is \LUPS\ then
we have to multiply by the clock speed. Note that the ECM model as
presented here assumes a fully inclusive cache hierarchy. It can be
adapted to exclusive caches as well~\cite{th09}.

For multi-core scalability We assume that the performance
is linear in the number of cores until the maximum bandwidth of a
bottleneck data path is exhausted.
On Intel processors the
memory bandwidth is the only bottleneck, so that the absolute maximum
performance is the Roof\/line prediction for memory-bound
execution: $P_\mathrm{BW} = I\cdot b_\mathrm S$,
with $I$ being the computational intensity.
Thus, the model allows to predict the number of cores
required to reach bandwidth saturation.

Although it is possible in many situations to analyze the data
transfer properties of a stencil code and construct the ECM model from
first principles, in practice one faces difficulties with this
approach when looking at temporally blocked codes. The reason is that
the model requires an accurate analysis of the data transfer volume in
different cache levels, which becomes very difficult with small,
non-rectangular block shapes (e.g., diamonds). In such situations
the model can only give very rough upper limits. However, it is still
possible to apply the principles of the model in a phenomenological
way by \emph{measuring} the data transfers by hardware performance
monitoring. If the model thus constructed agrees with the performance
measurements, this is an indication that the code utilizes the
hardware in an optimal way, and especially that the low-level
machine code produced by the compiler does not incur any
significant overhead. This phenomenological modeling approach
will be applied later when analyzing the performance of the MWD
code.

\begin{lstlisting}[style=lstc, float=tb,label=lst:7const,caption={$1^{st}$-order-in-time 7-point constant-coefficient isotropic stencil in three dimensions, with symmetry. The code shows single time iteration, where this code is repeated many times in the time loop, with arrays pointer swapping after each iteration.}]
for(int k=1; k < N-1; k++) {
 for(int j=1; j < N-1; j++) {
  for(int i=1; i < N-1; i++) {
   U[k][j][i] = c0 *  V[ k ][ j ][ i ]
              + c1 * (V[ k ][ j ][i+1] + V[ k ][ j ][i-1])
              + c1 * (V[ k ][j+1][ i ] + V[ k ][j-1][ i ])
              + c1 * (V[k+1][ j ][ i ] + V[k-1][ j ][ i ]);
}}}
\end{lstlisting}

\begin{lstlisting}[style=lstc, float=tb,label=lst:7var,caption={$1^{st}$-order-in-time 7-point variable-coefficient stencil in three dimensions, with no coefficient symmetry. The code shows single time iteration, where this code is repeated many times in the time loop, with arrays pointer swapping after each iteration.}]
for(int k=1; k < N-1; k++) {
 for(int j=1; j < N-1; j++) {
  for(int i=1; i < N-1; i++) {
   U[k][j][i] = C0[k][j][i] * V[ k ][ j ][ i ]
              + C1[k][j][i] * V[ k ][ j ][i+1]
              + C2[k][j][i] * V[ k ][ j ][i-1]
              + C3[k][j][i] * V[ k ][j+1][ i ]
              + C4[k][j][i] * V[ k ][j-1][ i ]
              + C5[k][j][i] * V[k+1][ j ][ i ]
              + C6[k][j][i] * V[k-1][ j ][ i ];
}}}
\end{lstlisting}

\begin{lstlisting}[style=lstc, float=tb,label=lst:25const,caption={$2^{nd}$ order in time 25-point constant-coefficient isotropic stencil in three dimensions, with symmetry across each axis. The code shows single time iteration, where this code is repeated many times in the time loop, with arrays pointer swapping after each iteration.}]
for(int k=4; k < N-4; k++) {
 for(int j=4; j < N-4; j++) {
  for(int i=4; i < N-4; i++) {
   U[k][j][i] = 2*V[k][j][i] - U[k][j][i] + C[k][j][i] * [
               +c0 *  V[ k ][ j ][i  ]
               +c1 * (V[ k ][ j ][i+1]+V[ k ][ j ][i-1]
                     +V[ k ][j+1][ i ]+V[ k ][j-1][ i ]
                     +V[k+1][ j ][ i ]+V[k-1][ j ][ i ])
               +c2 * (V[ k ][ j ][i+2]+V[ k ][ j ][i-2]
                     +V[ k ][j+2][ i ]+V[ k ][j-2][ i ]
                     +V[k+2][ j ][ i ]+V[k-2][ j ][ i ])
               +c3 * (V[ k ][ j ][i+3]+V[ k ][ j ][i-3]
                     +V[ k ][j+3][ i ]+V[ k ][j-3][ i ]
                     +V[k+3][ j ][ i ]+V[k-3][ j ][ i ])
               +c4 * (V[ k ][ j ][i+4]+V[ k ][ j ][i-4]
                     +V[ k ][j+4][ i ]+V[ k ][j-4][ i ]
                     +V[k+4][ j ][ i ]+V[k-4][ j ][ i ])];
}}}
\end{lstlisting}

\begin{lstlisting}[style=lstc, float=tb,label=lst:25var,caption={$1^{st}$-order-in-time 25-point variable-coefficient anisotropic stencil in three dimensions, with symmetry across each axis. The code shows single time iteration, where this code is repeated many times in the time loop, with arrays pointer swapping after each iteration.}]
for(int k=4; k < N-4; k++) {
 for(int j=4; j < N-4; j++) {
  for(int i=4; i < N-4; i++) {
   U[k][j][i] = C00[k][j][i]*  V[ k ][ j ][ i ]
               +C01[k][j][i]*( V[ k ][ j ][i+1]+V[ k ][ j ][i-1])
               +C02[k][j][i]*( V[ k ][j+1][ i ]+V[ k ][j-1][ i ])
               +C03[k][j][i]*( V[k+1][ j ][ i ]+V[k-1][ j ][ i ])
               +C04[k][j][i]*( V[ k ][ j ][i+2]+V[ k ][ j ][i-2])
               +C05[k][j][i]*( V[ k ][j+2][ i ]+V[ k ][j-2][ i ])
               +C06[k][j][i]*( V[k+2][ j ][ i ]+V[k-2][ j ][ i ])
               +C07[k][j][i]*( V[ k ][ j ][i+3]+V[ k ][ j ][i-3])
               +C08[k][j][i]*( V[ k ][j+3][ i ]+V[ k ][j-3][ i ])
               +C09[k][j][i]*( V[k+3][ j ][ i ]+V[k-3][ j ][ i ])
               +C10[k][j][i]*( V[ k ][ j ][i+4]+V[ k ][ j ][i-4])
               +C11[k][j][i]*( V[ k ][j+4][ i ]+V[ k ][j-4][ i ])
               +C12[k][j][i]*( V[k+4][ j ][ i ]+V[k-4][ j ][ i ]);
}}}
\end{lstlisting}

\section{Motivation: On temporal blocking practical performance limits} \label{sec:motivation}

Temporal blocking techniques are commonly used to improve the performance of memory bandwidth-starved stencil computations.
In particular, stencil computation are usually memory bandwidth-bound, even with efficient spatial blocking techniques~\cite{doi:10.1137/140991133}.

In this section we study the performance of highly efficient state-of-the-art temporal blocking schemes, while keeping the implementation practical and efficient for contemporary processors.
We contribute accurate models to estimate the cache block size and the memory traffic of these schemes to show their limits.
These models are validated using the corner-case stencils shown in Code Listings~\ref{lst:7const},~\ref{lst:7var},~\ref{lst:25const}, and~\ref{lst:25var}.
This work is particularly important to show the shortcomings of using single-thread per tile in contemporary processors, even when best tiling techniques are used.

\subsection{Testbed}
We conducted all our experiments on two recent microprocessors: an Intel
Ivy Bridge (10-core Xeon E5-2660v2, 2.2\,\GHZ, 25\,\MB\ L3 cache,
40\,\GBS\ of applicable memory bandwidth) and an Intel Haswell
(18-core Xeon E5-2699 v3, 2.3\,\GHZ, 45\,\MB\ L3 cache, 50\,\GBS\ of
applicable memory bandwidth). The ``Turbo Mode'' feature was disabled,
i.e., the CPUs ran at their nominal clock speeds, to avoid performance
fluctuations. The Haswell chip was set up in a standard configuration
with ``Cluster on Die'' (CoD) mode disabled, meaning that the full
chip was a single ccNUMA memory domain with 18 cores and a shared L3
cache. The documented clock slowdown feature with highly optimized 
AVX code \cite{hackenberg15} was not observed on this machine with 
any of our codes. Simultaneous multi-threading (SMT) was enabled 
on both machines but not used in our tests because it would have 
adverse effects on the cache size per thread, which is crucial 
for the type of optimizations we are targeting.
More details on the hardware architectures can be found in the 
vendor documents~\cite{intelopt}. 

Performance counter measurements were done with the \verb.likwid-perfctr.
tool from the LIKWID multicore tool suite~\cite{likwid} in version 4.0. 
The compiler versions used for different code variants are described
in Sect.~\ref{sec:results}.
Intel C 15 compiler is used everywhere in the experiments.
In the distributed memory results, we use the Intel MPI library 4.1.3.

\subsection{Single-thread wavefront diamond blocking}
\begin{figure}[tbp]
    \centering
    \includegraphics[width=8cm]{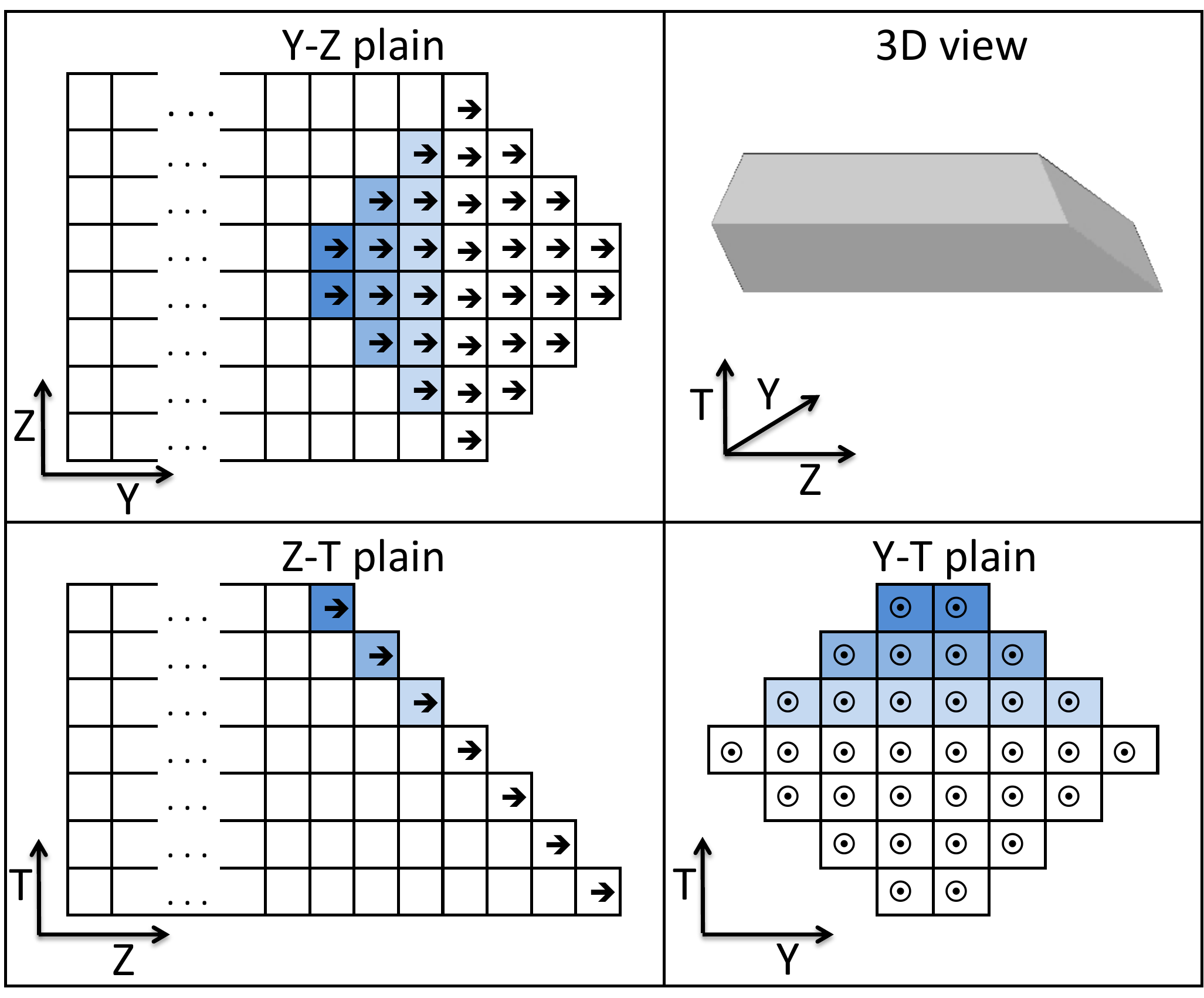}
    \caption{Single-core wavefront diamond tiling (1WD) in three-dimensional grid, with wavefront traversal along the $z$ dimension and diamond tiling along the $y$ dimension, cf.~\protect\cite{malas_sisc2015}.}
    \label{fig:1wd_2d}
\end{figure}

In Sections~\ref{sec:wavefront}~and~\ref{sec:diamond_tiling} we showed that wavefront blocking and diamond tiling techniques maximize the data reuse in the cache memory.
We find it reasonable to use these techniques over the outer dimensions ($y$ and $z$) in three-dimensional grids.
We prefer to leave the $x$ dimension intact, at reasonable grid sizes, for efficient hardware data prefetching, minimum \gls{tlb} misses, and longer strides for vectorization.
In fact, recent works are adopting these techniques in their implementations. The work in~\cite{Strzodka:2011:CAT} performs the diamond tiling along the $y$-axis and the wavefront blocking along the $z$-axis.
The work in~\cite{Bandishti6468470} performs the diamond tiling along the $z$-axis and the wavefront blocking along the $y$-axis in PLUTO framework.
As will be shown in Sect.~\ref{sec:results}, the auto-tuning of PLUTO tiling parameters shows that longer strides along the $x$-axis achieves the best performance, which supports our argument.

We implement a \gls{swd} scheme in this work, as shown in Figure~\ref{fig:1wd_2d}.
The wavefront traverses along the $z$-axis.
The diamond tiling is performed along the $y$-axis.
Hence, each grid point in Fig.~\ref{fig:1wd_2d} represents full stride along the $x$ dimension.

The \gls{swd} scheme is an important ingredient of this work.
We construct and validate cache block size and memory traffic models in the following subsections for the \gls{swd} implementation to show its requirements and limits on contemporary processors.

\subsection{Cache block size model}\label{sec:blk_size_model}

We construct a cache block size model of \gls{swd}, validate its correctness, and study its impact on the code balance at different diamond sizes.
The model calculations require four parameters: the diamond width $D_w$ in the $y$ axis, the wavefront tile width $N_{F}$, the bytes number in the leading dimension $N_{xb}$, the stencil radius $R$, and the number of domain-sized streams in the stencil operator, $N_D$. 
Examples of stencil radius are $R\! =\! 1$ and $R\! =\!4$ at the 7- and 25-point stencils, respectively.
The 7-point constant-coefficient stencil has $N_D=2$ (Jacobi-like update). The 7-point variable-coefficient stencil uses seven
additional domain-sized streams to hold the coefficients.
For a stencil with $R\! =\! 1$, the wavefront width $W_w$ has the size: $W_w\!=\!D_w\!+\!N_{F}\!-\!2$ and the total required bytes in the wavefront cache block $C_S$, with some approximations, is:
\bq\label{eq:7pt_cache_blk_size}
C_S = N_{xb} \cdot \left[N_D\cdot \left(\frac{D_w^2}{2}\!+\! D_w\! \cdot\!(N_{F}\!-\!1)\right) 
  + 2 \cdot(D_w + W_w)\right] \eos
\eq
The equation is composed of three parts: ``$N_{xb}$'' factor is the size
of the leading dimension tile size, ``${D_w^2}/{2} + D_w\cdot (N_{F}-1)$'' term is the
diamond area in the $y$-$z$ plane (shown in the top
view of Fig.~\ref{fig:1wd_2d}), and 
the halo region of the wavefront is the ``$2 \cdot(D_w + W_w)$'' term.

For example, $D_w = 8$ and $N_{F} = 1$ in Fig.~\ref{fig:1wd_2d}, so 
$W_w = 8 + 1 - 2 = 7$ and the total block size at 7-point constant-coefficient stencil is $N_{xb} \cdot ( 2 \cdot ( 8^2 / 2 + 8 \cdot 0) + 2 \cdot(8 + 7)) = 94 \cdot  N_{xb}$~\bytes.

The steeper wavefront in higher-order stencils results in different wavefront width ($W_w= D_w - 2\cdot
R + N_{F}$) and different $C_S$ as follows:
\bq\label{eq:cache_blk_size}
C_S\!=\!N_{xb}\cdot\!\left[\!N_D\!\cdot\!D_w\!\cdot\! \left(\frac{D_w}{2}\!-\!R\!+\!N_{F}\right) 
  + 2 R (D_w + W_w)\right]\eos
\eq

It is worth mentioning that each thread requires a dedicated $C_S$ in the blocked cache level. For example, using a 16-core Intel Haswell socket requires fitting $16\cdot C_S$ \bytes\ in the L3 cache memory.

\subsection{Memory traffic model}\label{sec:mem_model}

In order to validate the effectiveness of the
bandwidth pressure reduction on the memory interface, 
we set up a model to estimate the code balance for the temporally
blocked case. If the wavefront fits completely in the L3
cache, each grid point is loaded once from main memory and is stored once after updating it during the
extruded diamond update.  In this case, the
amount of data transfers during the extruded diamond update consists of  $(2D_w-2)$ data writes plus $(N_D \cdot D_w+2)$ data reads, all multiplied by $N_z$. The number of total \LUPs\ performed through the diamond volume is:
$N_z \cdot D^2_w/2$. The code balance at double precision of a stencil with $R=1$ is thus:
\bq \label{eq:mwd_bw_req}
B_\mathrm{C} = \frac{16 \cdot \left[(2D_w-2)+
    (N_D \cdot D_w + 2)\right]}{D^2_w}\frac{\bytes}{\LUP}\eos
\eq

When $R\!>\! 1$ the amount of data transfers becomes $N_z \cdot \left[(2D_w-2R) +(N_D \cdot D_w +2R)\right]$ and the extruded diamond volume becomes $N_z \cdot D^2_w/(2\cdot R )$. In total, the equation becomes:

\bq \label{eq:mwd_bw_req_ho}
B_\mathrm{C} = \frac{16R \cdot \left[(2D_w-2R)+
    (N_D \cdot D_w + 2R)\right]}{D^2_w}\frac{\bytes}{\LUP}\eos
\eq

\subsection{Model verification}

We verify the correctness of our memory traffic and cache block size models of the \gls{swd} scheme.
Our models and measurements prove the limitation of using separate cache block per thread, which is common in the literature.
The desired code balance to decouple from the main memory bandwidth requires larger cache block size than the available cache memory in contemporary processors when separate cache block is used per thread.

We use the four stencils described in Listing~\ref{lst:7const}, ~\ref{lst:7var}, ~\ref{lst:25const}, and ~\ref{lst:25var}.
The grid sizes are larger than the cache memory size and fit in the main memory of the processor, which is typical in real applications.
We minimize the cache block size in our experiments by using a unity wavefront tile width ($N_F=1$), where larger $N_F$ would increase the cache block size without decreasing the code balance.
We perform our experiments using a single thread in the 18-core Haswell processor to dedicate the $45$ MiB cache memory for its use.
The single thread experiment allows us to test larger cache block sizes without having cache capacity misses.

\begin{figure}[tbp]
    \newcommand*{\sfwidtha}{5cm}
    \centering
    \subfloat[7-point constant-coefficient stencil using grid size $N=960^3$]{
        \centering
        \includegraphics[width=\sfwidtha]{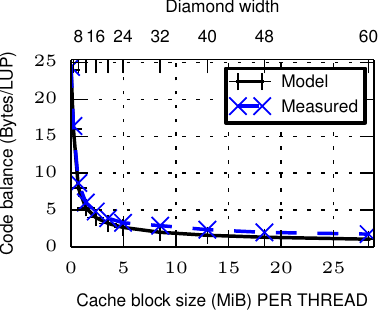}
        \label{fig:7pt_const_code_balance}
	}
	\enskip
    \subfloat[7-point variable-coefficient stencil using grid size $N=680^3$]{
        \centering
        \includegraphics[width=\sfwidtha]{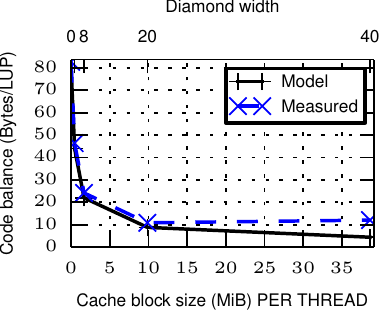}
        \label{fig:7pt_var_code_balance}
	}
	\enskip
	\subfloat[25-point constant-coefficient stencil using grid size $N=960^3$]{
        \centering
        \includegraphics[width=\sfwidtha]{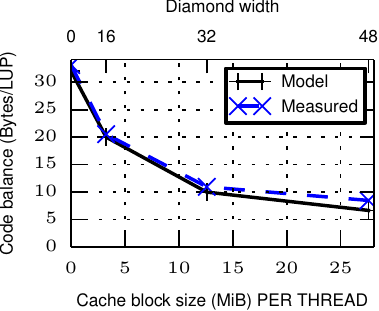}
        \label{fig:25pt_const_code_balance}
	}
	\enskip
	\subfloat[25-point variable-coefficient stencil using grid size $N=480^3$]{
        \centering
        \includegraphics[width=\sfwidtha]{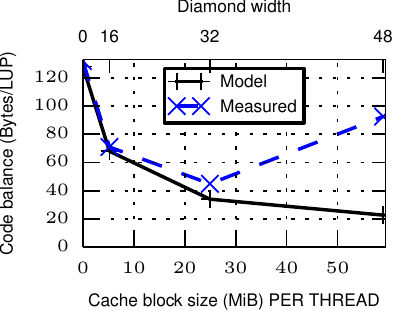}
        \label{fig:25pt_var_code_balance}
	}
	\caption{Cache block size vs.\ modeled and measured code balance using four corner-case stencil operators in Intel 18-core Haswell processor.
	Several diamond tile sizes are evaluated using unit wavefront tile width. Cache block size and code balance are computed using the models in sections~\ref{sec:blk_size_model} and~\ref{sec:mem_model}, respectively. All cases show accurate prediction of the code balance when the cache block size falls within the usable cache size (i.e., half the cache size of the processor).}
    \label{fig:code_balance}
\end{figure}

Figure~\ref{fig:code_balance} shows the cache block size vs.\ the code balance at various diamond tile sizes (top $x$-axis).
We compute the ``Model'' data using our cache block size model in Eq.~\ref{eq:cache_blk_size} (bottom $x$-axis) and the code balance model in Eq.~\ref{eq:mwd_bw_req_ho}.
The ``Measured'' data is the measured code balance in our experiments, which is computed by dividing the total measured memory traffic by the total updated grid points.
We set the diamond tiles' widths to multiples of 4 and 16 for the 7-point and 25-point stencils, respectively.
Data points at zero diamond width correspond to an efficient spatial blocking scheme.

Our models are very accurate in predicting the code balance of corner-case stencil operators.
There is a strong agreement between the model and the empirical results when the cache block fits in the L3 cache (below $22.5$ \MiB).
These findings show that our implementation of the \gls{swd} blocking scheme can achieve the theoretical memory traffic reductions.
The measured code balance in Fig.~\ref{fig:code_balance} starts to deviate from the model at cache blocks larger than about half the Intel Haswell's L3 cache size (i.e., $22.5$ \MiB). 
The deviation at this point can be predicted from our cache block size model,  considering the rule-of-thumb that half the cache size is usually usable for blocking~\cite{stengel14}.

The results in Fig.~\ref{fig:code_balance} prove how using separate temporally blocked tile per thread leads to starvation in the cache size requirement.
The minimum diamond width ($D_W\!=\!16$) of the 25-point constant-coefficient stencil, in Fig.~\ref{fig:25pt_const_code_balance}, requires a tile size of $\sim\!3$ MiB/thread.
To run all the $18$ cores of the processor with efficient temporal blocking, the processor has to provide a minimum of $3*18=54$ MiB of cache memory, which is far from the available one.
We observe more starvation in the cache memory in the 25-point variable-coefficient stencil in Fig~\ref{fig:25pt_var_code_balance}.
The 7-point stencils in Figs.~\ref{fig:7pt_var_code_balance}~and~\ref{fig:7pt_const_code_balance} can fit $18$ tiles in the cache memory, but the diamond width has to be limited  (i.e., provide limited data reuse) and does not reduce the code balance sufficiently to prevent main memory bandwidth saturation.
We investigate these observations in more details in the results in Sect.~\ref{sec:results}.

Our accurate cache block size model can be used to tune the tiling parameters to achieve high performance.
We can use our model to select the largest tile size that fits in the usable cache memory size to replace auto-tuning, as performed in~\cite{Strzodka:2011:CAT}.
On the other hand, even an accurate cache block size model is not sufficient to select the tuning parameters for the best performance.
For example, Fig.~\ref{fig:25pt_const_code_balance} shows a case when a tile size larger than the usable cache memory (i.e., $D_W\!=\!48$ requires $\sim\! 27$ MiB cache block) can still achieve better code balance than the maximum tile size that fits entirely in the cache memory (i.e., $D_W\!=\!32$ requires  $\sim\! 12.5$ MiB cache block).
This observation shows the importance of auto-tuning even when an accurate model for the cache block size is available.
In fact, we use auto-tuning in our work, assisted with our modeling techniques to narrow down the parameters search space.

In this section, we showed how the best practices temporal blocking are not sufficient to overcome the main memory bandwidth limitation in contemporary processors.
These approaches will become less efficient in future processors as the machine balance decrease, cache size per thread decrease, and concurrency increase.
In particular, variable-coefficient and high-order stencils suffer the most from these limitations as they demand more cache memory to decouple from the main memory bandwidth bottleneck.

\section{Approach: Multi-dimensional intra-tile parallelization} \label{sec:approach}

In this section, we describe our intra-tile multi-dimensional parallelization algorithm in details and its wavefront diamond tiling implementation in a structured grid.
We also describe the components of our open-source testbed framework, called \emph{Girih}~\cite{malas_girih}.

\subsection{Algorithm}

We shown in Section \ref{sec:motivation} that using single-thread per cache block is not sufficient to decouple from main memory in several situations, even with very efficient temporal blocking approaches.
Larger cache block sizes than available cache memory are required to reduce the main memory bandwidth requirements.
To resolve these issues, we introduce an advanced cache block sharing scheme.
It alleviates the pressure on the cache size to provide sufficient data reuse that decouples the computations from the main memory bandwidth bottleneck through larger shared cache blocks.

We introduce a (d+1)-dimensional intra-tile parallelization algorithm for tiled d-dimensional grids.
Each Cartesian dimension, $i$, of the tiles is divided equally into $T_i$ chunks that can be updated concurrently. Additionally, the components of each grid cell may be divided into $T_c$ chunks, when at least $T_c$ equations per grid cell can be updated concurrently.
The number of threads updating a tile (``thread group'') are equal to $T_c \times \prod_{n=1}^{d}T_n$.
Multiple \glspl{tg} may exist to update tiles concurrently.
The data reuse in the threads' private caches can be improved by making the boundaries of the sub-tiles parallel to the time dimension, where each thread would be reusing its local grid points across the time steps in the tile.

Our multi-dimensional cache block sharing algorithm has several advantages. 
First, it requires less main memory bandwidth by enabling more in-cache data reuse through larger cache blocks in the shared cache levels.
It has less penalty on the cache block size requirement compared to other cache block sharing approaches, where they require more space to provide sufficient concurrency, as will be shown in the related work, Sect.~\ref{sec:related_work}.

Cache block sharing along the leading dimension reduces the need for tiling along it, resulting in better utilization of the loaded \gls{tlb} data and the hardware data prefetching to the \gls{llc}.
That is, long contiguous strides are utilized in the shared cache level, while each thread updates a block of the unit stride.
On the other hand, tiling along the leading dimension would cause the hardware prefetching unit to load data beyond the cache block boundaries along the leading dimension, which will occupy space in the cache memory and will be evicted from the cache memory before being used.
Tiling along the leading dimensions would also reduce the utilization of the loaded \gls{tlb} pages.


Finally, coupled with auto-tuning, our cache block sharing algorithm provides a rich set of run-time configurable options that allow architecture-friendly data access patterns for various setups.
For example, the auto-tuner finds cases when sharing the leading dimension among multiple threads is beneficial to reduce the cache memory pressure, as will be shown in the results Sect.~\ref{sec:results}.

\subsection{Girih framework}
\begin{figure}[tbp]
\centering
\includegraphics[width=10cm]{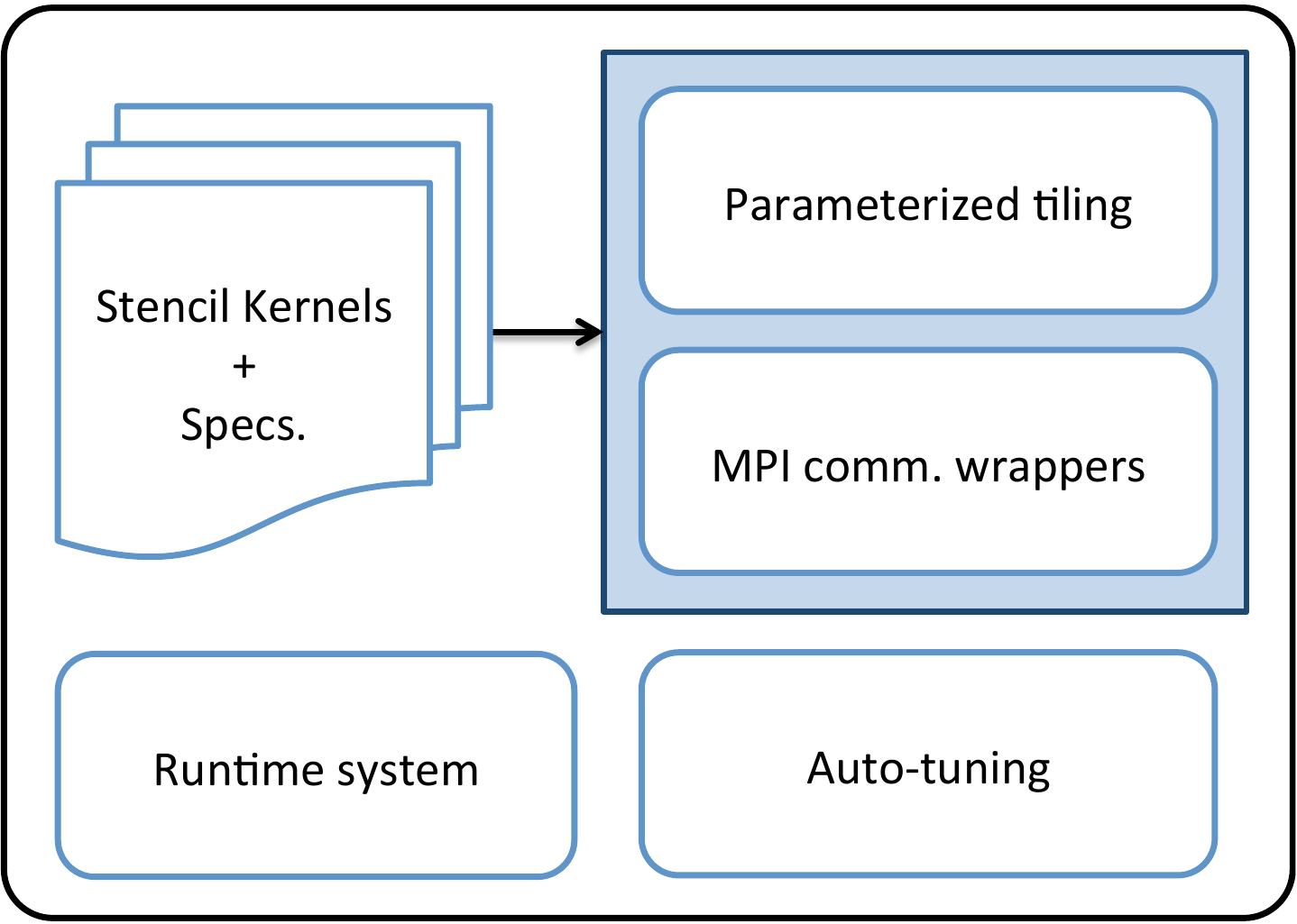}
  \caption{Overview of the Girih framework.}
\label{fig:girih_diagram}
\end{figure}

Our system diagram is shown in Fig.~\ref{fig:girih_diagram}.
It consists of our parametrized tiling kernels that use the loop body of the stencil computation and its specifications, for example, stencil radius, as described in Sect.~\ref{sec:mwd_implementation}.
In Sect.~\ref{sec:runtime} we describe our runtime system, which dynamically schedules thread groups to the tiles.
We use auto-tuning that searches for the best performing parameter set, as described in Sect.~\ref{sec:autotuning}.
Finally, we use \gls{mpi} wrappers to handle the distributed memory communication for the tiles at the boundaries of the subdomain, which is described in more details in~\cite{doi:10.1137/140991133}.

\subsubsection{Multi-core wavefront temporal blocking} \label{sec:mwd_implementation}
\begin{figure}[tbp]
\centering
\includegraphics[width=14cm]{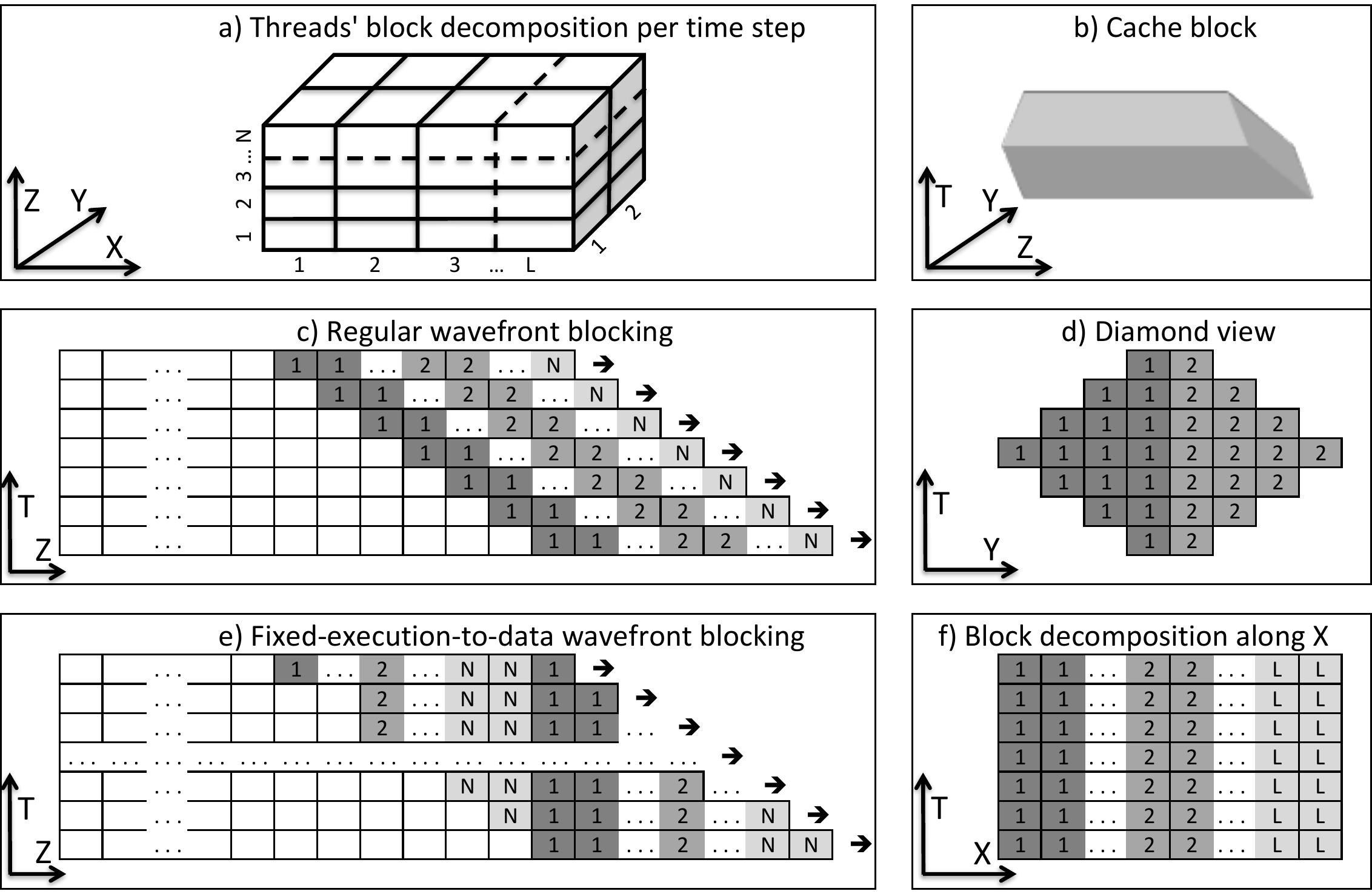}
  \caption{Example of implementing multi-dimensional intra-tile parallelization over wavefront diamond tiles in a three-dimensional grid.}
\label{fig:mwd}
\end{figure}

We present a practical implementation of our approach using \glsreset{mwd}\gls{mwd} in Fig.~\ref{fig:mwd}.
Diamond tiling is performed along the $y$-axis and wavefront blocking is performed along the $z$-axis.
Fig.~\ref{fig:mwd}a shows the tile decomposition among the threads in three spatial dimensions at a single time step of the tile update.
The thread group size is equal to the product of the threads number in all dimensions.
Fig.~\ref{fig:mwd}b shows extruded diamond tile, and Fig. \ref{fig:mwd}c and \ref{fig:mwd}d show two perspectives of the extruded diamond along the $z$- and $y$-axes, respectively.
We use this wavefront-diamond tiling implementation because it uses provably-optimal tiling techniques, as described earlier in Sect.~\ref{sec:motivation}.
Slicing the $x$-axis into small tiles is known to be impractical as it would negatively affect the \gls{tlb} misses, hardware prefetching, and the control-flow overhead.
We replaced tiling along $x$-axis with our intra-tile parallelization along $x$-axis, although it may be useful to combine them in very large grid sizes.

At each time step in the tile update, each thread performs updates of multiple grid points before proceeding to the next time step.
All threads update the equal number of grid points to maintain load-balanced work, and they synchronize at certain points to ensure correctness. 

We present two wavefront parallelization schemes in Fig. \ref{fig:mwd}c and \ref{fig:mwd}e.
The first one assigns a fixed location for the thread in the wavefront tile, allowing the use of a simple relaxed-synchronization scheme in the thread group, where each thread has data dependency over its left neighbor only.
This wavefront approach has the disadvantage of pipelining the data across the cores, resulting in larger data volume transfers in the cache hierarchy.
The parallelization scheme in Fig. \ref{fig:mwd}e resolves this issue by using a \glsreset{fed} \gls{fed} scheme that updates each subset of grid points by a single thread, regardless of the wavefront tile location, allowing wavefront cache block to span multiple cache domains.
The limitation of this scheme is the more complex synchronization within the thread group, so we use OpenMP barrier after each time step update in this \gls{fed} wavefront.
The \gls{fed} wavefront scheme is useful for stencils with high number of bytes per grid cell, such as the variable coefficient and high-order stencils.
It is particularly important for high-order stencils because almost no reuse is usually achieved between the time steps of the wavefront due to the steep temporal blocking slope. 


Our intra-tile parallelization scheme along the wavrfront in Fig.~\ref{fig:mwd}c and \ref{fig:mwd}e allows an arbitrary number of threads to be assigned along the $z$-axis, with the cost of increasing the cache block size, without increasing the data reuse.
An arbitrary number of threads can also be assigned along the $x$-axis of the tile.
The parallelization along the $x$-axis replaces the tiling along this dimension, when tiling would be useful to reduce the cache block size of the thread.
The diamond tile, shown in Fig.  \ref{fig:mwd}d, uses only one or two threads of parallelization to maintain load-balance and assign the same data to each thread (i.e., use a tile hyperplane parallel to the time dimension).

Relaxed synchronization is used along the $z$-axis in the regular wavefront approach (Fig.~\ref{fig:mwd}c).
A software sub-group barrier is used to synchronize threads along $y$- and $x$-axes for each sub-tile along the $z$-axis.

Listing~\ref{lst:mwd_alg} shows a reduced version of our \gls{mwd} tiling implementation.
The code synchronizes the thread group using an OpenMP barrier after each time step and uses the regular wavefront blocking strategy (described in Fig.~\ref{fig:mwd}c).
The full kernel, the relaxed-synchronization variant, and the \gls{fed} variant are available online in~\cite{malas_girih}.
We use an OpenMP parallel region to spawn the threads of the thread group.
The three-dimensional coordinates of the threads are calculated in line 4.
The tile is decomposed among the threads along the $y$-, $x$-, and $z$-axes in lines 7-8, 10-12, and 14, respectively.

\begin{minipage}[th]{\linewidth}
\begin{lstlisting}[style=lst_alg,label=lst:mwd_alg,caption={A variant of MWD for wavefront steady-state update.
 Tile's bounds along $x$, $z$, and time are \texttt{xb-xe, zb-ze, tb-te}, respectively.
  $y$ base bound is \texttt{yb-ye}.}]
#pragma omp parallel for ...
{ tid = omp_get_thread_num();
  //Thread 3D coordinates: tid = tid_z*(th_x*th_y) + tid_y*th_x + tid_x
  tid_x=tid%th_x;  tid_y=(tid/th_x)%th_y; tid_z=tid/(th_x*th_y);
  //Decompose the tile along the y-axis
  //b_inc and e_inc update tile size for each time step using stencil radius (R)
  if (tid_y == 0){ ye=(yb+ye)/2; b_inc=R; e_inc=0;}
  else           { yb=(yb+ye)/2; b_inc=0; e_inc=R;}
  //Decompose the tile along the x-axis
  q=(xe-xb)/th_x; r=(xe-xb)%th_x;
  if(tid_x<r) { ib=xb+tid_x*(q+1); ie=ib+(q+1);}
  else        { ib=xb+r*(q+1)+(tid_x-r)*q; ie=ib+q;}
  //Decompose the tile  along the z-axis, of size bs_z
  zbi = bs_z/th_z * tid_z;  zei = bs_z/th_z * (tid_z+1);
  for(zi=zb; zi<ze; zi+=bs_z) { //wavefront loop along z-axis
    ybi=yb; yei=ye; kt=zi; //Tile's Base index init. along y- and z-axes
    for(t=tb; t<te; t++){ //Tile loop in time
      for(k=kt+zbi; k<kt+zei; k++){ //Tile loop in z
        for(j=ybi; j<yei; j++) { //Tile loop in y
#pragma simd
          for(i=ib; i<ie; i++) { //Tile loop in x
            stencil_computations_macro()
      }}}
      //Block size update along y-axis
      if(t < diam_height/2) ybi-=b_inc; yei+=e_inc; //Lower half of the diamond
      else                  ybi+=e_inc; yei-=e_inc; //Upper half of the diamond
      kt -= R; // Update wavefront base index for each time step
#pragma omp barrier //Synchronize after each time step
      tmpp=u;  u=v;  v=tmpp; // Pointers swapping
}}}
\end{lstlisting}
\end{minipage}

\subsubsection{Auto-tuning} \label{sec:autotuning}

We perform the auto-tuning as a preprocessing step, once the user selects the problem details, for example, the stencil type and grid size.
The auto-tuner allocates and initialize the required arrays for its own use and deallocate them once it is completed.
This saves significant tuning time when the grid array is very large, in the order of 100 GiB.

\begin{figure}[tbp]
\centering
\includegraphics[width=14cm]{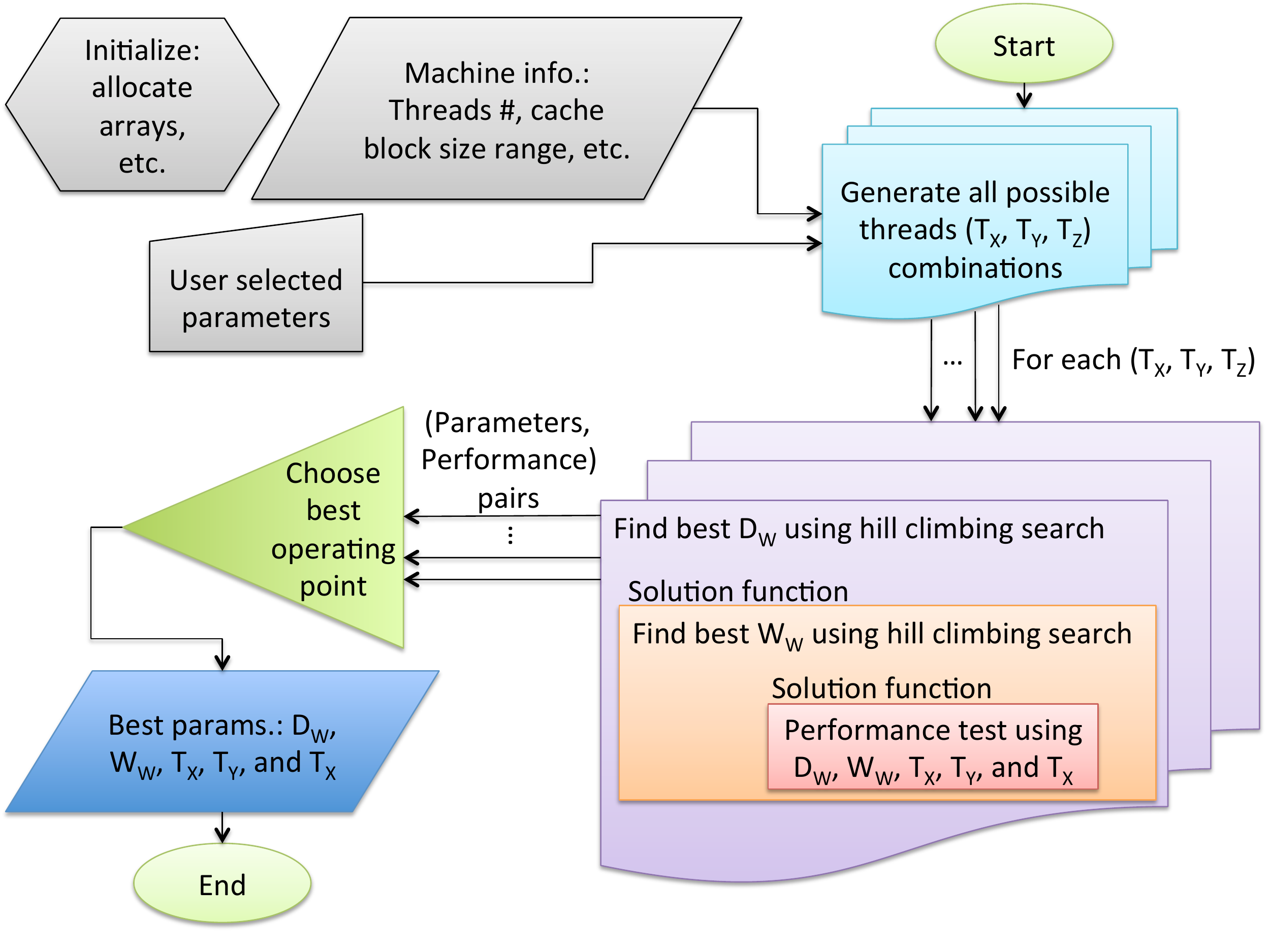}
  \caption{Girih auto-tuner flow chart.}
\label{fig:tuner}
\end{figure}

Fig. \ref{fig:tuner} shows the details of our auto-tuning approach.
It tunes several parameters: diamond width, wavefront tile width, threads along $x$-, $y$-, and $z$-axes.
The auto-tuner start with fixed user-selected parameters, then determine the feasible set of of intra-tile threads dimensions by factorizing the available number of threads.
It tests the performance of all valid TGS combinations.
A local search hill-climbing algorithm is applied to tune the diamond and wavefront tile widths for each TGS case.
The auto-tuner uses our cache block size model in Section~\ref{sec:blk_size_model} and the processor's available cache size (specified by the user) to reduce the parameter search space.

Selecting proper test size, i.e., number of time steps, for auto-tuning test cases is challenging.
A very small test may be affected by the system jitter and other sources of noise, which produces a false indication of the achievable performance of the test case.
A large test, on the other hand, may increase the tuning time significantly.
We use multiples of diamond rows to set the number of time steps.
The run time of a single diamond row has many dependencies.
It relies on the grid size, stencil type, processor speed, tiling efficiency, etc., so a priori setting of the test size is not practical.
To resolve this issue, for each test case, we dynamically change the test size until ``acceptable performance'' is obtained.
We repeat running the test case with increasing number of diamond rows.
Once the performance variation between two repetitions falls within certain threshold, we consider that the largest test case size produces ``acceptable performance'' and we use it to report the test case performance.

\subsubsection{Runtime system} \label{sec:runtime}
Threads scheduling to diamond tiles can be handled in a variety of ways.
Since each row of diamond tiles can be updated concurrently, a barrier can be placed after updating each row of diamond tiles to avoid violating the data dependencies~\cite{orozco2009ICPP}.
The runtime can schedule the diamond tiles to threads statically before running the code, with considerations to the diamond tiles data dependency during runtime~\cite{Strzodka:2011:CAT}.
When the workload is balanced, these approaches can be sufficient, as all the threads complete their tile updates at the same time.
On the other hand, when the workload is not balanced, for example, due to MPI communication of the subdomain boundaries, dynamic load balancing becomes necessary to maximize utilizing the threads time.
We use a multi-producer multi-consumer \gls{fifo} queue to perform the dynamic scheduling of tiles to thread groups.
The queue holds the available tiles to update.
When a thread group updates a tile, it pushes the next tiles that have no more unmet dependencies to the queue.
When a tile from the queue is assigned to a thread group, the runtime performs ``pop'' operation to it.
A critical region is used to prevent race conditions.
The synchronization cost is negligible because each extruded diamond update involves updating millions of grid points.


Our implementation relies on OpenMP nested parallelization.
The outer parallel region spawns one thread per thread group (we call them ``group masters''), where the tiles are scheduled to thread groups at this level.
The inner parallel region spawns the threads of each group to update the grid cells in the tile.
Spawning the inner parallel region happens in undefined order many time during runtime.
As a result, threads are grouped in different subsets during runtime.

For example, let us consider a case of running six threads in two groups, with pinning order \{0,3,1,2,4,5\}.
The outer parallel region uses threads 0 and 3 for the group master threads.
When thread 0 spawns its threads first, the thread are divided into \{0,1,2\} and \{3,4,5\} subsets.
Threads subset may also be divided into \{0,4,5\} and \{3,1,2\} subsets.

Grouping threads in different subsets does not affect the performance in \gls{uma} setups, as in single socket Ivy Bridge processor.
On the other hand, this issue has to be resolved in \gls{numa} cases, such as in the Intel \gls{knc} or when running systems with multiple processors.

OpenMP 4.0 provides thread affinity features.
The \verb.OMP_PLACES. environment variable allows the user to set thread groups or pools of threads.
It also supports thread groups binding in nested parallel regions through the \verb.proc_bind. clause in the parallel region OpenMP pragma.
The \verb.proc_bind. feature in the parallel region initialization allows for setting a certain affinity.
For example, it is possible to use \verb.scatter. option at the outer parallel region, which uses 1 thread from each thread group. 
Then use \verb.master. option at the inner parallel region, so that each thread is spawned from the same thread group as its parent thread. 
Unfortunately, using this feature in The Intel C compiler 15 degraded the performance of our code to 80\% at a single socket.
The performance degradation may be a result of sub-optimal use of these features or the Intel implementation is not well optimized yet for these new features.

To resolve this issue efficiently, we use the UNIX low-level interface of \verb.sched_setaffinity. to set the threads affinity manually. 
Thread affinity is set inside the inner parallel region every time it is executed. It does not hurt the performance much, as there core regions have work in order of hundreds of micro seconds.

It is possible to avoid the thread groups affinity issue by using the pthreads library instead of OpenMP, where thread pinning is needed only once at the initialization stage.
We believe the pinning overhead in OpenMP is negligible, so we use it to exploit its portability and usability features.

Another option would be to use single level OpenMP parallel region, but this adds complexity to the thread group scheduling.
It also requires creating point-to-point synchronization constructs within each thread group, instead of using OpenMP barriers at the thread groups.

We use memory aligned allocation with array padding at the leading dimension using a parameterized runtime variable.
This is mainly targeted for Intel \gls{mic} processor, which has long vector units and the data alignment has high impact on the performance.

\section{Results} \label{sec:results}

We perform our experiments over the four stencils under consideration using Intel 10-core Ivy Bride and 18-core Haswell processors, over a broad range of grid sizes (cubic domain).
Our \gls{mwd} approach is faster than \gls{swd}/CATS~\cite{Strzodka:2011:CAT}, PLUTO~\cite{Bandishti6468470,Bondhugula2008}, and Pochoir~\cite{Tang2011} for all stencils on both architectures with most grid sizes.
\gls{mwd} is also the only approach that provides a significant improvement over an efficient spatial blocking implementation in all cases.
To understand the strength of our approach in potentially memory-starved future processors, we study the impact of MWD cache block sharing over the performance, code balance, memory transfer volumes, and energy consumption.
Finally, we show significant memory bandwidth and memory transfer volumes saving.

\subsection{Frameworks setup}
We set up both PLUTO and Pochoir frameworks in the presented results.
To ensure fair comparison, we investigated and used the best setup of the tested frameworks.

In all the experiments, we run each test case twice and report the performance of the faster one.
Since the test cases are large enough, the repeated tests achieve very similar performance.
In Sects.~\ref{sec:igs_results} and~\ref{sec:tgs_results}, we run the experiments several times to measure different hardware counter groups.
To demonstrate the stability of our results, we plot the reported performance of the repeated experiments in our performance figures.
In most cases, the results are aggregated in a single point.
At some small grid sizes, where the noise is relatively more significant, the performance points are stretched vertically.

\subsubsection{PLUTO setup}
PLUTO framework uses \texttt{polycc} executable to perform the source-to-source transformation.
We use the flags ''\texttt{--tile --parallel --pet --partlbtile}'' to generate efficient codes for our stencil computation kernels.
The Intel compiler flags are:''-O3 -xHost -ansi-alias -ipo -openmp'', which are also the default configuration in PLUTO examples.
The tests use PLUTO development version with commit tag \texttt{82b03fbc0d734a19fd122cbe568167cd911068ea} on 28 July 2015\footnote{The commit link: http://repo.or.cz/pluto.git/commit/82b03fbc0d734a19fd122cbe568167cd911068ea}.

Compiling the tiled stencil codes using the Intel C compiler 15 achieves twice the performance compared to the Intel C compiler 12.
The more recent compiler optimizes PLUTO tiled codes more efficiently and have improvements in vectorizing and using prefetching in the compiled codes.

The selected tiling transformations perform diamond tiling along the $z$-axis and parallelepiped tiling along the $y$- and $x$-axes.
Since parameters tuning is essential to run the tiled code efficiently, we implemented a Python script to tune the parameters of each test case in the presented results.
We performed brute-force parameters search in diverse setups (i.e., different stencils, processors, and grid sizes) to ensure fair and efficient tuning.
We found that the parameters search space is convex, where only single maximum exist in the tested processors and stencil kernels.
In the results presented here, we use a recursive local search algorithm, in the same manner discussed earlier in \emph{Girih} auto-tuner for the \gls{mwd} code, to tune the three tiling parameters.

\subsubsection{Pochoir setup}

Pochoir does not have performance-critical tuning parameters for tiling, as it relies on cache-oblivious algorithms.
The default Intel compiler flags in Pochoir examples are used: ''-O3 -funroll-loops -xHost -fno-alias -openmp''.

\subsubsection{Girih setup}

We use Intel compiler options: ''-O3 -xAVX -fno-alias -openmp'' in \emph{Girih} codes.
Our auto-tuning code is used to select the diamond and wavefront tiles widths and thread group parallelization parameters in the results.

The wavefront implementation parameter is selected according to the stencil type.
A relaxed-synchronization wavefront scheme is used for the 7-point stencils, as this implementation has lower synchronization overhead.
\gls{fed} is not important here as sufficient data reuse is possible using reasonable wavefront width.
On the other hand, \gls{fed} is used in the 25-point stencils, to allow sufficient data reuse in the wavefront update.
The efficiency of these options are also confirmed by manually testing several cases.

We also tune the 1WD case separately.
This is the nearest implementation we have to CATS2 algorithm of \cite{Strzodka:2011:CAT}.

\subsection{Performance at increasing grid size} \label{sec:igs_results}

In this section we present a performance comparison between MWD,
Pluto, Pochoir, 1WD/CATS, and optimal spatial blocking over a wide
range of problem sizes (cubic domain) for the four stencils under
consideration.  For a better identification of relevant bottlenecks we
also show memory bandwidth and memory transfer volume per \LUP\ (see
Sect.~\ref{sec:code_balance_energy_analysis} for a thorough code
balance analysis).

Our major finding is that MWD is faster than 1WD, PLUTO, and Pochoir
for all stencils on both architectures with most problem sizes. MWD is
also the only approach that provides a significant improvement over
optimal spatial blocking in all cases. Especially for the high-order
(25-point) stencils it is the only efficient solution.  In general, the
Haswell CPU shows better speedups for temporal blocking vs.\ optimal
spatial blocking due to its large number of cores (18), which leads to
a low machine balance of $0.15\,\bytes/\flop$ (assuming fused
multiply-add is not used; with FMA, the machine balance goes down further to
$0.075\,\bytes/\flop$). In contrast, the Ivy Bridge CPU only has 10
cores, a 5\% lower clock speed, and a 20\% lower memory bandwidth for
a better machine balance of $0.22\,\bytes/\flop$. A low machine
balance generally indicates a more ``bandwidth-starved'' situation with
a higher potential for temporal blocking techniques, although
a quantitative analysis requires a more elaborate performance model.

At large problem sizes, where boundary effects become negligible, 
it is possible to construct a phenomenological
ECM performance model by
combining measured data traffic per \LUP\ in all memory levels with
code composition characteristics such as the number of load
instructions per \LUP, as described in Sect.~\ref{sec:models}.
Tables~\ref{tab:ecm-ivb} and \ref{tab:ecm-hsw} summarize the comparison
between the phenomenological models and the measurements. Details are
discussed below.
\begin{table}[tbp]
\centering\renewcommand{\arraystretch}{1.2}
\begin{tabular}{c|ccc}
Stencil            & Model [\cycles] & Pred. [\GLUPS] & Meas. [\GLUPS] \\\hline
7-pt const. coeff. & \ecm{12}{14}{14}{8.3}{2.2}{} & 4.6 & 4.1 \\
7-pt var. coeff.   & \ecm{14}{28}{30}{24}{11}{}   & 1.6 & 1.4    \\
25-pt const. coeff.& \ecm{12}{56}{40}{28}{11}{}   & 1.3 & 1.2     \\
25-pt var. coeff.  & \ecm{12}{76}{115}{50}{40}{}  & 0.44& 0.36   \\ 
\end{tabular}
\caption{\label{tab:ecm-ivb}Phenomenological ECM models, predictions, and performance measurements for the four 
  stencils under investigation with MWD at large grid sizes on the Intel Ivy Bridge CPU.}
\end{table}
\begin{table}[tbp]
\centering\renewcommand{\arraystretch}{1.2}
\begin{tabular}{c|ccc}
Stencil            & Model [\cycles] & Pred. [\GLUPS] & Meas. [\GLUPS] \\\hline
7-pt  const. coeff. &  \ecm{12}{14}{7}{7.5}{1.8}{} & 10 & 8.0 \\
7-pt  var. coeff.   &  \ecm{14}{21}{14}{25}{4.8}{} & 3.9 & 2.6   \\
25-pt  const. coeff.&  \ecm{12}{56}{20}{30}{7.4}{} & 2.5 & 2.2    \\
25-pt  var. coeff.  &  \ecm{12}{38}{56}{50}{26}{}  & 0.71& 0.65    \\ 
\end{tabular}
\caption{\label{tab:ecm-hsw}Phenomenological ECM models, predictions, and performance measurements for the four 
  stencils under investigation with MWD at large grid sizes on the Intel Haswell CPU.}
\end{table}

\subsubsection{7-point stencil with constant coefficients}\label{sec:7pt_const_igs_all}

\begin{figure}[tbp]
    \centering
    \subfloat[Performance.]{
        \centering
        \includegraphics[width=4.2cm]{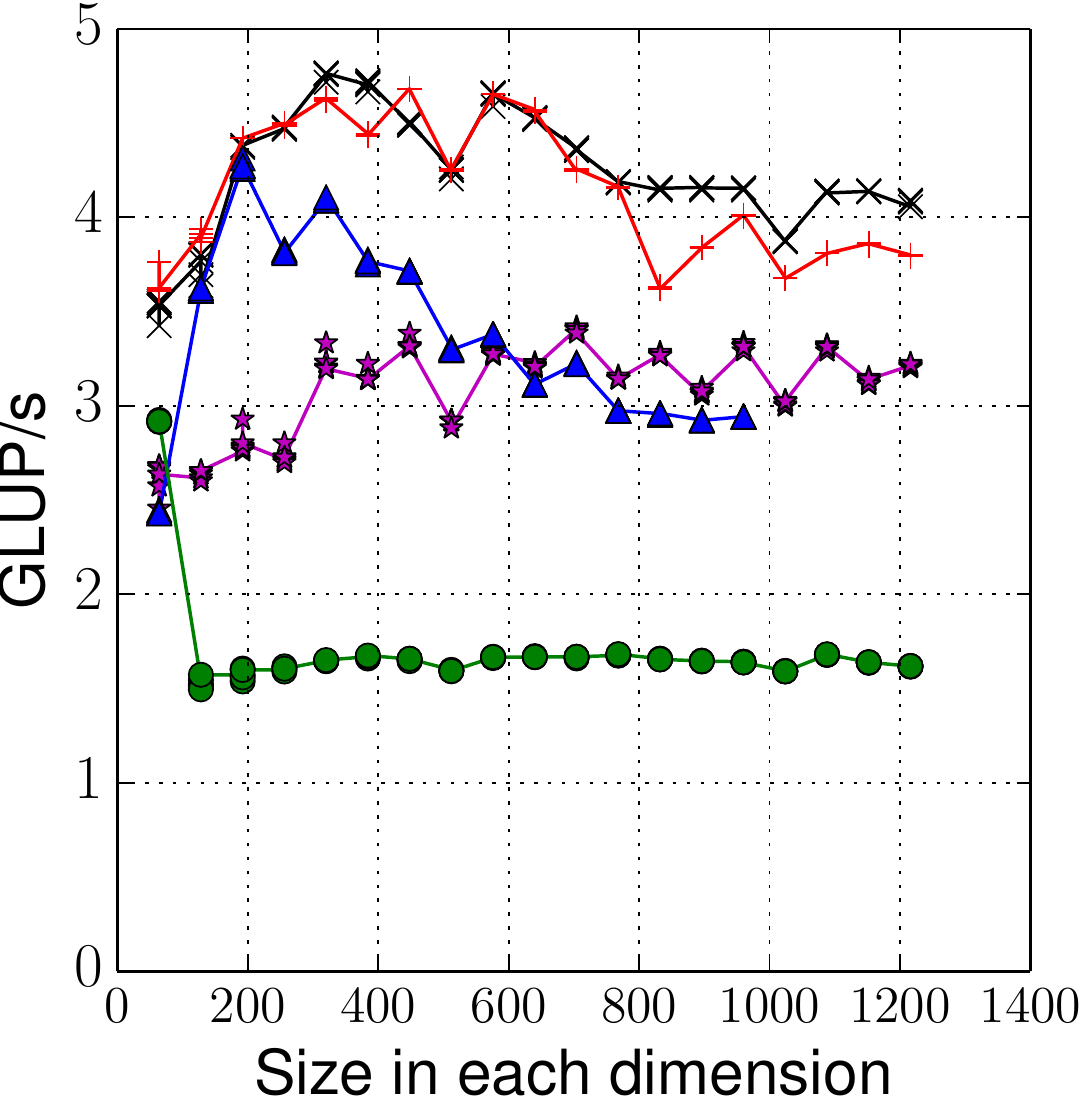}
        \label{fig:ivb_7_pt_const_perf_igs_all}
    }
    \enskip
    \subfloat[Memory bandwidth.]{
        \centering
        \includegraphics[width=\sfwidth]{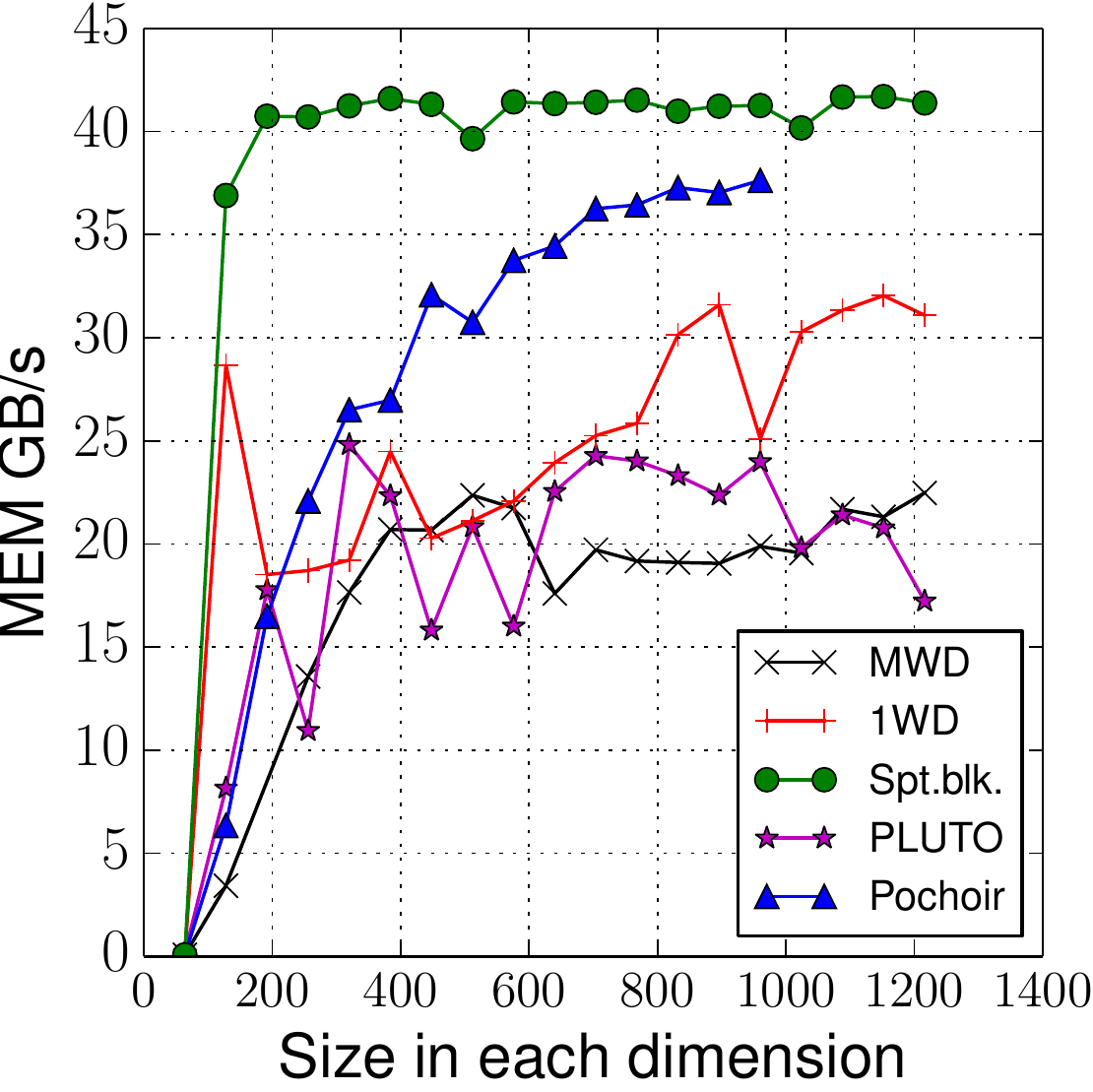}
        \label{fig:ivb_7_pt_const_mem_bw_igs_all}
    }
    \enskip
    \subfloat[Memory transfer volume.]{
        \centering
        \includegraphics[width=\sfwidth]{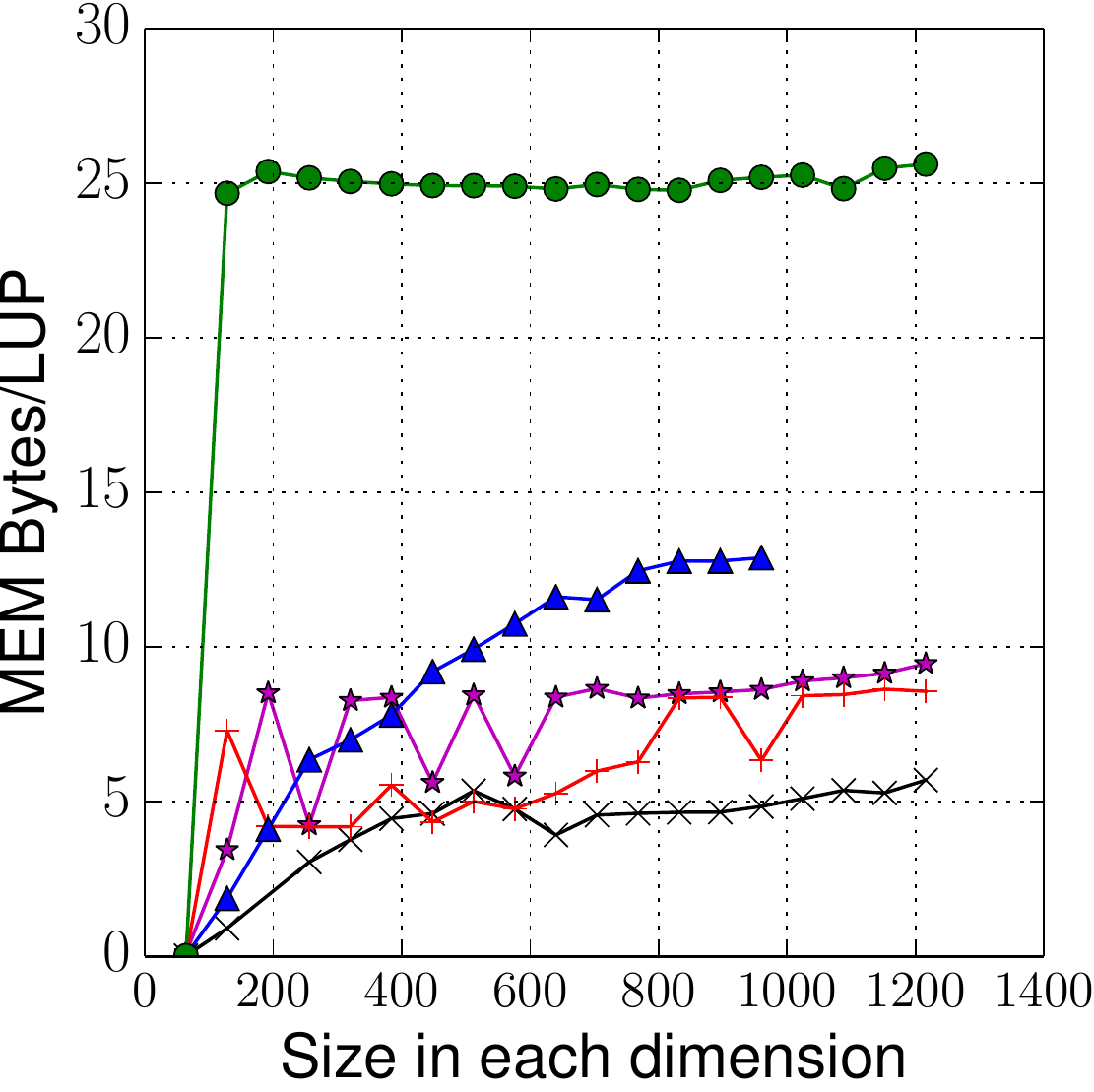}
        \label{fig:ivb_7_pt_const_mem_vol_igs_all}
    }
    \caption{Ivy Bridge 7-point constant-coefficient stencil results, using increasing cubic grid size. Showing performance and memory transfer measurements of \gls{mwd}, PLUTO, Pochoir, \gls{swd}, and spatial blocking.}
    \label{fig:ivb_7_pt_const_igs_all}
\end{figure}

\begin{figure}[tbp]
    \centering
    \subfloat[Performance.]{
        \centering
        \includegraphics[width=\sfwidth]{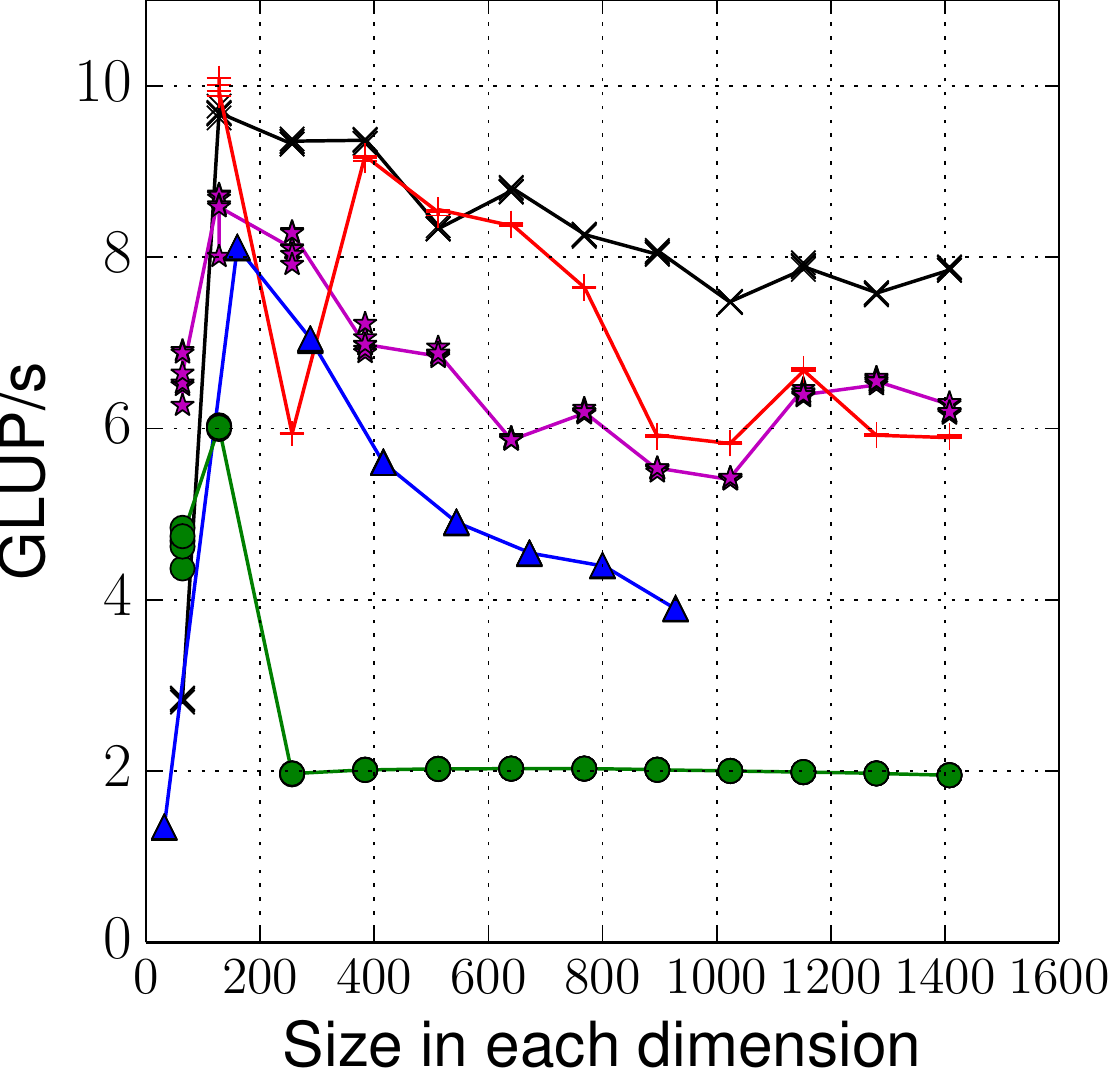}
        \label{fig:hw_7_pt_const_perf_igs_all}
    }
    \enskip
    \subfloat[Memory bandwidth.]{
        \centering
        \includegraphics[width=\sfwidth]{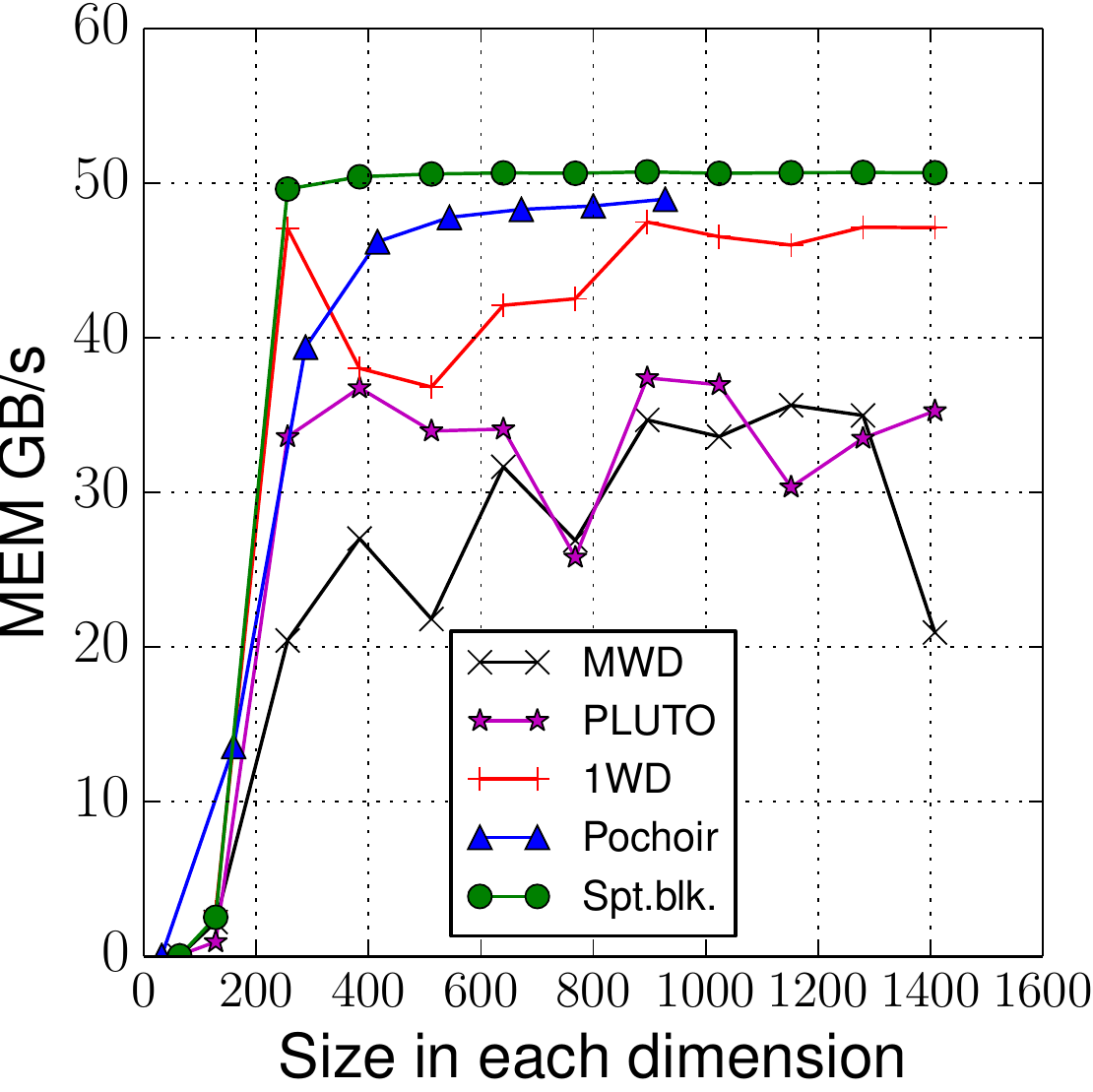}
        \label{fig:hw_7_pt_const_mem_bw_igs_all}
    }
    \enskip
    \subfloat[Memory transfer volume.]{
        \centering
        \includegraphics[width=\sfwidth]{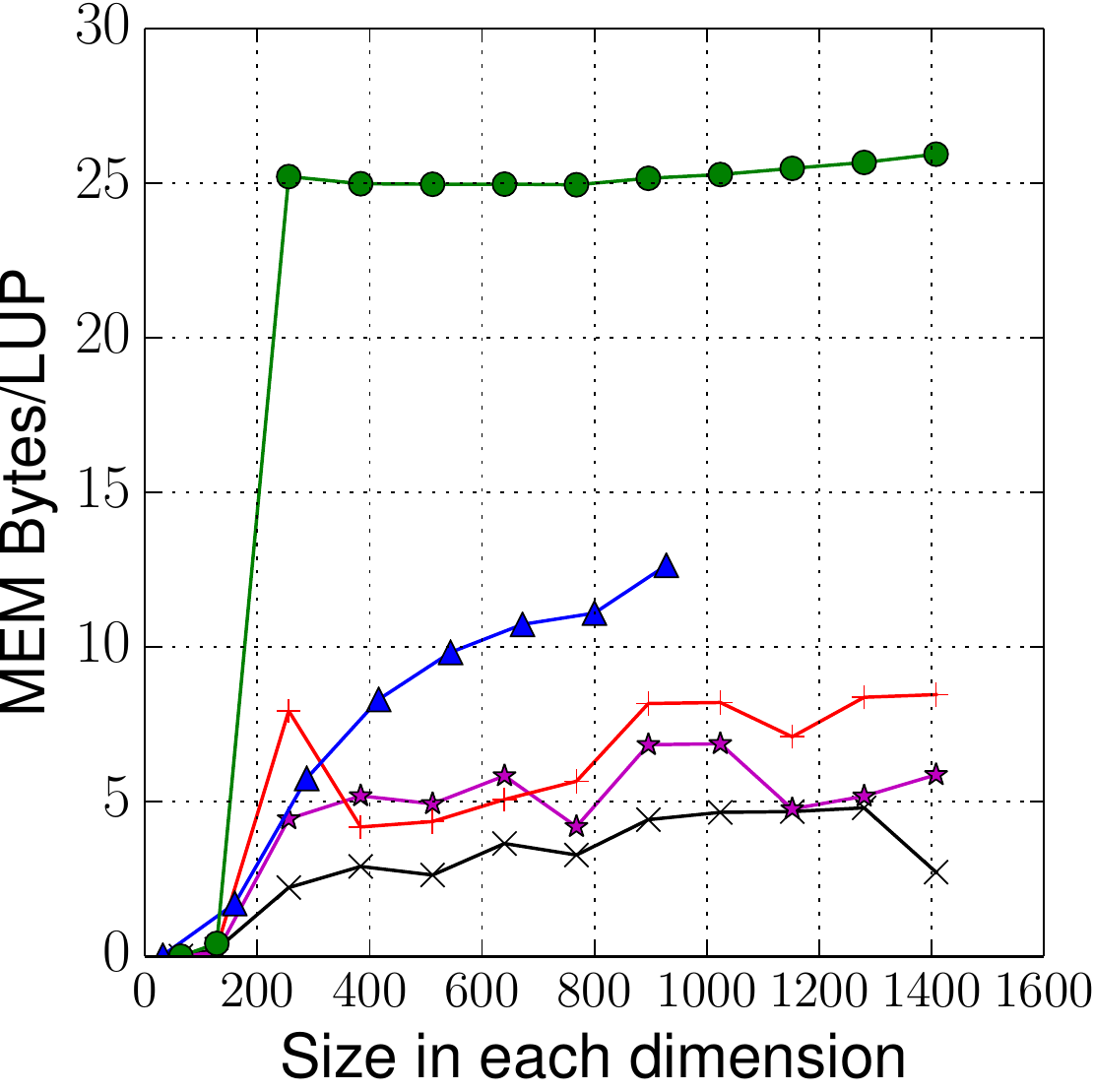}
        \label{fig:hw_7_pt_const_mem_vol_igs_all}
    }
    \caption{Haswell 7-point constant-coefficient stencil results, using increasing cubic grid size. Showing performance and memory transfer measurements of \gls{mwd}, PLUTO, Pochoir, \gls{swd}, and spatial blocking.}
    \label{fig:hw_7_pt_const_igs_all}
\end{figure}

The results for the 7-point stencil with constant coefficients
are shown in Figs. \ref{fig:ivb_7_pt_const_perf_igs_all},
\ref{fig:ivb_7_pt_const_mem_bw_igs_all}, and
\ref{fig:ivb_7_pt_const_mem_vol_igs_all} for the Ivy Bridge
CPU and in 
Figs. \ref{fig:hw_7_pt_const_perf_igs_all},
\ref{fig:hw_7_pt_const_mem_bw_igs_all}, and
\ref{fig:hw_7_pt_const_mem_vol_igs_all} for the Haswell CPU.

This stencil performs seven \flops\ per \LUP\ and has a minimum
code balance for pure spatial blocking of $24\,\bytes/\LUP$ in
double precision. The memory-bound maximum performance in the latter 
case is thus $(41\,\GBS)/(24\,\bytes/\LUP)\approx1.7\,\GLUPS$. If non-temporal stores
were used the code balance would go down to $16\,\bytes/\LUP$, leading to a 
maximum performance of about $2.6\,\GLUPS$. However, this is the only
stencil of the four that can profit from non-temporal stores
in a significant way. 

All temporal blocking variants outperform optimal spatial blocking by far.
MWD and 1WD are consistently faster than PLUTO and Pochoir, with MWD taking
a clear lead for larger problems. 
MWD also exhibits the lowest memory bandwidth and the lowest code balance
(memory traffic per \LUP). For most problem sizes the autotuner
selects a thread group size of 2 or 6, except for very large and very
small problems.

 The 7-point stencil with constant coefficients
uses 28 half-wide (16-byte) load instructions per 8\,\LUPs, which
require 14 cycles to execute on both architectures. The required data
traffic in the L2 cache is $56\,\bytes/\LUP$, since the L1 cache is
too small to hold three consecutive rows of the source array. Hardware
counter measurements confirm this estimate for MWD. On Ivy Bridge
(Haswell) this transfer takes 14 (7) cycles. The L3 cache traffic is
hard to predict due to the blocking strategy; we use the measured
value of 35 (30) \bytes/\LUP\ on Ivy bridge (Haswell), leading to a
transfer time of about 9 (4) cycles. In memory we observe a code
balance of about $5\,\bytes/\LUP$. With the chosen input data the ECM
model predicts a socket-level performance of about $4.5\,\GLUPS$
on Ivy Bridge and $10\,\GLUPS$ on Haswell. While there is no memory 
bandwidth bottleneck on Ivy Bridge (expected saturation at 18 cores), 
the model predicts bandwidth saturation at 17 cores on Haswell. 
Looking at the measured performance data we see that the prediction 
is within a 10\% margin of the measurements on Ivy Bridge and about 
20\% above the measurements on Haswell. The larger deviation on
Haswell is expected because the ECM model is known to be over-optimistic
near the saturation point. Overall the ECM model describes the
performance characteristics of the MWD code at larger problem sizes
quite well, proving that the MWD code is operating at the limits
of the hardware. This is also true for the other stencils described below.
Further substantial performance improvements could only be expected
from optimizations that reduce the amount of work. 

\subsubsection{7-point stencil with variable coefficients}\label{sec:7pt_var_igs_all}

\begin{figure}[tbp]
    \centering
    \subfloat[Performance.]{
        \centering
        \includegraphics[width=\sfwidth]{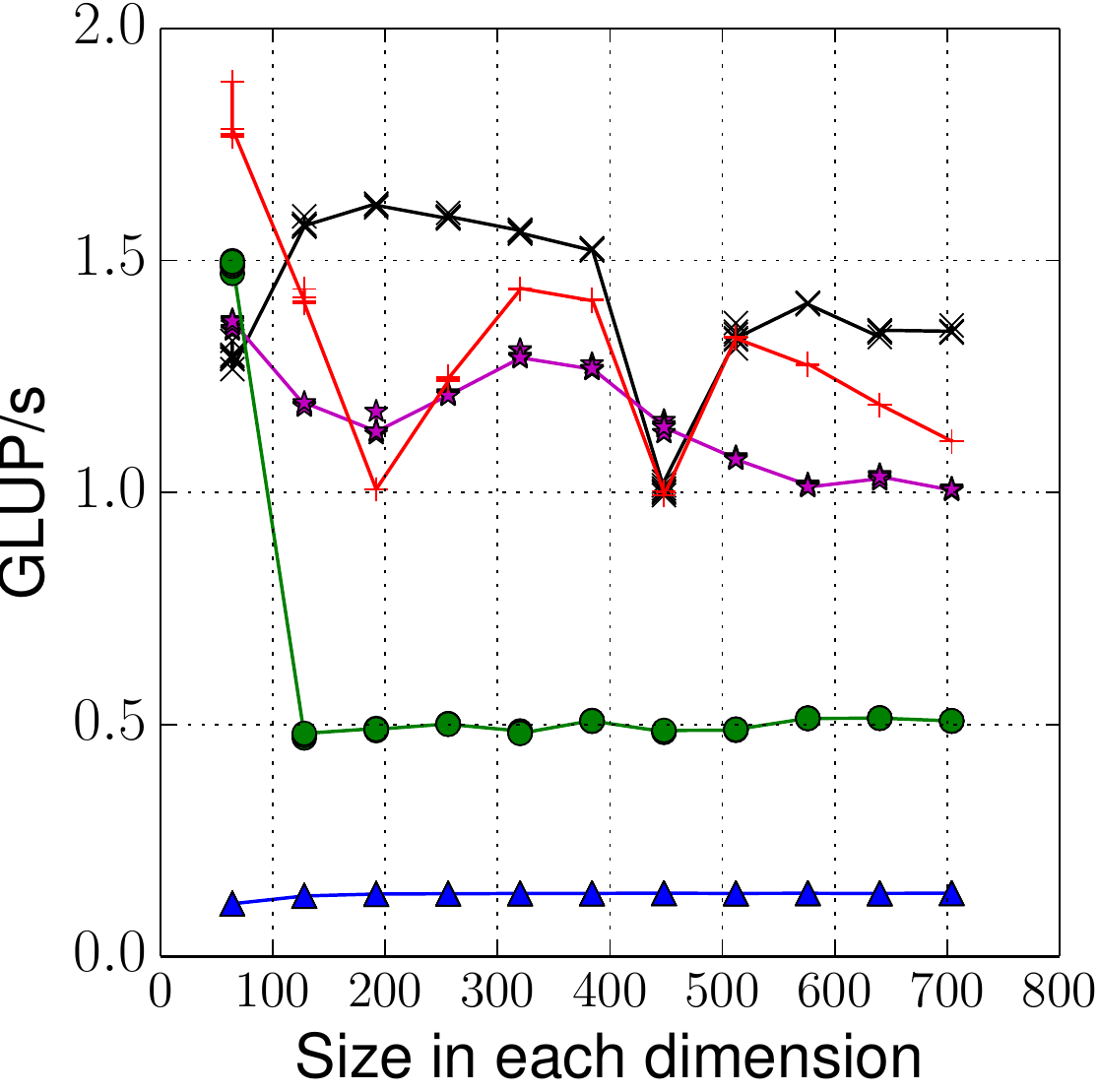}
        \label{fig:ivb_7_pt_var_perf_igs_all}
    }
    \enskip
    \subfloat[Memory bandwidth.]{
        \centering
        \includegraphics[width=\sfwidth]{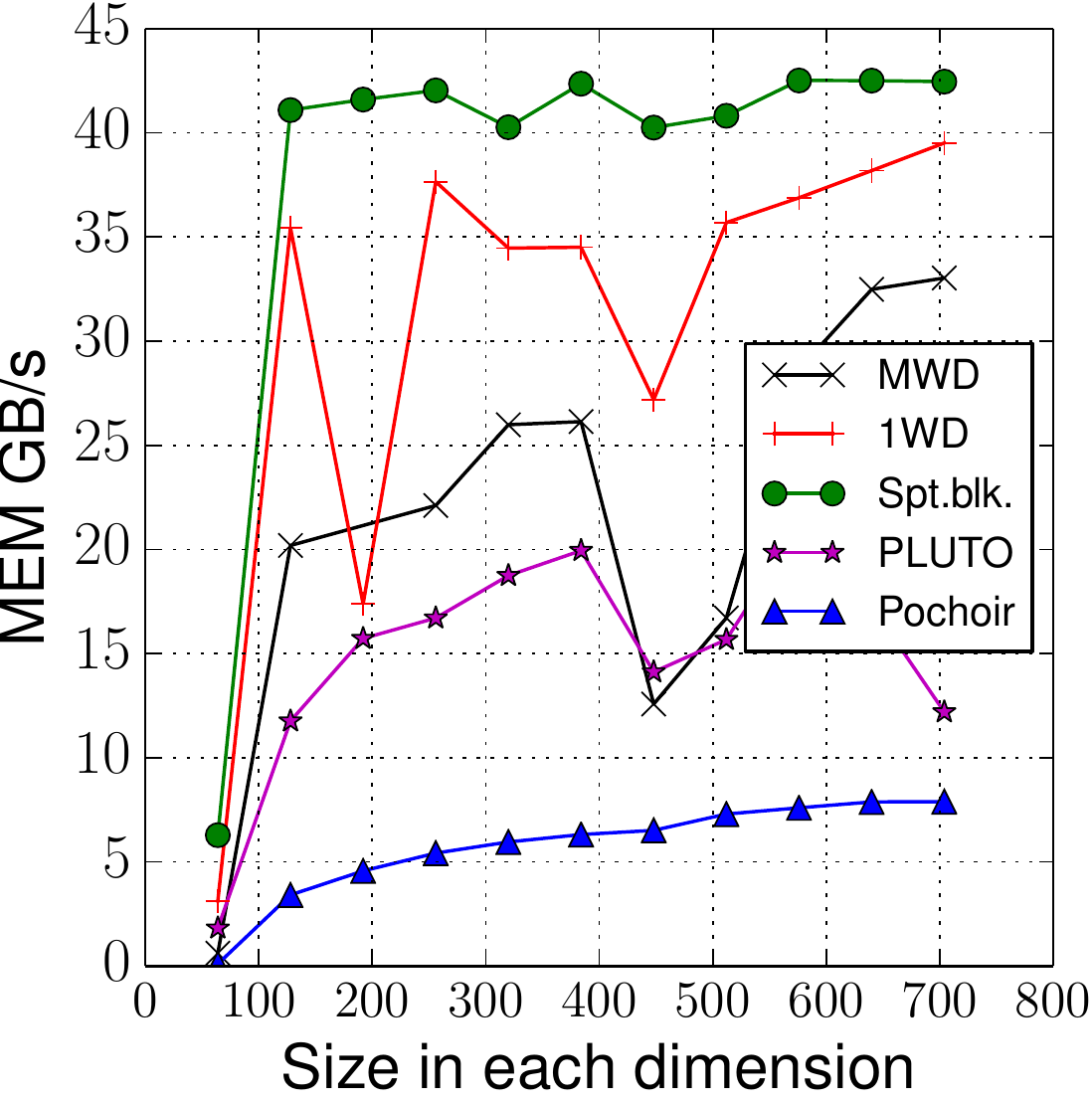}
        \label{fig:ivb_7_pt_var_mem_bw_igs_all}
    }
    \enskip
    \subfloat[Memory transfer volume.]{
        \centering
        \includegraphics[width=\sfwidth]{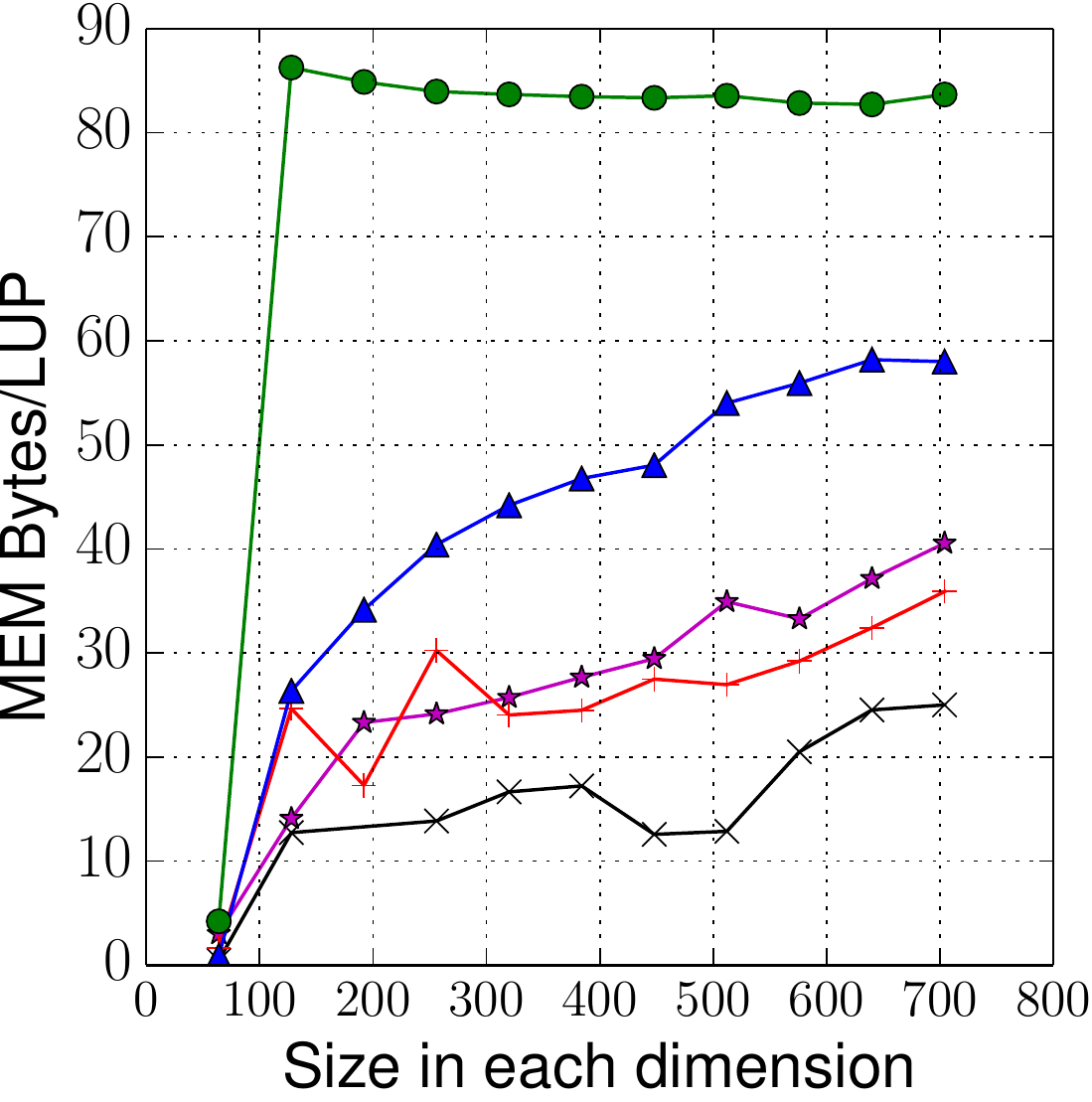}
        \label{fig:ivb_7_pt_var_mem_vol_igs_all}
    }
    \caption{Ivy Bridge 7-point variable-coefficient stencil results, using increasing cubic grid size. Showing performance and memory transfer measurements of \gls{mwd}, PLUTO, Pochoir, \gls{swd}, and spatial blocking.}
    \label{fig:ivb_7_pt_var_igs_all}
\end{figure}

\begin{figure}[tbp]
    \centering
    \subfloat[Performance.]{
        \centering
        \includegraphics[width=\sfwidth]{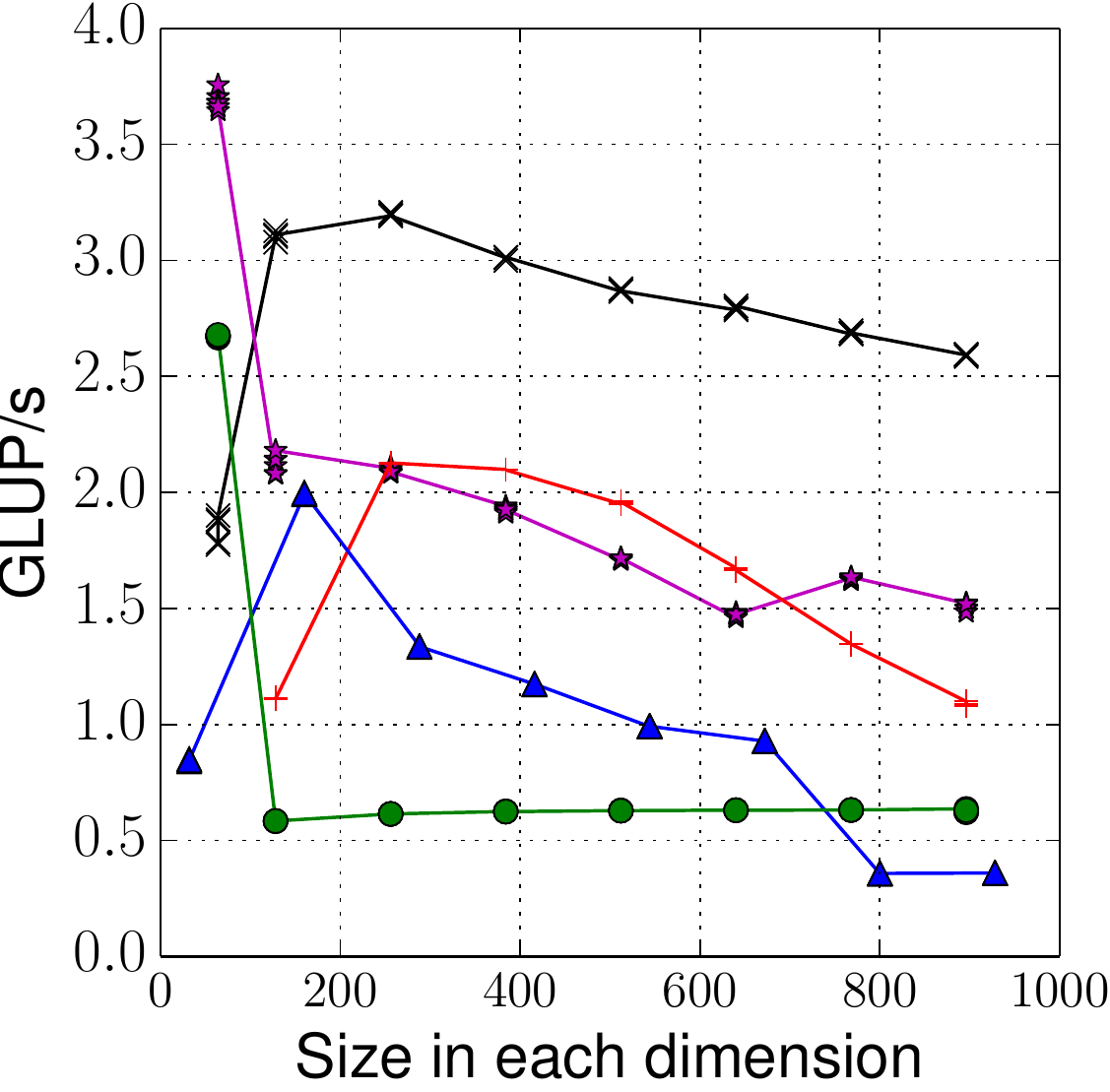}
        \label{fig:hw_7_pt_var_perf_igs_all}
    }
    \enskip
    \subfloat[Memory bandwidth.]{
        \centering
        \includegraphics[width=\sfwidth]{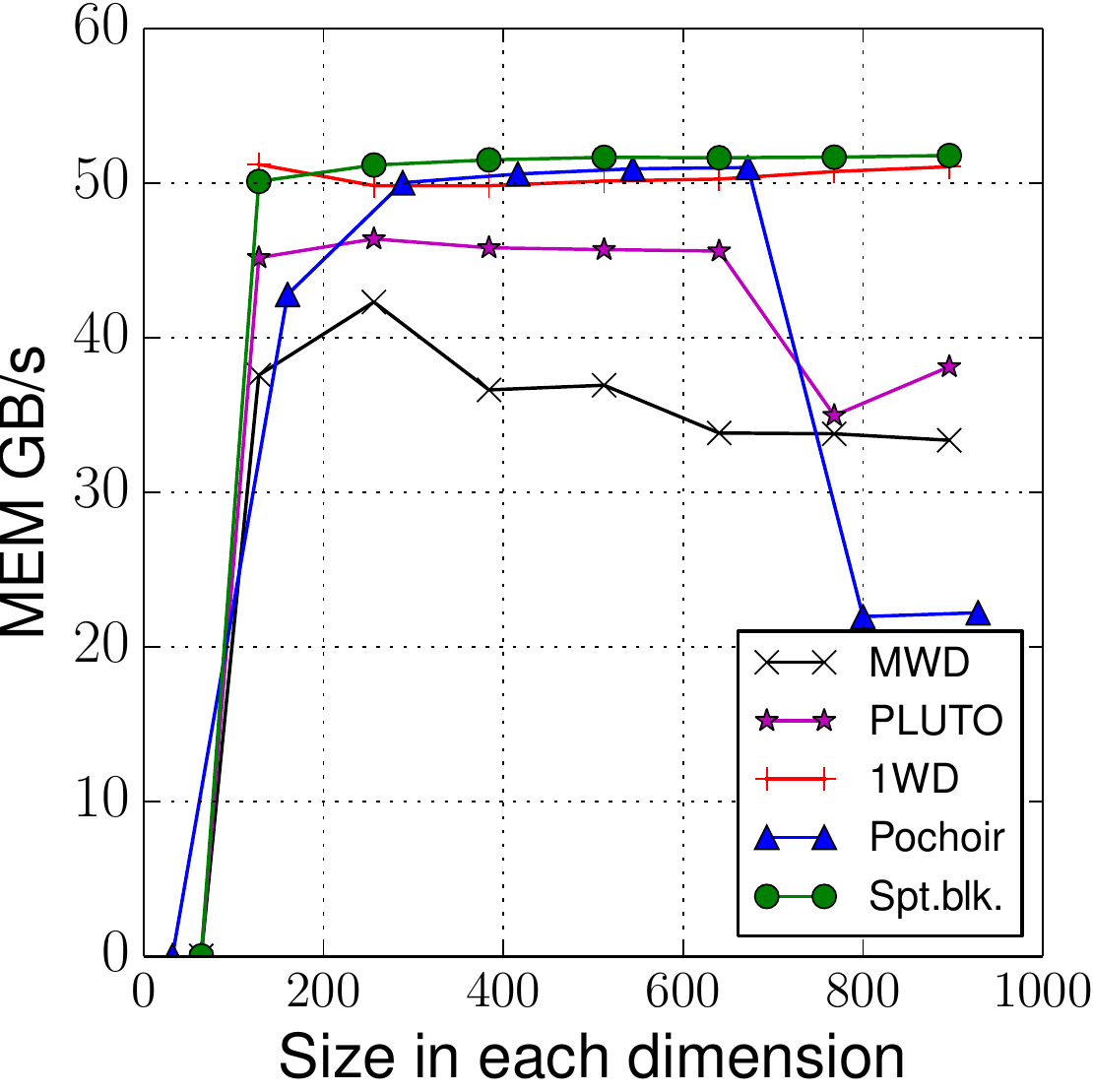}
        \label{fig:hw_7_pt_var_mem_bw_igs_all}
    }
    \enskip
    \subfloat[Memory transfer volume.]{
        \centering
        \includegraphics[width=\sfwidth]{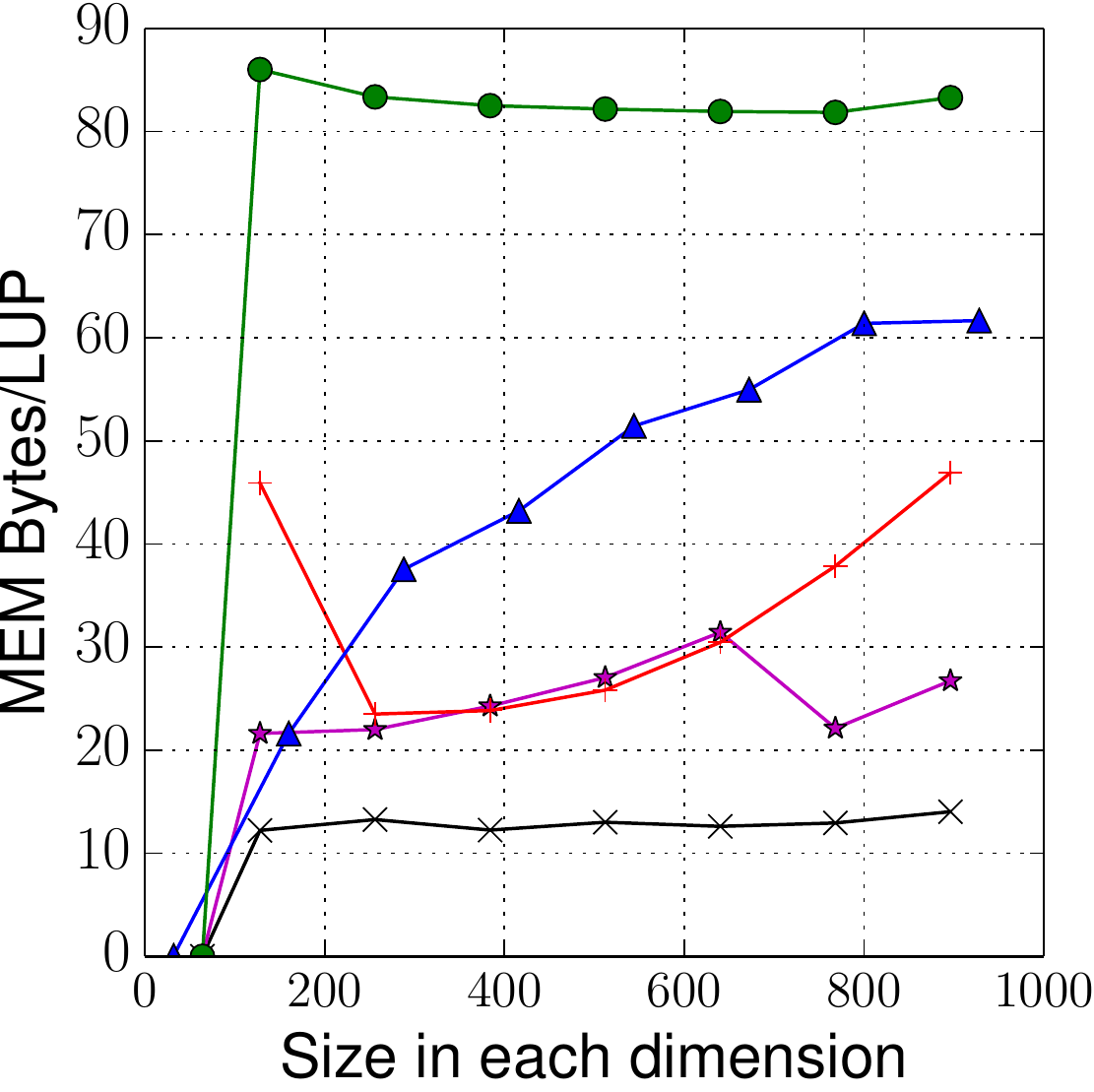}
        \label{fig:hw_7_pt_var_mem_vol_igs_all}
    }
    \caption{Haswell 7-point variable-coefficient stencil results, using increasing cubic grid size. Showing performance and memory transfer measurements of \gls{mwd}, PLUTO, Pochoir, \gls{swd}, and spatial blocking.}
    \label{fig:hw_7_pt_var_igs_all}
\end{figure}

The results for the 7-point stencil with constant coefficients are
shown in Figs.~\ref{fig:ivb_7_pt_var_perf_igs_all},
\ref{fig:ivb_7_pt_var_mem_bw_igs_all}, and
\ref{fig:ivb_7_pt_var_mem_vol_igs_all} for Ivy Brige and in
Figs.~\ref{fig:hw_7_pt_var_perf_igs_all},
\ref{fig:hw_7_pt_var_mem_bw_igs_all}, and
\ref{fig:hw_7_pt_var_mem_vol_igs_all} for Haswell.  
This stencil performs $13\,\flops/\LUP$, but it has a higher pressure on
memory for purely spatial blocking (80\,\bytes/\LUP\ instead of 24
without non-temporal stores). Hence, we expect 
more potential for temporal blocking and accordingly a higher
speedup for MWD, which exhibits the lowest in-memory code balance of all.
Indeed the speedup of MWD compared to spatial blocking is 
$2.8${}$\times$--$3.2${}$\times$
on Ivy Bridge and $4.5${}$\times$--$5.2${}$\times$ on the more bandwidth-starved
Haswell. 

On Ivy Bridge, Pochoir has the lowest memory bandwidth at large
problem sizes, but it is five times slower than even spatial blocking.
This shows that low memory bandwidth does not guarantee high
performance. In this particular case the compiler appears to have
generated exceptionally slow code, but even if the memory bandwidth 
were fully utilized ($40\,\GBS$ instead of $8\,\GBS$) 
the performance would hardly surpass spatial
blocking.

The phenomenological ECM model for MWD with  stencil predicts a performance
of $1.6\,\GLUPS$ ($3.8\,\GLUPS$) at large problem sizes for the full
Ivy Bridge (Haswell) socket assuming perfect scalability. The
deviation from the measurement is larger than for the constant-coefficient
7-point variant, because the large data traffic requirements lead to 
a breakdown of parallel efficiency due to cache size constraints.

\subsubsection{25-point stencil with constant coefficients}\label{sec:25pt_const_igs_all}

\begin{figure}[tbp]
    \centering
    \subfloat[Performance.]{
        \centering
        \includegraphics[width=\sfwidth]{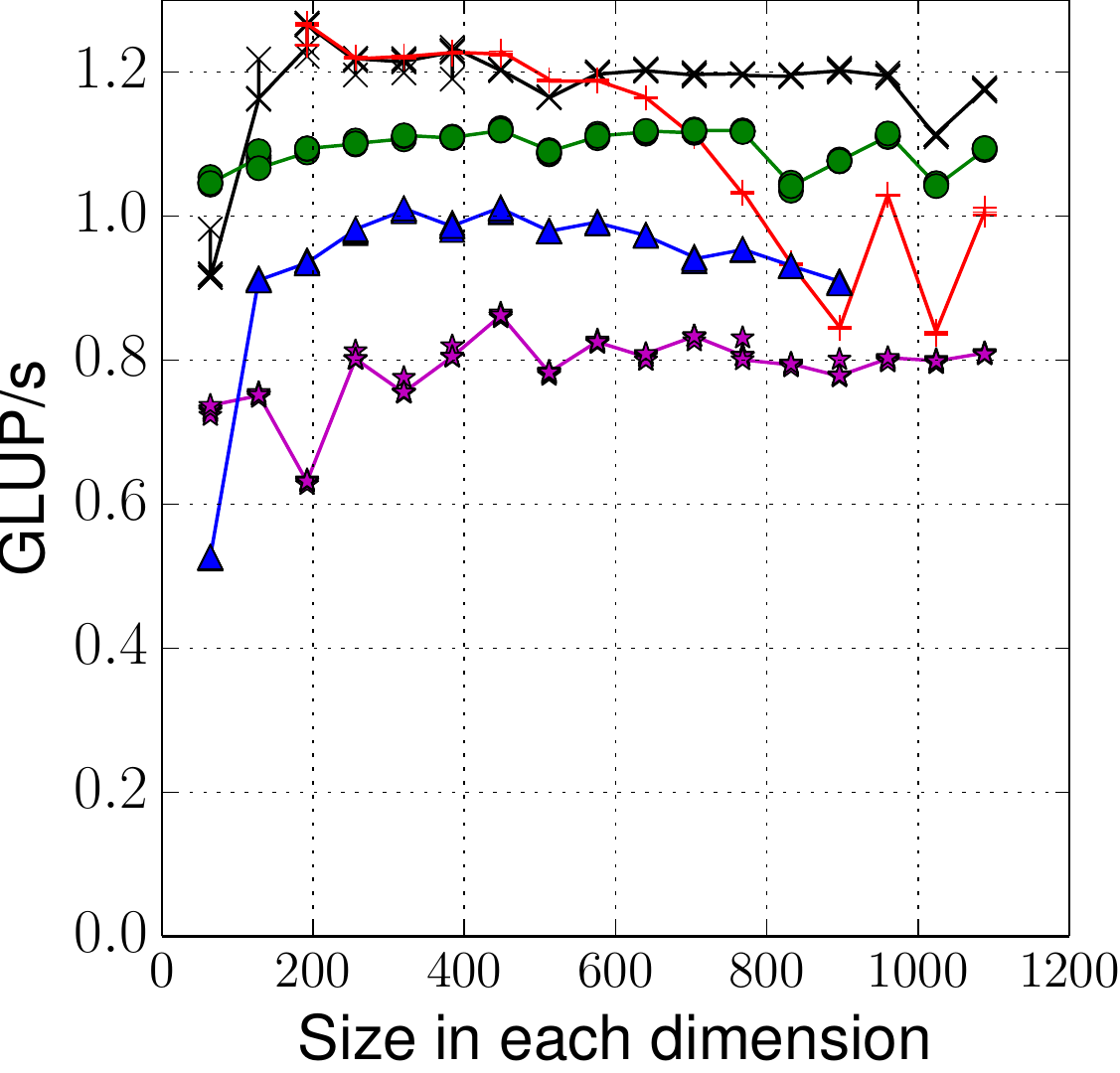}
        \label{fig:ivb_25_pt_const_perf_igs_all}
    }
    \enskip
    \subfloat[Memory bandwidth.]{
        \centering
        \includegraphics[width=\sfwidth]{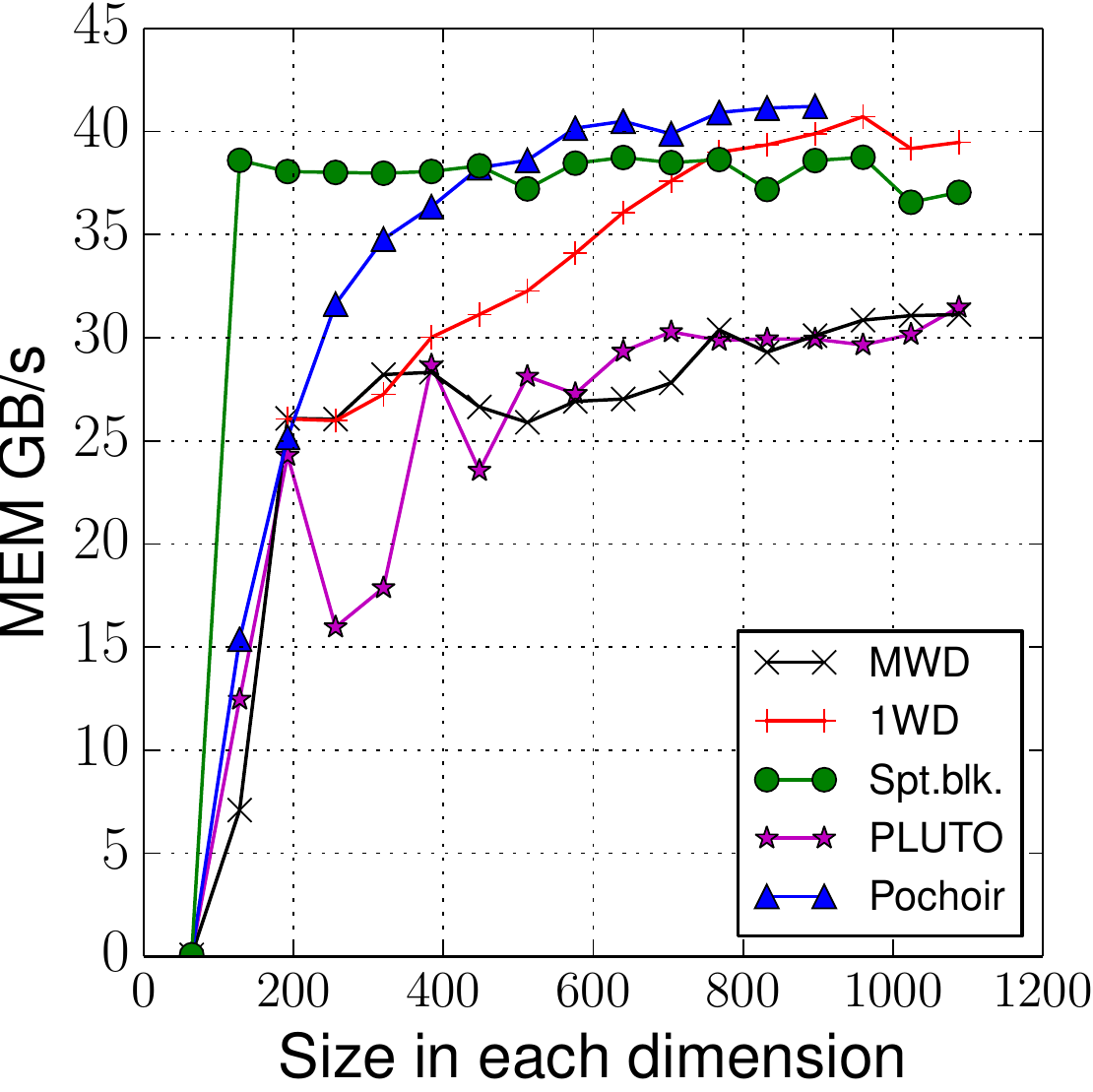}
        \label{fig:ivb_25_pt_const_mem_bw_igs_all}
    }
    \enskip
    \subfloat[Memory transfer volume.]{
        \centering
        \includegraphics[width=\sfwidth]{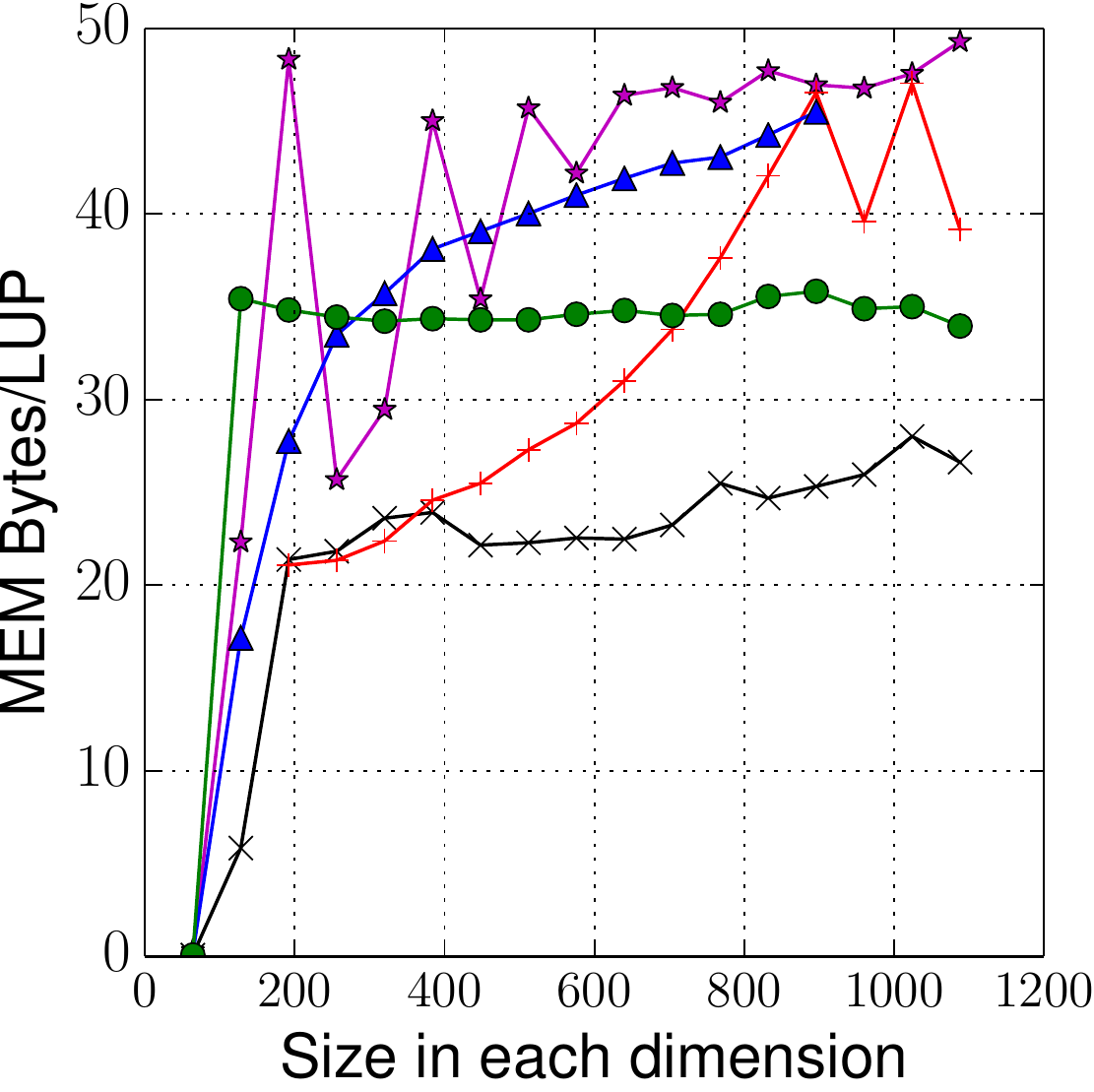}
        \label{fig:ivb_25_pt_const_mem_vol_igs_all}
    }
    \caption{Ivy Bridge 25-point constant-coefficient stencil results, using increasing cubic grid size. Showing performance and memory transfer measurements of \gls{mwd}, PLUTO, Pochoir, \gls{swd}, and spatial blocking.}
    \label{fig:ivb_25_pt_const_igs_all}
\end{figure}

\begin{figure}[tbp]
    \centering
    \subfloat[Performance.]{
        \centering
        \includegraphics[width=\sfwidth]{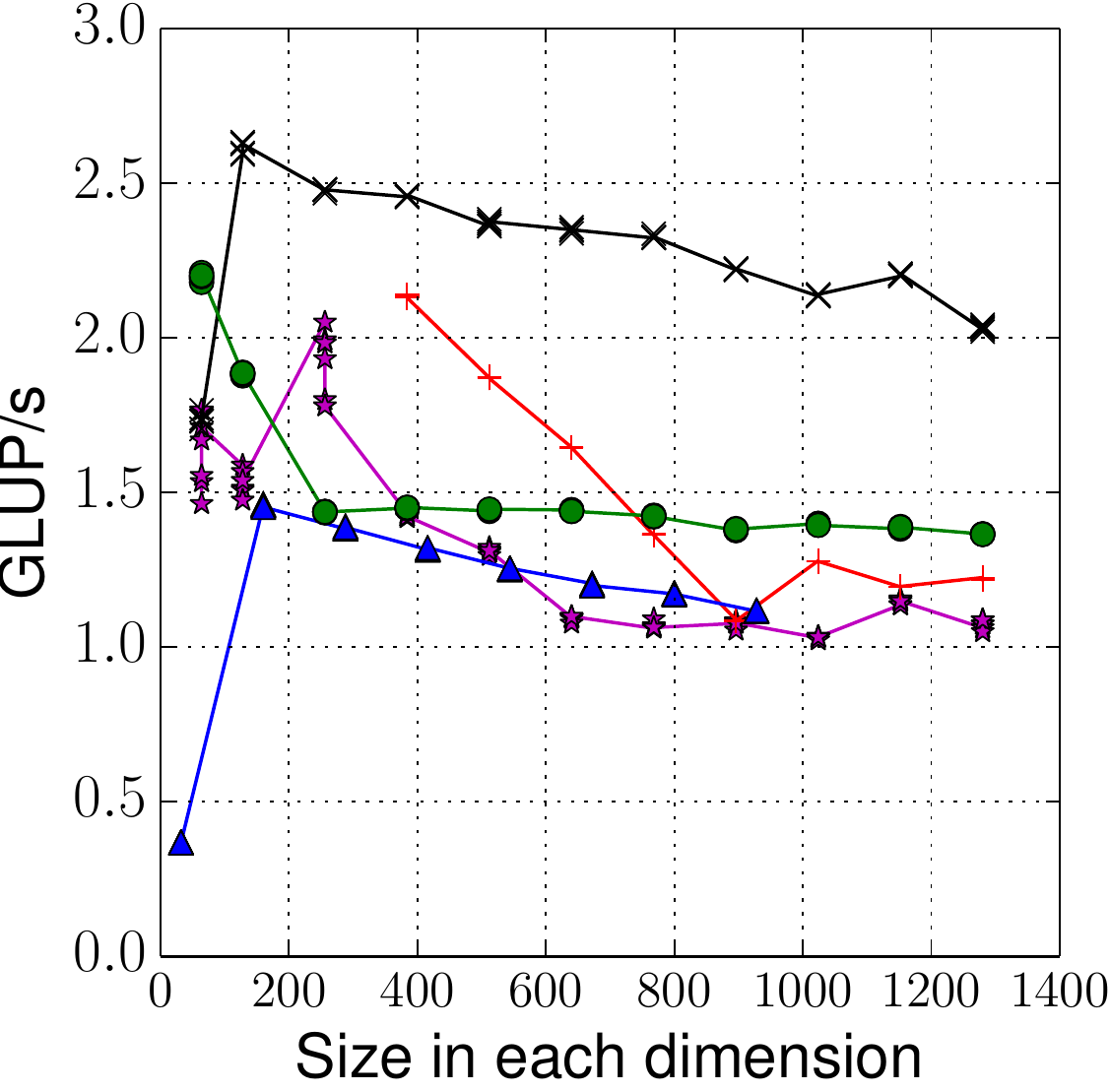}
        \label{fig:hw_25_pt_const_perf_igs_all}
    }
    \enskip
    \subfloat[Memory bandwidth.]{
        \centering
        \includegraphics[width=\sfwidth]{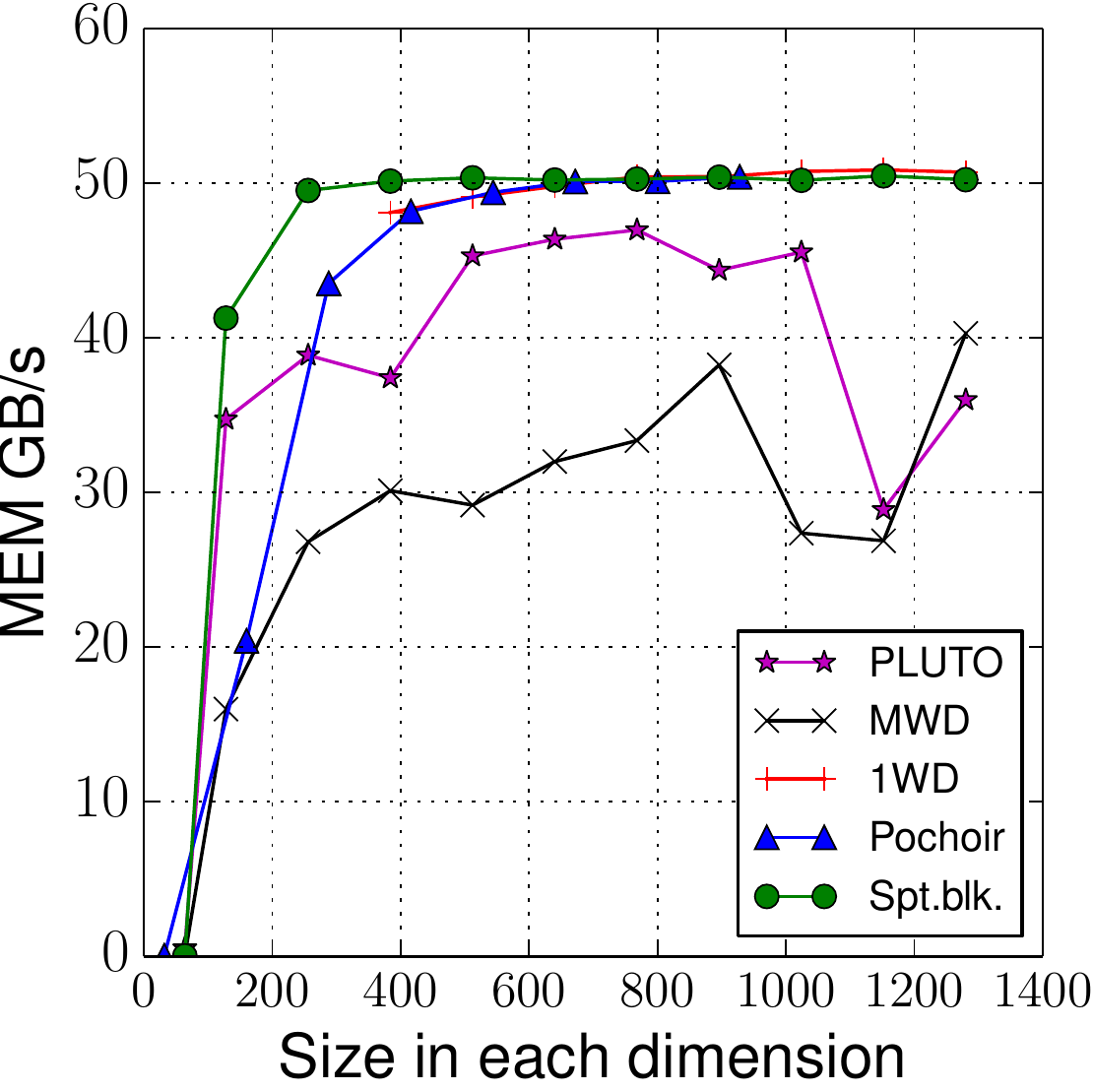}
        \label{fig:hw_25_pt_const_mem_bw_igs_all}
    }
    \enskip
    \subfloat[Memory transfer volume.]{
        \centering
        \includegraphics[width=\sfwidth]{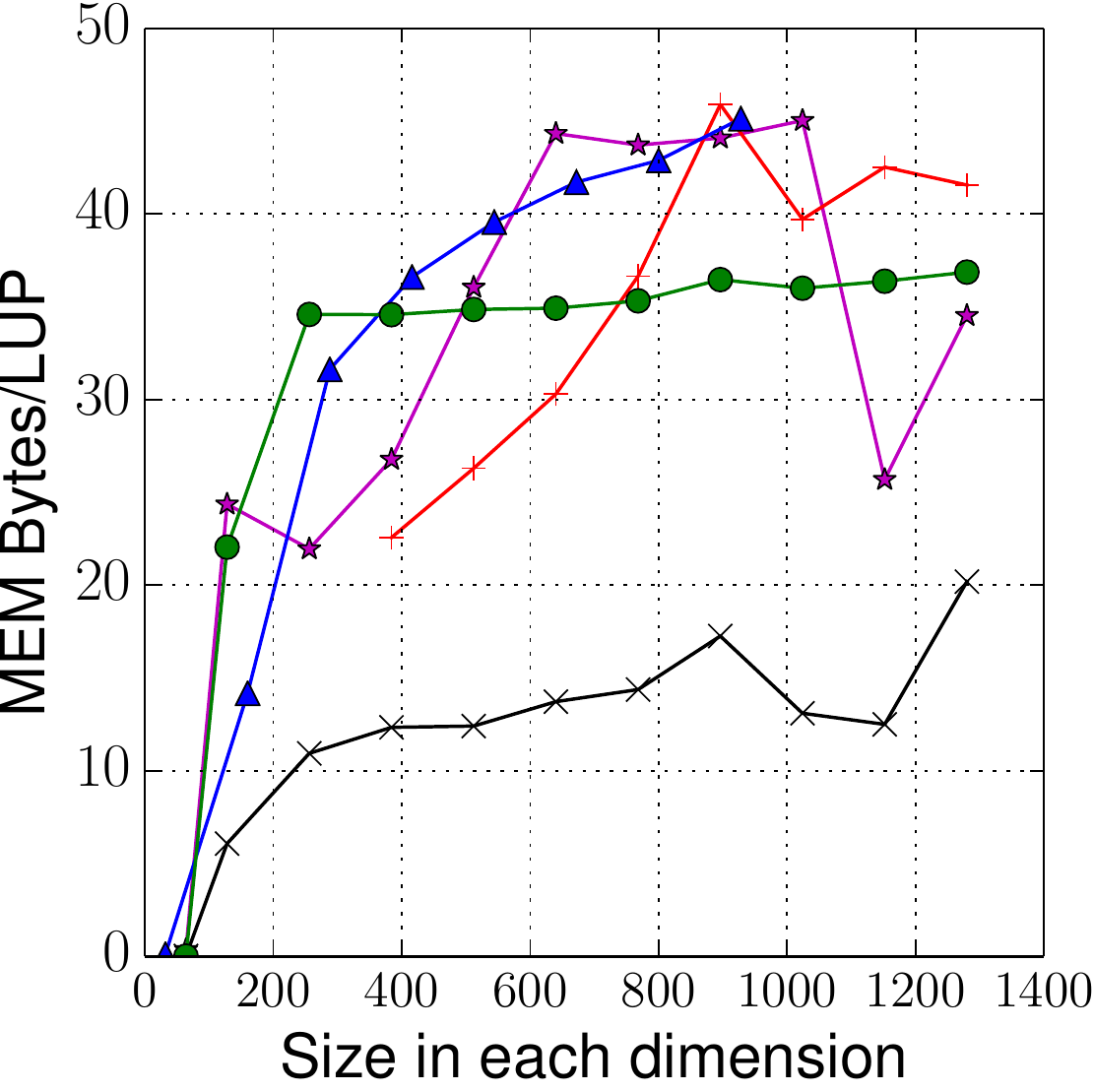}
        \label{fig:hw_25_pt_const_mem_vol_igs_all}
    }
    \caption{Haswell 25-point constant-coefficient stencil results, using increasing cubic grid size. Showing performance and memory transfer measurements of \gls{mwd}, PLUTO, Pochoir, \gls{swd}, and spatial blocking.}
    \label{fig:hw_25_pt_const_igs_all}
\end{figure}

The results for the 25-point stencil with constant coefficients are
shown in Figs.~\ref{fig:ivb_25_pt_const_perf_igs_all}, 
\ref{fig:ivb_25_pt_const_mem_bw_igs_all}, and 
\ref{fig:ivb_25_pt_const_mem_vol_igs_all} for Ivy Bridge, and in 
Figs.~\ref{fig:hw_25_pt_const_perf_igs_all}, 
\ref{fig:hw_25_pt_const_mem_bw_igs_all}, and 
\ref{fig:hw_25_pt_const_mem_vol_igs_all} for Haswell. The stencil
has a minimum code balance for optimal spatial blocking
of $32\,\bytes/\LUP$.

Due to its high ratio of $33\,\flops/\LUP$ and its large radius
this stencil poses a challenge for temporal blocking schemes. Many
layers of grid points have to be supplied by the caches, so the
time needed for memory transfers is rather small \cite{stengel14} and
the cache size needed for temporal blocking is large. Only MWD reduces
the memory pressure significantly, since it can leverage the
multi-threaded wavefronts to reduce the need for cache size. As a
consequence, only MWD is consistently faster than pure spatial
blocking by a factor of $1.1${}$\times$ on Ivy Bridge
and by $1.5${}$\times$--$1.7${}$\times$ on Haswell.  
Pochoir and 1WD even exhibit a memory code balance larger than spatial 
blocking for most problem sizes.

With MWD the compiler produces some register spilling, adding
to the dominance of the in-core and in-cache contributions to the runtime.
At large problem sizes the ECM model predicts a full-socket MWD performance
of $1.3\,\GLUPS$ on Ivy Bridge and of $2.5\,\GLUPS$ on Haswell. Both
predictions are reasonably close to the measurements.

\subsubsection{25-point stencil with variable coefficients}\label{sec:25pt_var_igs_all}

\begin{figure}[tbp]
    \centering
    \subfloat[Performance.]{
        \centering
        \includegraphics[width=\sfwidth]{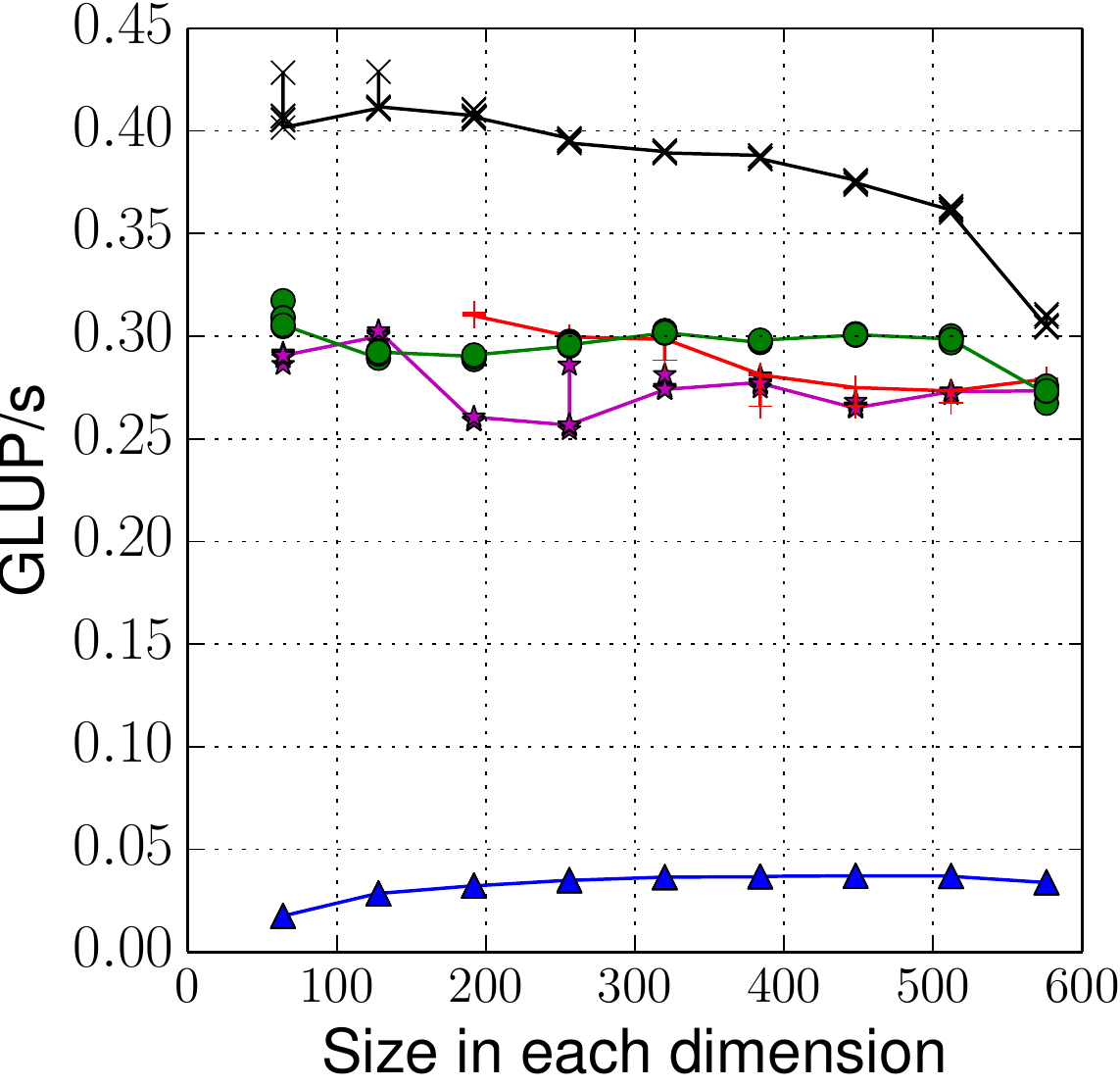}
        \label{fig:ivb_25_pt_var_perf_igs_all}
    }
    \enskip
    \subfloat[Memory bandwidth.]{
        \centering
        \includegraphics[width=\sfwidth]{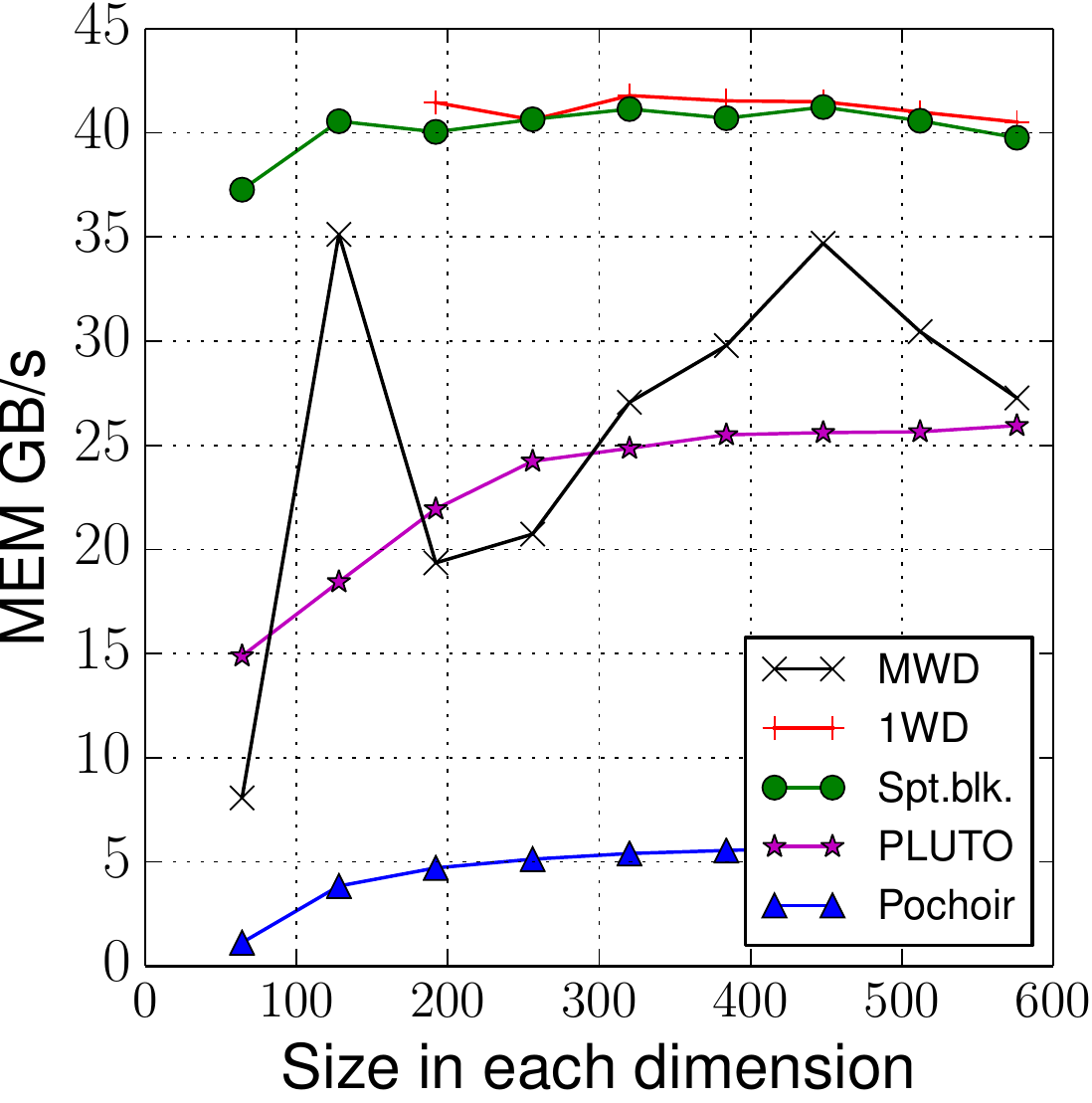}
        \label{fig:ivb_25_pt_var_mem_bw_igs_all}
    }
    \enskip
    \subfloat[Memory transfer volume.]{
        \centering
        \includegraphics[width=\sfwidth]{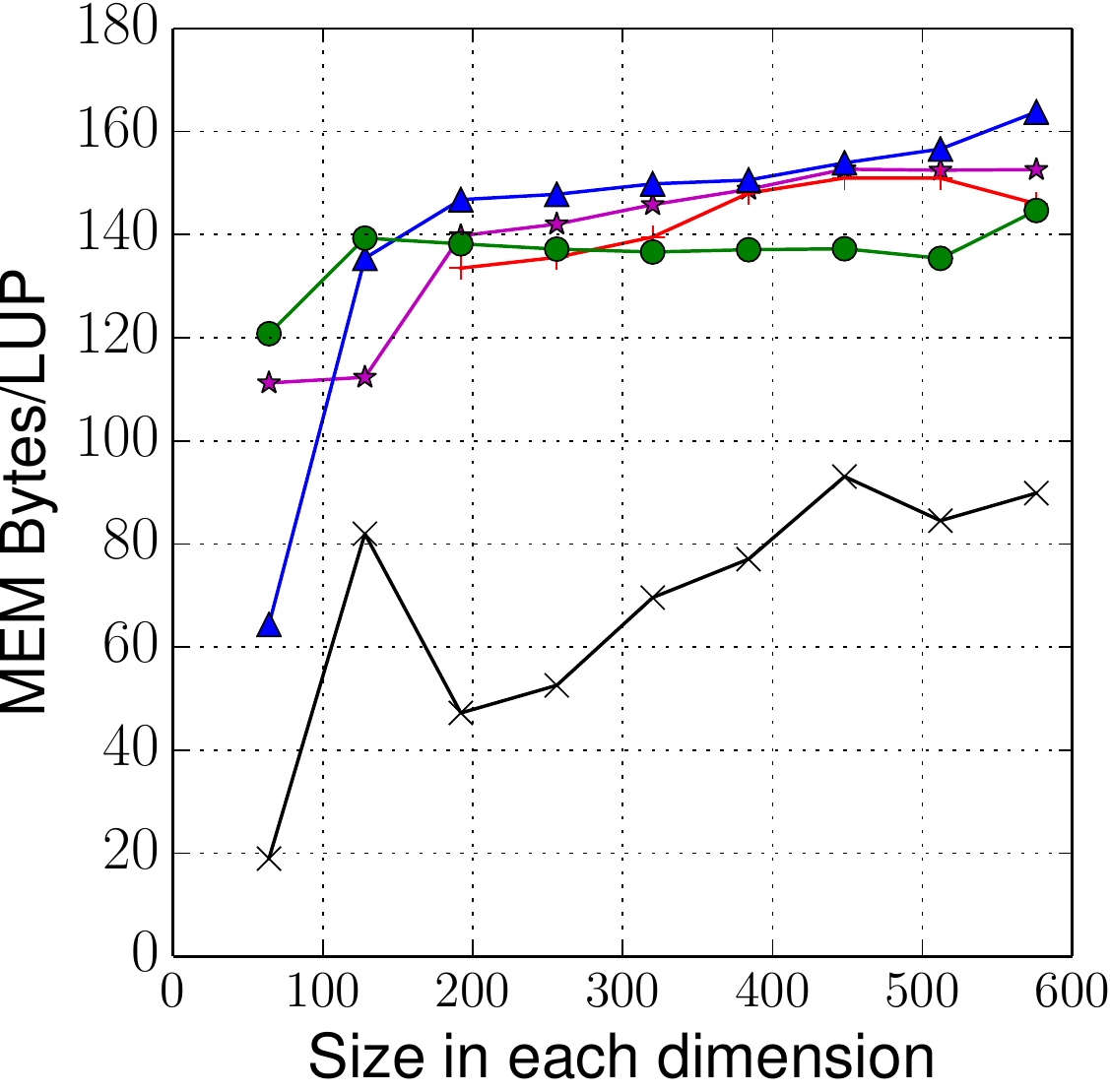}
        \label{fig:ivb_25_pt_var_mem_vol_igs_all}
    }
    \caption{Ivy Bridge 25-point variable-coefficient stencil results, using increasing cubic grid size. Showing performance and memory transfer measurements of \gls{mwd}, PLUTO, Pochoir, \gls{swd}, and spatial blocking.}
    \label{fig:ivb_25_pt_var_igs_all}
\end{figure}

\begin{figure}[tbp]
    \centering
    \subfloat[Performance.]{
        \centering
        \includegraphics[width=\sfwidth]{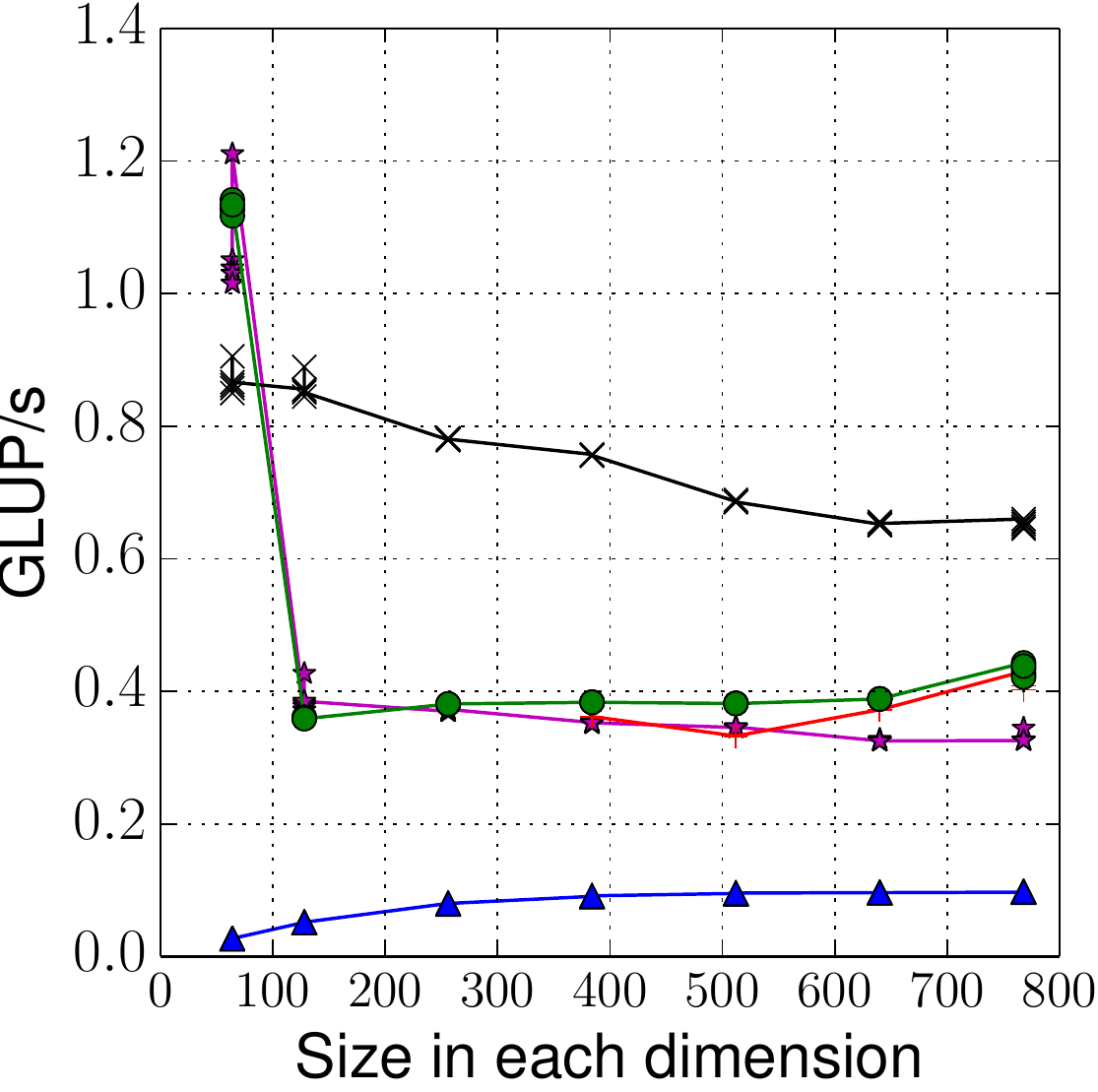}
        \label{fig:hw_25_pt_var_perf_igs_all}
    }
    \enskip
    \subfloat[Memory bandwidth.]{
        \centering
        \includegraphics[width=\sfwidth]{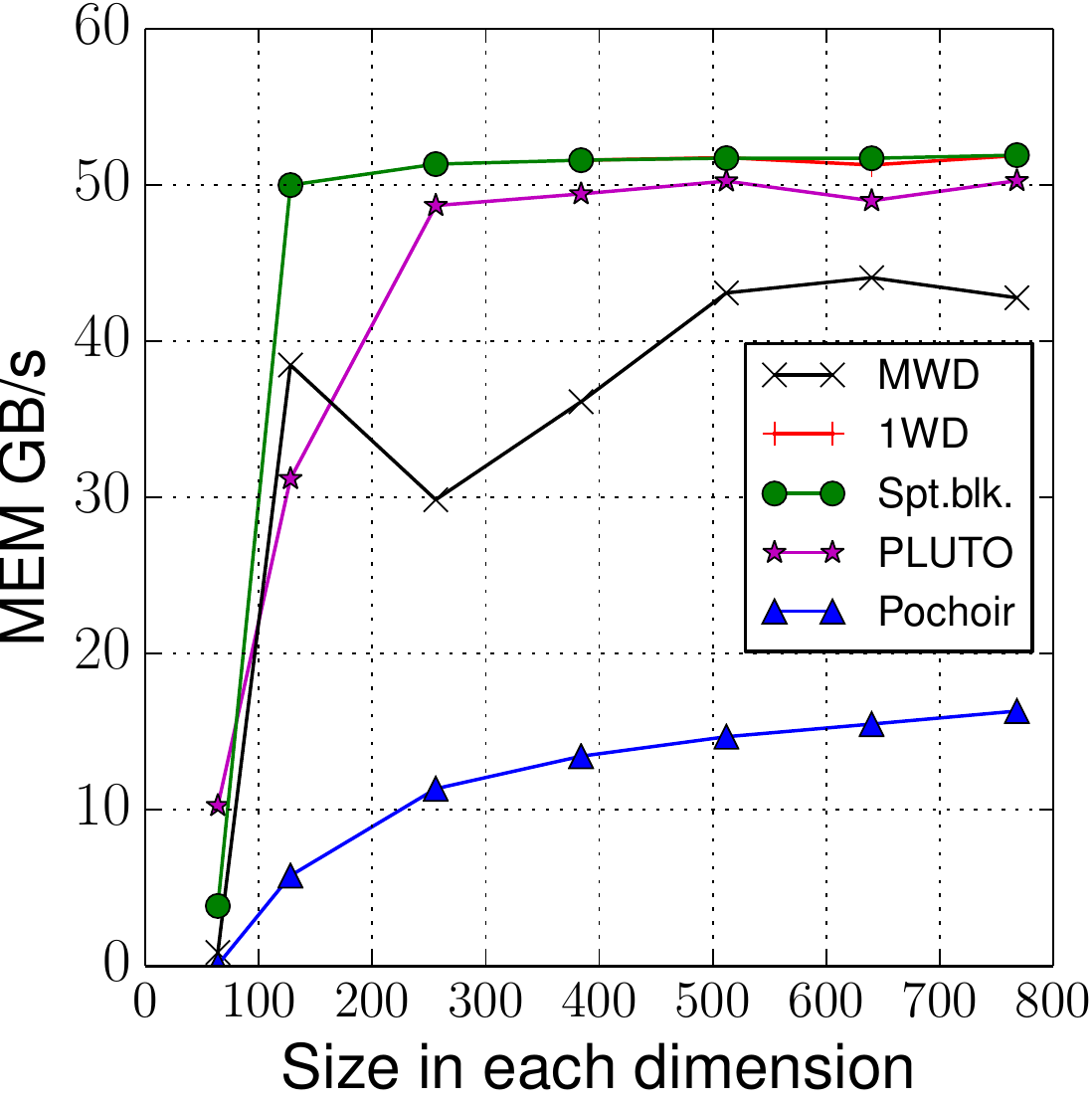}
        \label{fig:hw_25_pt_var_mem_bw_igs_all}
    }
    \enskip
    \subfloat[Memory transfer volume.]{
        \centering
        \includegraphics[width=\sfwidth]{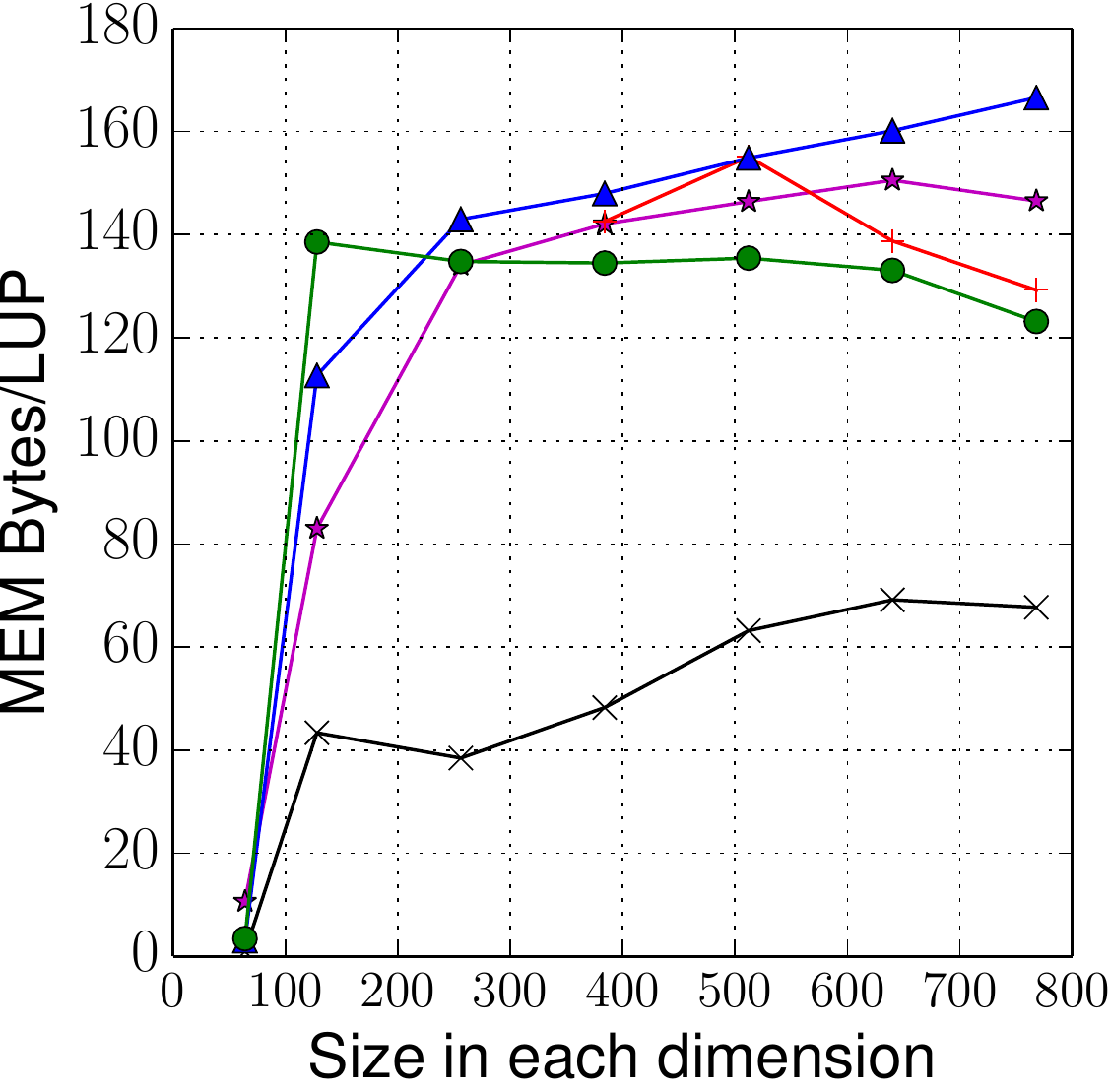}
        \label{fig:hw_25_pt_var_mem_vol_igs_all}
    }
    \caption{Haswell 25-point variable-coefficient stencil results, using increasing cubic grid size. Showing performance and memory transfer measurements of \gls{mwd}, PLUTO, Pochoir, \gls{swd}, and spatial blocking.}
    \label{fig:hw_25_pt_var_igs_all}
\end{figure}

The results for the 25-point stencil with variable coefficients are
shown in Figs.~\ref{fig:ivb_25_pt_var_perf_igs_all}, 
\ref{fig:ivb_25_pt_var_mem_bw_igs_all}, 
and \ref{fig:ivb_25_pt_var_mem_vol_igs_all} for Ivy Bridge, and in 
Figs.~\ref{fig:hw_25_pt_var_perf_igs_all}, 
\ref{fig:hw_25_pt_var_mem_bw_igs_all}, 
and \ref{fig:hw_25_pt_var_mem_vol_igs_all} for Haswell.

This stencil performs $37\,\flops/\LUP$ but requires much more data
due to the variable coefficients array ($128\,\bytes/\LUP$ for optimal
spatial blocking). Although it has a lower intensity compared to the
25-point constant coefficient stencil, it poses the same cache size
problems. Again, only MWD can reduce the code balance significantly,
and it is the only scheme that is faster than spatial blocking at all
(by $1.2${}$\times$--$1.3${}$\times$ on Ivy Bridge and by
$1.5${}$\times$--$2${}$\times$ on Haswell).  Pochoir shows low memory
bandwidth at the same memory code balance as spatial
blocking. Measurements show that Pochoir requires a massive amount of
data traffic between the L1 and L2 cache for this stencil, which is a
contributor (in addition to slow low-level code) to its extremely low
performance. The phenomenological ECM model for MWD yields socket-level
estimates of $0.44\,\GLUPS$ on Ivy Bridge and $0.71\,\GLUPS$ on Haswell,
both of which are in line with the measurements.

\subsection{MWD tile sharing impact on performance, memory transfer, and energy consumption} \label{sec:tgs_results}

While the cache block sharing reduces the memory bandwidth requirements of the stencil codes, it increases the overhead by performing fine-grained synchronization among more threads.
As a result, the auto-tuner would select the minimum thread group size that sufficiently decouples from the main memory bandwidth, when allowed to tune all the parameters.

In this section we run the \gls{mwd} approach using fixed thread group sizes to study their impact on the performance, memory bandwidth and transfer, and energy consumption.
The auto-tuner selected the parallelization dimensions and tiling parameters in these experiments.
We limit the energy consumption analysis to the Intel Ivy Bridge processor.
The memory modules of the Haswell processor consume a similar amount of energy regardless of the memory bandwidth usage, making the energy consumption rate oblivious to our optimization techniques.

Since we observe similarities in the studied characteristics of the four stencils under investigation here, we describe a representative subset of these results.
We use a range of cubic grid sizes, where each dimension is set to multiples of 64 in the Ivy Bridge processor and multiples of 128 in the Haswell processor.

We describe the Haswell results of the 7-point constant-coefficient and the 25-point variable-coefficient stencils, to show \gls{mwd} behavior at two extremes in the code balance.
The energy consumption behavior is illustrated using the 25-point constant-coefficient stencil results on Ivy Bridge.

\subsubsection{7-point stencil with constant coefficients on Haswell}\label{sec:7pt_const_tgs}

\begin{figure}[tbp]
    \centering
    \subfloat[Performance.]{
        \centering
        \includegraphics[width=\sfwidth]{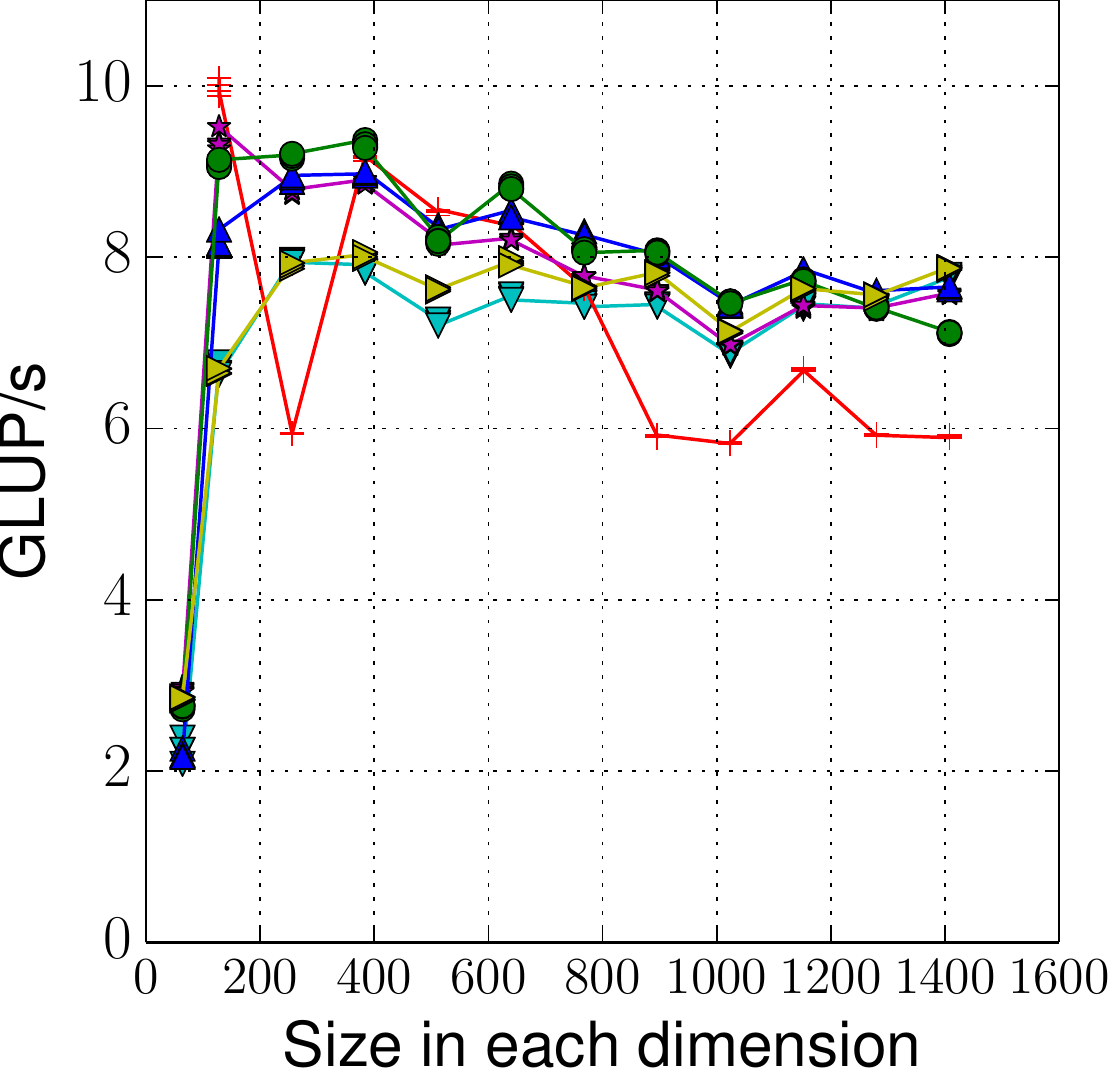}
        \label{fig:hw_7_pt_const_perf_tgs}
    }
    \enskip
    \subfloat[Measured memory bandwidth.]{
        \centering
        \includegraphics[width=\sfwidth]{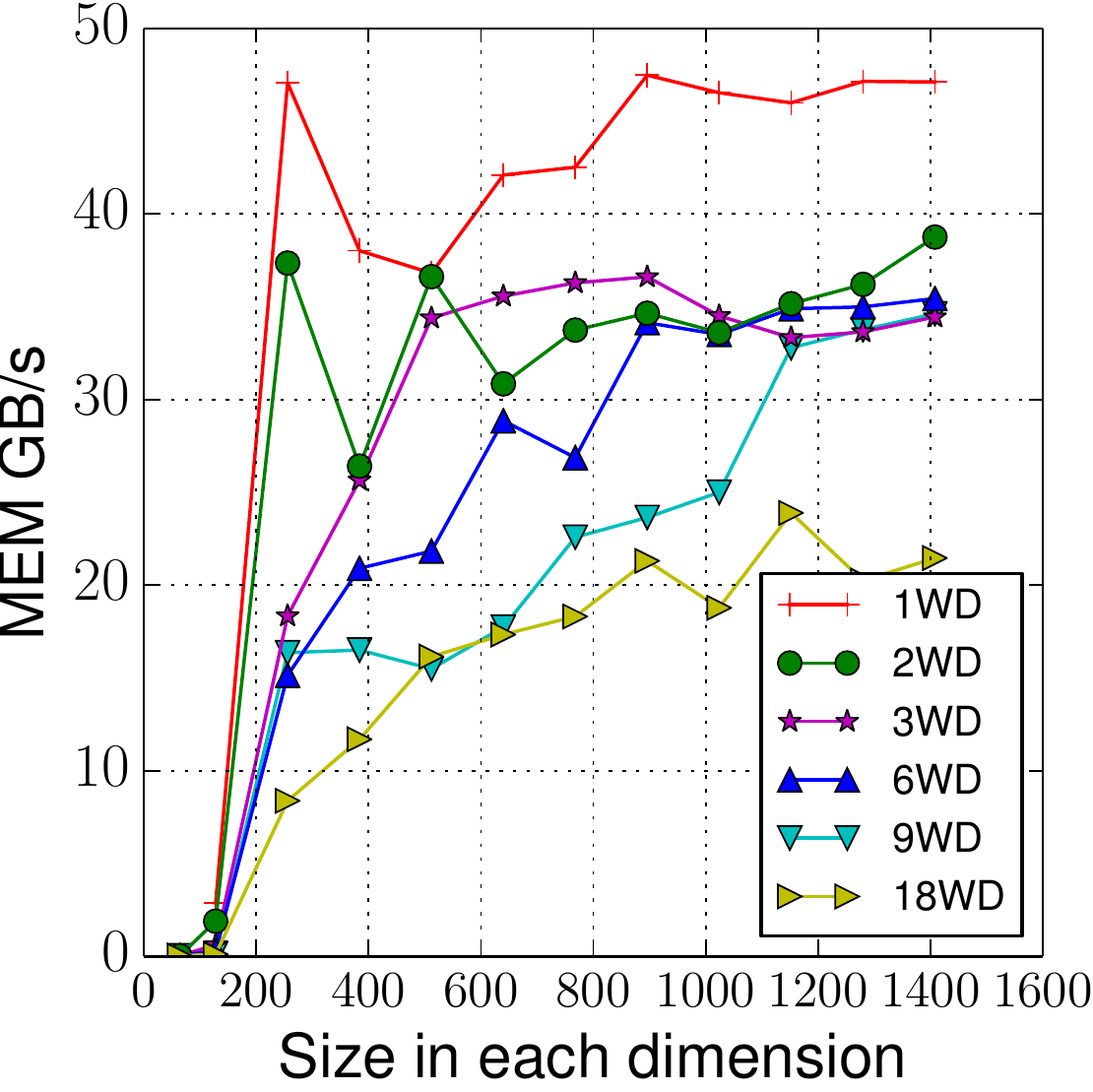}
        \label{fig:hw_7_pt_const_mem_bw_tgs}
    }
    \enskip
    \subfloat[Measured memory transfers per \gls{lup}.]{
        \centering
        \includegraphics[width=\sfwidth]{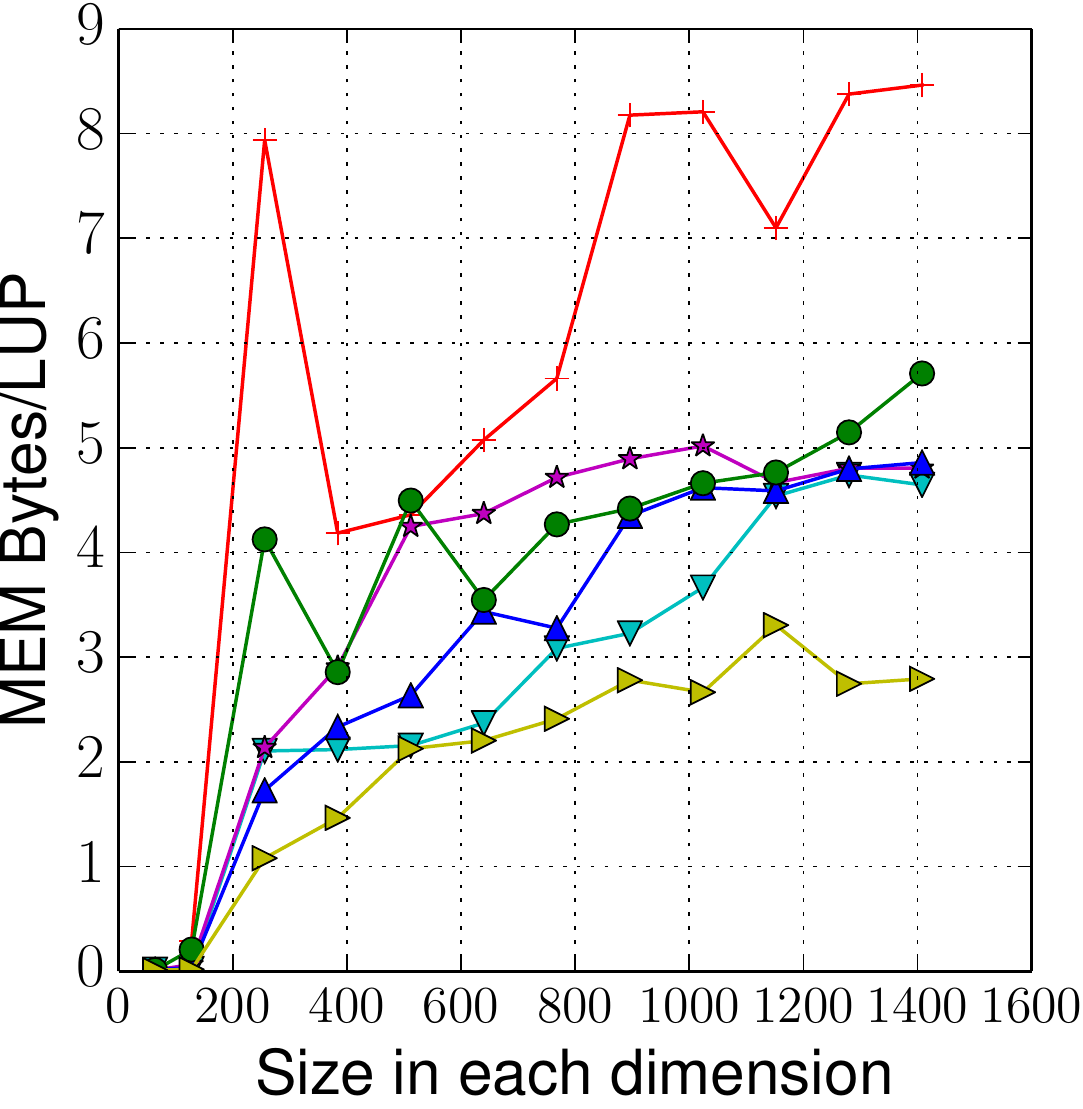}
        \label{fig:hw_7_pt_const_mem_vol_tgs}
    }
    \caption{Haswell performance and memory transfer measurements of the 7-point constant-coefficient stencil using increasing cubic grid size.
    We compare various thread group sizes in \gls{mwd}.}
    \label{fig:hw_7_pt_const_tgs}
\end{figure}

The Intel Haswell performance results using different thread group sizes are presented in Fig.~\ref{fig:hw_7_pt_const_perf_tgs}.
We also show the measured memory bandwidth in Fig.~\ref{fig:hw_7_pt_const_mem_bw_tgs}, and the memory transfer volumes normalized by the number of lattice site updates in Fig.~\ref{fig:hw_7_pt_const_mem_vol_tgs}.

The \gls{swd} implementation does not saturate the memory bandwidth at smaller grid sizes than $800^3$, as this stencil has small bytes requirements and moderate code balance.
When \gls{swd} runs at grid sizes larger than $800^3$ the memory bandwidth saturates and the performance drops, as larger cache blocks cannot fit in the L3 cache memory.
2WD and larger thread group sizes are sufficient to decouple from the main memory bandwidth bottleneck at large grid sizes.

At small grid sizes, the synchronization overhead of 9WD and 18WD is  significant compared to the computations, resulting in lower performance.
As the grid size becomes larger, the synchronization cost of 9WD and 18WD becomes less significant, and all \gls{mwd} variants achieve similar performance.

Larger thread group sizes save more cache memory as they keep a smaller number of tiles in the cache memory.
The cache size saving provides space for larger diamond tiles, which increases the in-cache data reuse and decreases the main memory bandwidth and data traffic.
Figures~\ref{fig:hw_7_pt_const_mem_bw_tgs} and~\ref{fig:hw_7_pt_const_mem_vol_tgs} show how larger thread group sizes consume less memory bandwidth and transfer less data per lattice update.
This shows that our \gls{mwd} approach is suitable for future processors, which are expected to be more starved for memory bandwidth.
Hence, although the synchronization overhead of \gls{mwd} impacts the performance on this particular processor, this is not a general result; a re-evaluation is needed on future architectures. Fortunately it can be left to the the auto-tuner to take such variations in hardware and software into account.

\subsubsection{25-point stencil with variable coefficients on Haswell}\label{sec:25pt_var_tgs}
\begin{figure}[tbp]
    \centering
    \subfloat[Performance.]{
        \centering
        \includegraphics[width=\sfwidth]{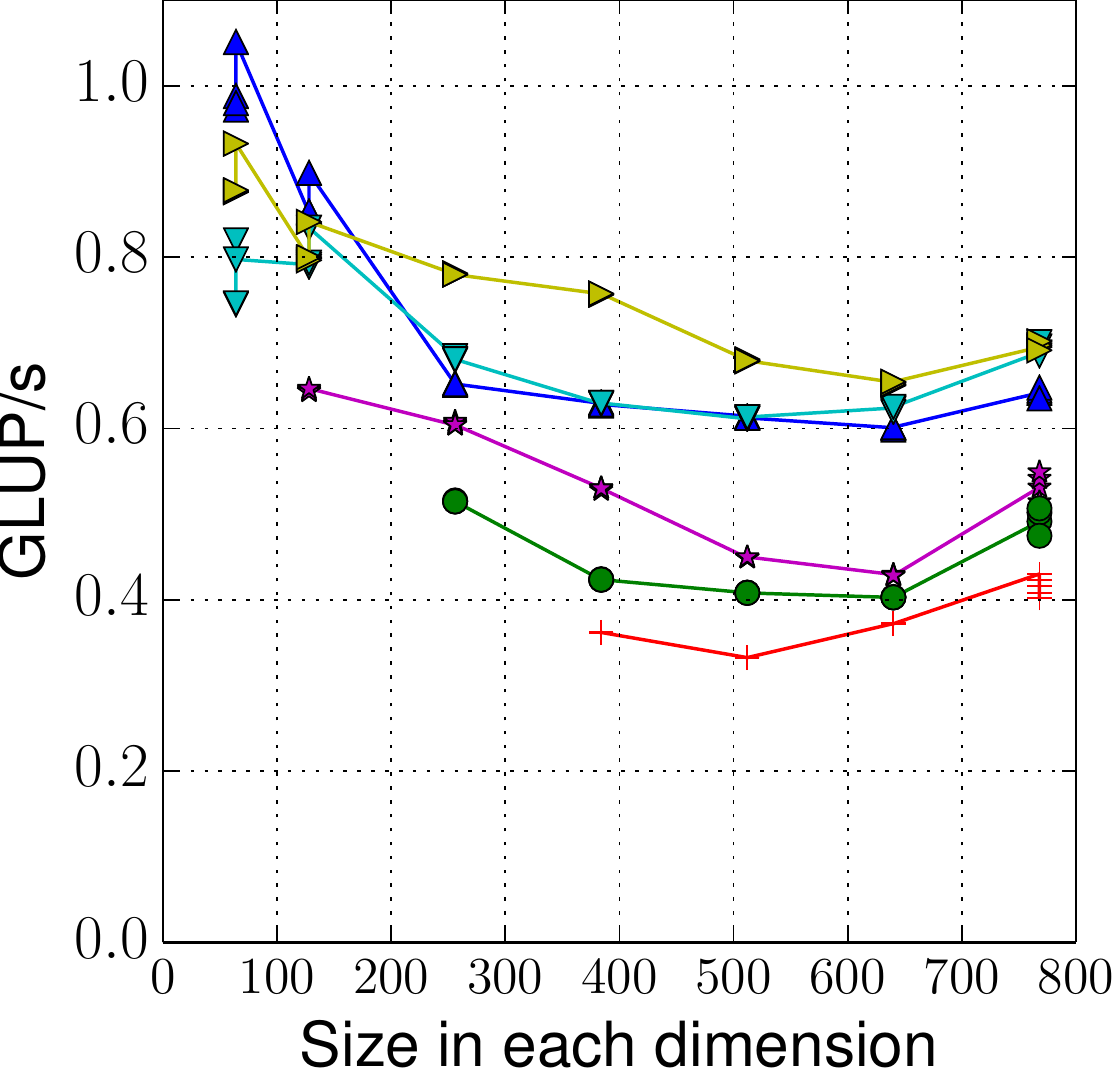}
        \label{fig:hw_25_pt_var_perf_tgs}
    }
    \enskip
    \subfloat[Measured memory bandwidth.]{
        \centering
        \includegraphics[width=\sfwidth]{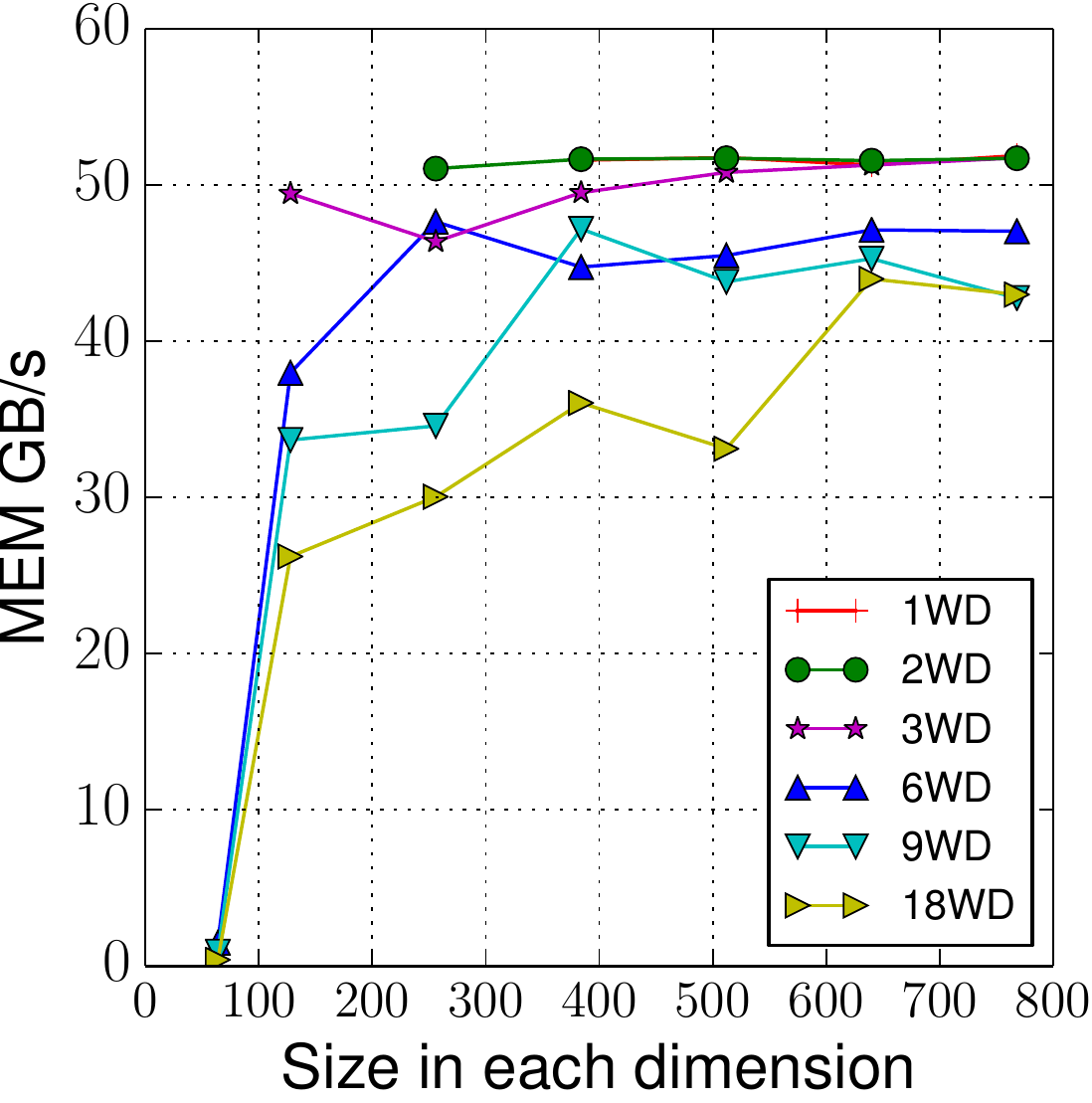}
        \label{fig:hw_25_pt_var_mem_bw_tgs}
    }
    \enskip
    \subfloat[Measured memory transfers per \gls{lup}.]{
        \centering
        \includegraphics[width=\sfwidth]{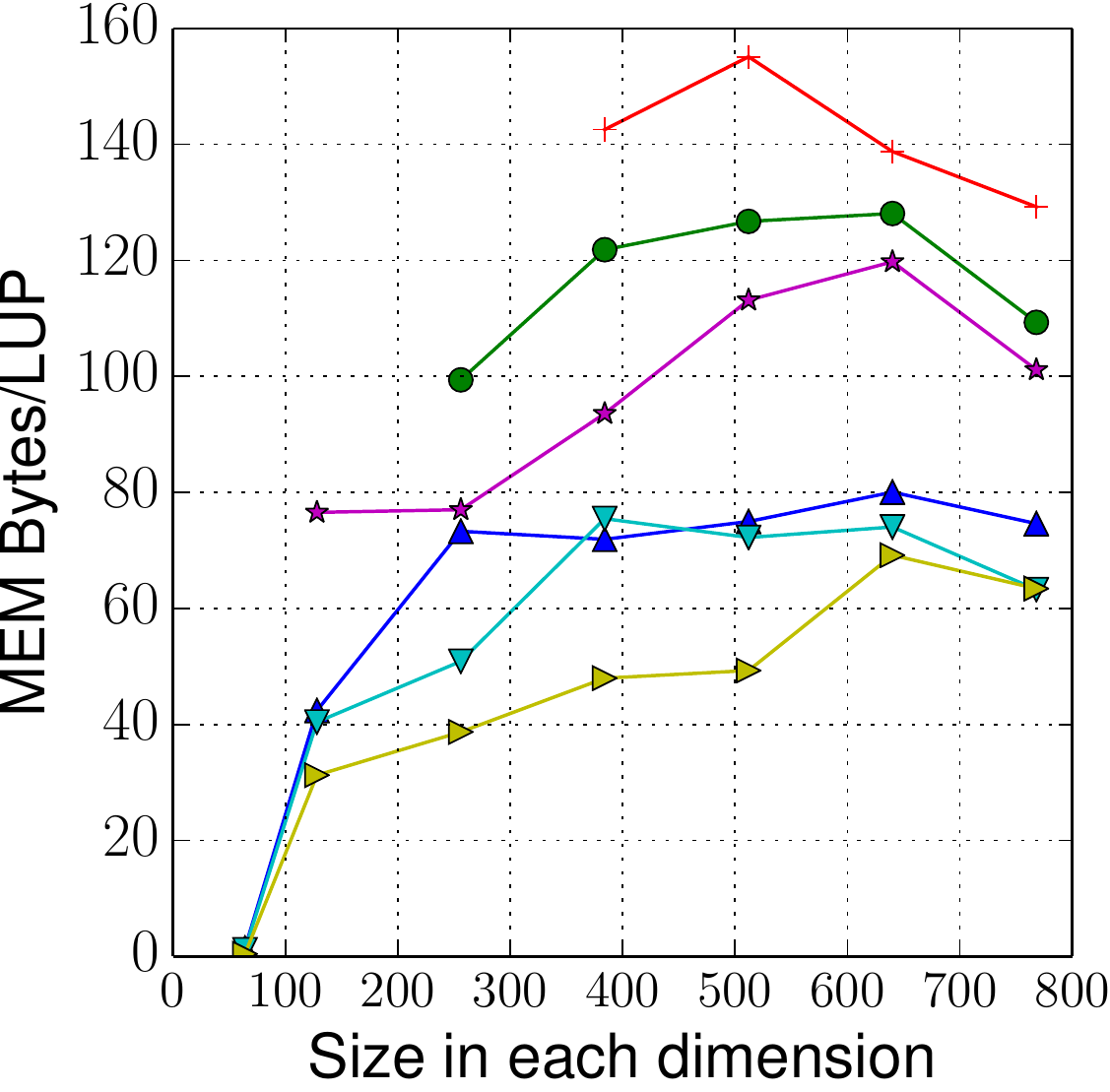}
        \label{fig:hw_25_pt_var_mem_vol_tgs}
    }
    \caption{Haswell performance and memory transfer measurements of the 25-point variable-coefficient stencil using increasing cubic grid size.
    We compare various thread group sizes in \gls{mwd}.}
    \label{fig:hw_25_pt_var_tgs}
\end{figure}

Haswell performance results using different thread group sizes are shown in Fig.~\ref{fig:hw_25_pt_var_tgs}.

The 25-point variable-coefficient stencil has high code balance and large cache block size requirements, so a large thread group size is necessary to decouple from the main memory bandwidth bottleneck.
As shown in Fig.~\ref{fig:hw_25_pt_var_perf_tgs}, 18WD achieves the best performance at most of the grid sizes.
Other \gls{mwd} variants saturate or nearly saturate the main memory bandwidth at larger grid sizes (Fig.~\ref{fig:hw_25_pt_var_mem_bw_tgs}) and show a much higher code balance (Fig.~\ref{fig:hw_25_pt_var_mem_vol_tgs}).

\subsubsection{25-point stencil with constant coefficients on Ivy Bridge}\label{sec:25pt_const_tgs}

\begin{figure}[tbp]
    \centering
    \subfloat[Performance.]{
        \centering
        \includegraphics[width=\sfwidth]{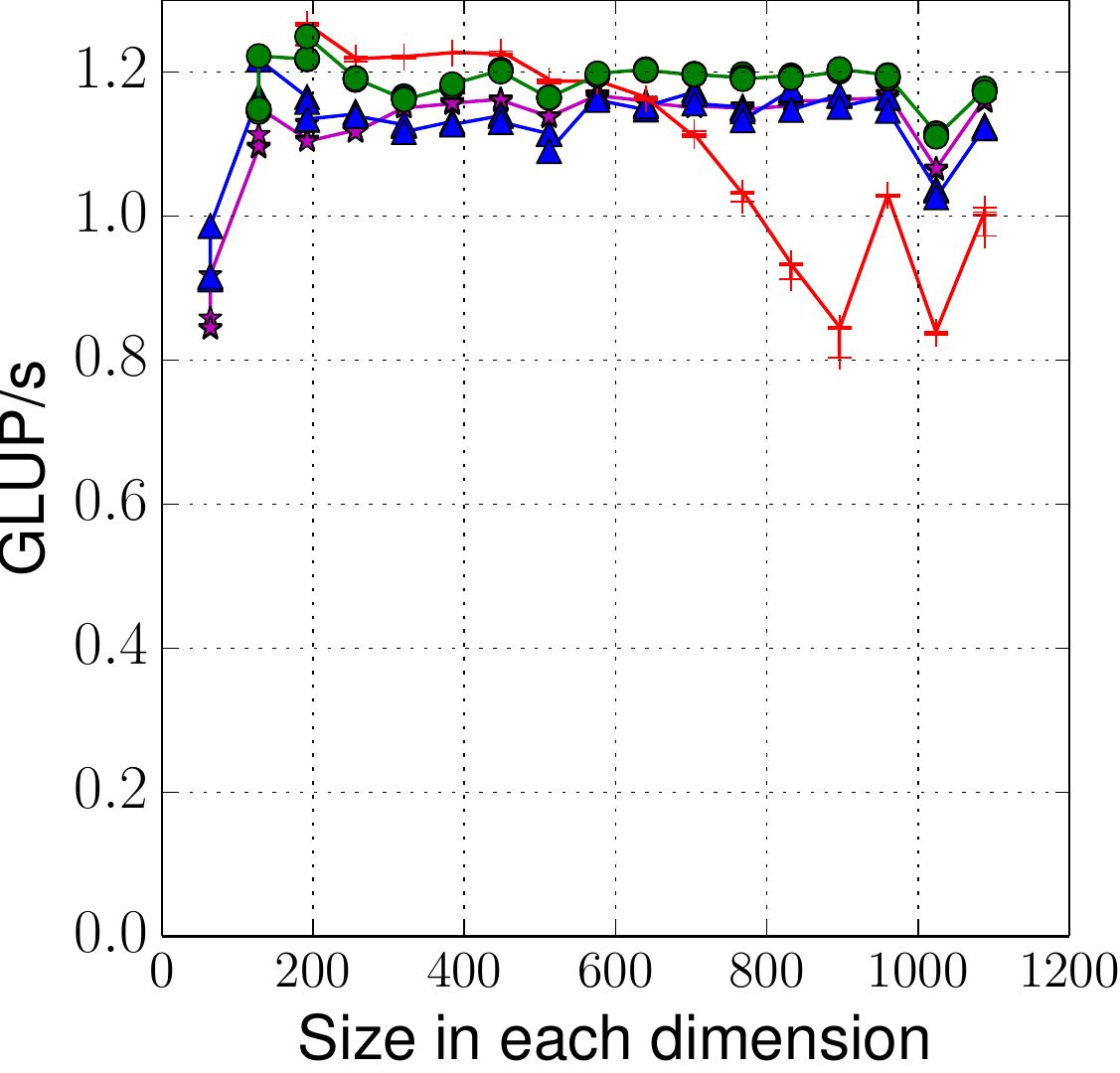}
        \label{fig:ivb_25_pt_const_perf_tgs}
    }
    \enskip
    \subfloat[Measured memory bandwidth.]{
        \centering
        \includegraphics[width=\sfwidth]{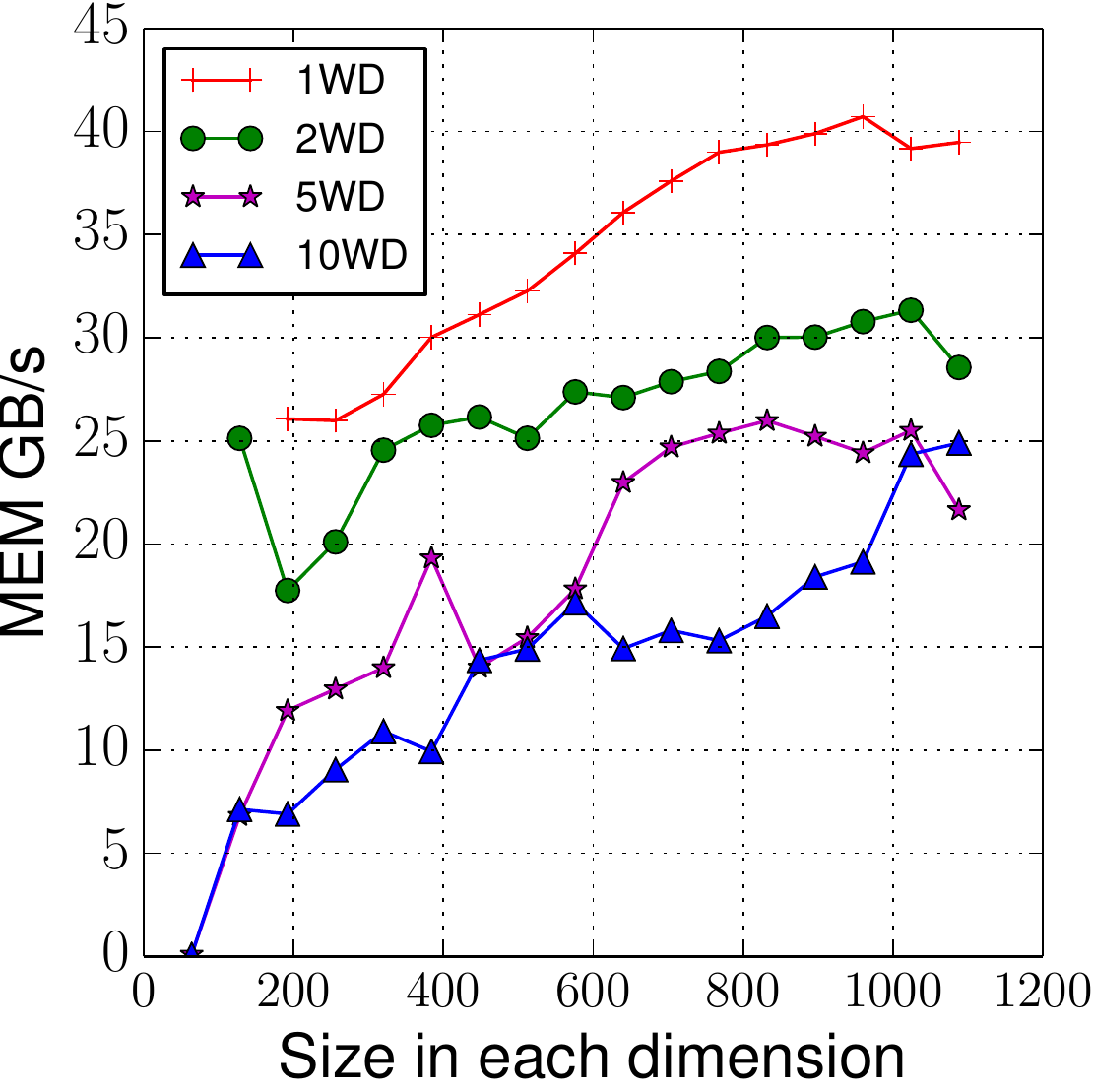}
        \label{fig:ivb_25_pt_const_mem_bw_tgs}
    }
    \enskip
    \subfloat[Measured memory transfers per \gls{lup}.]{
        \centering
        \includegraphics[width=\sfwidth]{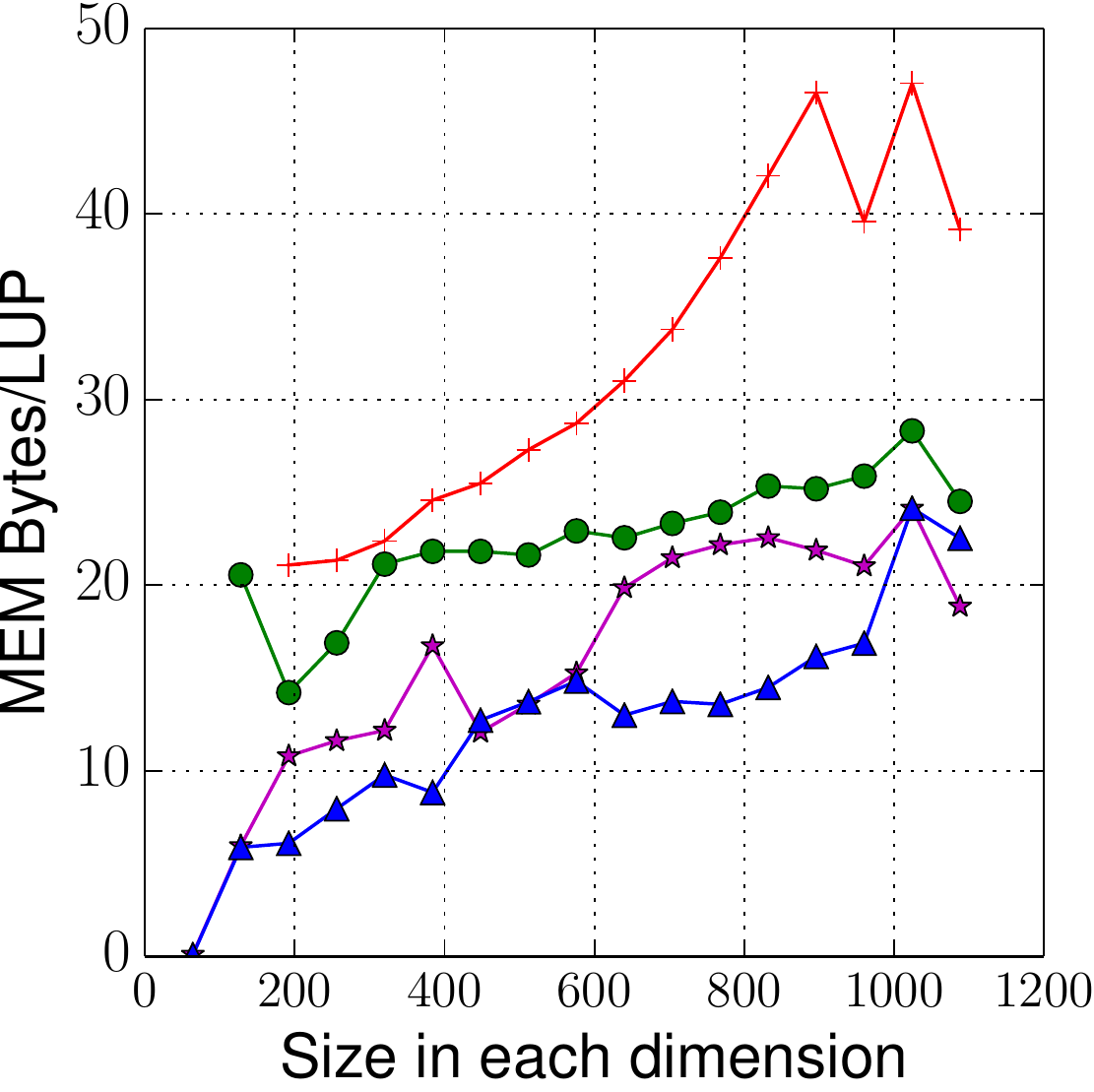}
        \label{fig:ivb_25_pt_const_mem_vol_tgs}
    }

    \subfloat[CPU energy consumption estimates.]{
        \centering
        \includegraphics[width=\sfwidth]{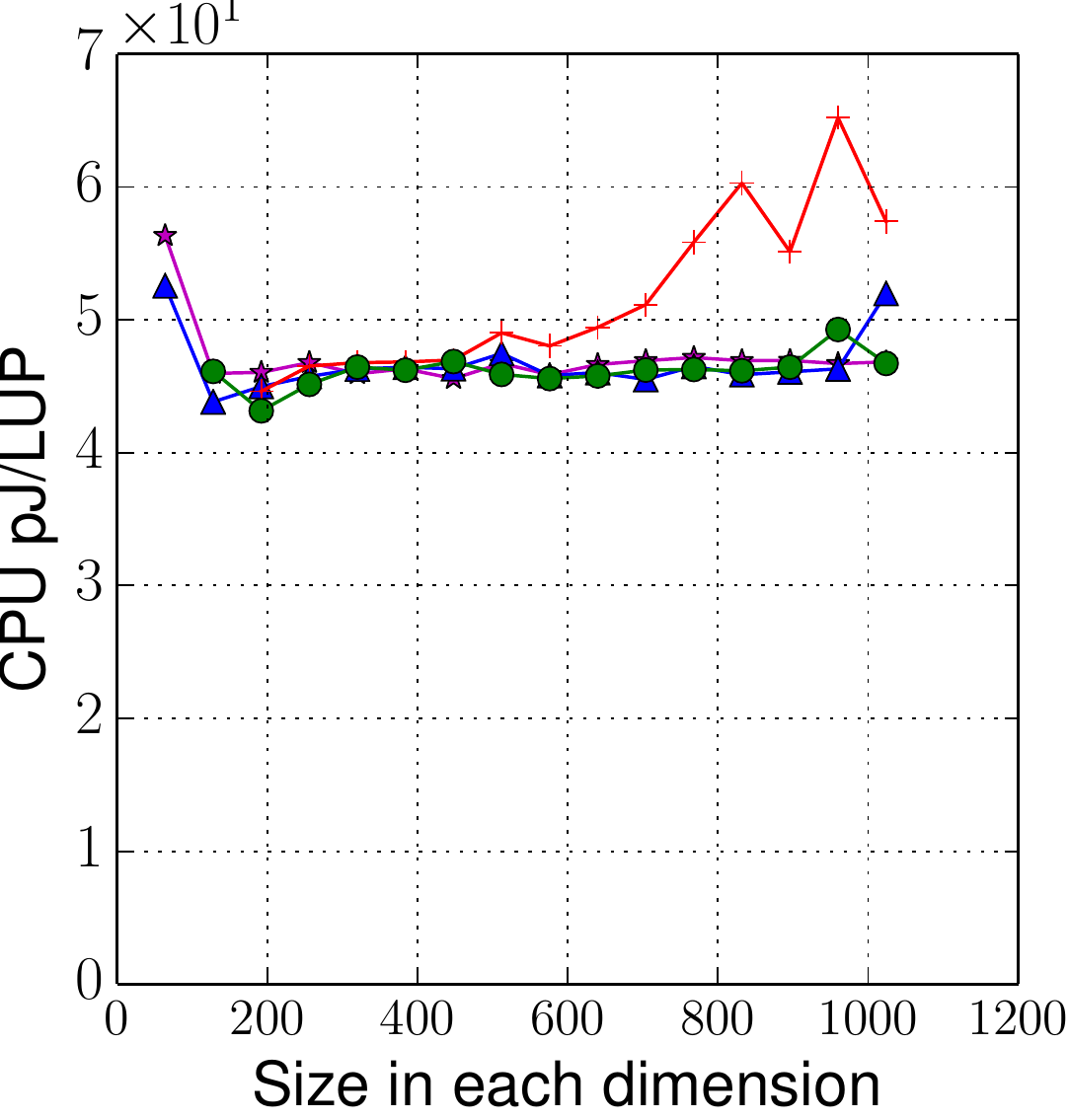}
        \label{fig:ivb_25_pt_const_energy_cpu_tgs}
    }
    \enskip
    \subfloat[DRAM energy consumption estimates.]{
        \centering
        \includegraphics[width=\sfwidth]{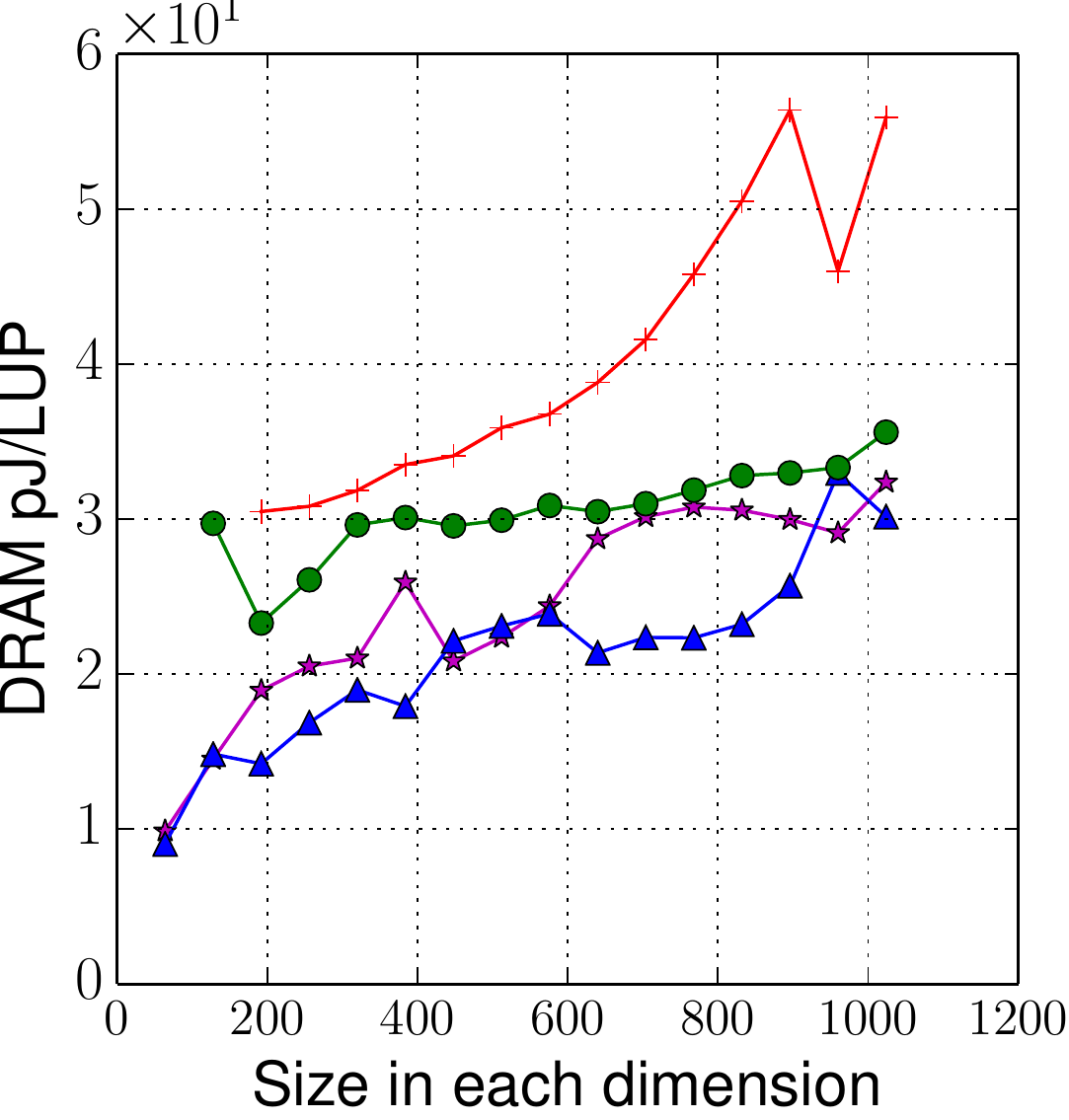}
        \label{fig:ivb_25_pt_const_energy_dram_tgs}
    }
    \enskip
    \subfloat[Total energy consumption estimates.]{
        \centering
        \includegraphics[width=\sfwidth]{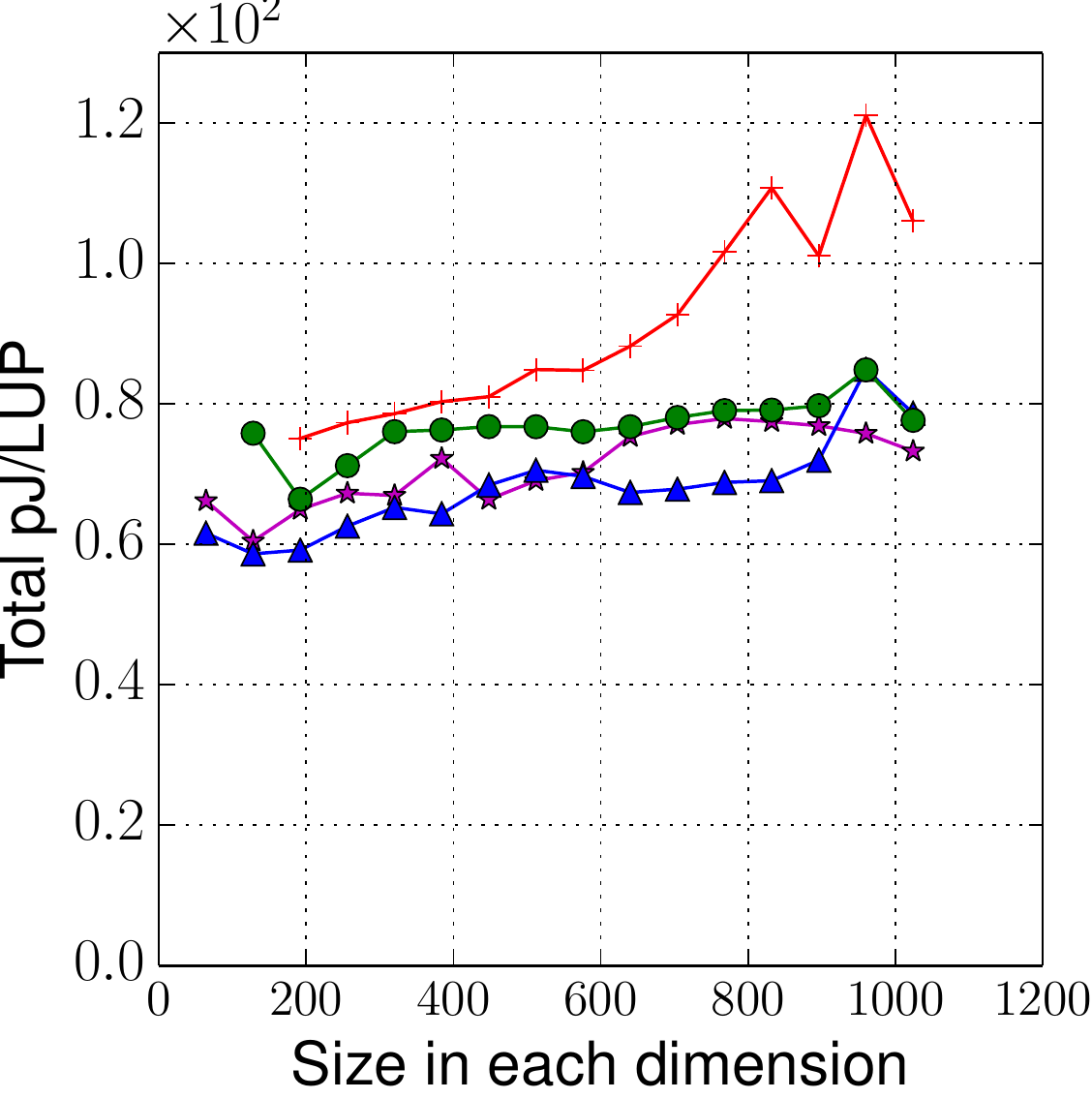}
        \label{fig:ivb_25_pt_const_energy_total_tgs}
    }
    \caption{Ivy Bridge performance, memory transfers, and energy consumption estimates of the 25-point constant-coefficient stencil using increasing cubic grid size.
    We compare various thread group sizes in \gls{mwd}.}
    \label{fig:ivb_25_pt_const_tgs}
\end{figure}

Ivy bridge performance results using different thread group sizes are presented in Fig.~\ref{fig:ivb_25_pt_const_perf_tgs}, along with the measured memory bandwidth in Fig.~\ref{fig:ivb_25_pt_const_mem_bw_tgs}.
The memory transfer volumes in Fig.~\ref{fig:ivb_25_pt_const_mem_vol_tgs} and the energy estimates in Figs.~\ref{fig:ivb_25_pt_const_energy_cpu_tgs},  \ref{fig:ivb_25_pt_const_energy_dram_tgs}, and \ref{fig:ivb_25_pt_const_energy_total_tgs} are normalized by the number of lattice site updates.

1WD achieves the best performance up a the grid size of $512^3$ before it starts saturating the memory bandwidth (Fig.~\ref{fig:ivb_25_pt_const_mem_bw_tgs}) and increasing its data transfer volume (Fig.~\ref{fig:ivb_25_pt_const_mem_vol_tgs}).

Although 2WD achieves the best performance with grid sizes larger than $512^3$, all \gls{mwd} variants show similar performance.
This similarity is caused by negligible synchronization overhead among all \gls{mwd} variants.

The energy consumption results in Fig.~\ref{fig:ivb_25_pt_const_energy_total_tgs} show that the usual law of ``faster code uses less energy'', or ``race-to-halt'' as defined in~\cite{hennessy2012computer}, is not always true.
For instance, we observe that 10WD achieves the lowest total energy to solution in most cases, although it does not achieve the best performance among all \gls{mwd} variants.
This is because the DRAM energy consumption is proportionally correlated with a linear function of its bandwidth usage.
The memory bandwidth saving of 10WD results in decreasing its DRAM energy consumption, as we observe in Fig.~\ref{fig:ivb_25_pt_const_energy_dram_tgs}, while consuming the same \gls{cpu} energy as other variants. We have to stress that these results, while being representative of a system with a relatively high
DRAM power dissipation, are not representative and will have to be
re-evaluated on later architectures. It is generally expected, however,
that future HPC systems will show this very characteristic~\cite{10.1109/MCSE.2013.95}.

\subsection{Code balance and energy consumption analysis} \label{sec:code_balance_energy_analysis}

The findings described in the previous section would not justify favoring the maximum thread group size over all other options on the Ivy Bridge processor.
However, they show clearly that if the future moves towards more memory bandwidth-starved systems and higher relative power dissipation in the memory subsystem, it should use algorithms that exhibit the lowest possible code balance.
This view is corroborated by another observation in our data: 
The \emph{overall} energy savings of temporal blocking vs.\ standard spatial blocking are roughly accompanied by equivalent runtime savings,
but when the energy consumption of \gls{cpu} and DRAM are inspected separately it is evident that this equivalence emerges from the mutual cancellation of two opposing effects: While the \gls{cpu} energy is less strongly correlated with the code performance, the DRAM energy shows an over-proportional reduction for temporal blocking. 

\begin{figure}[tbp]
    \centering
    \subfloat[7-point constant-coefficient stencil at grid size $N=960^3$.]{
        \centering
        \includegraphics[width=6.5cm]{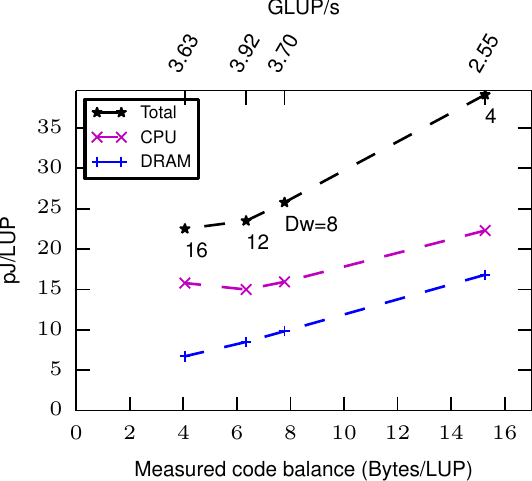}
        \label{fig:7_pt_const_code_balance_energy}
	}
	\enskip
    \subfloat[7-point variable-coefficient stencil at grid size $N=480^3$.]{
        \centering
        \includegraphics[width=6.7cm]{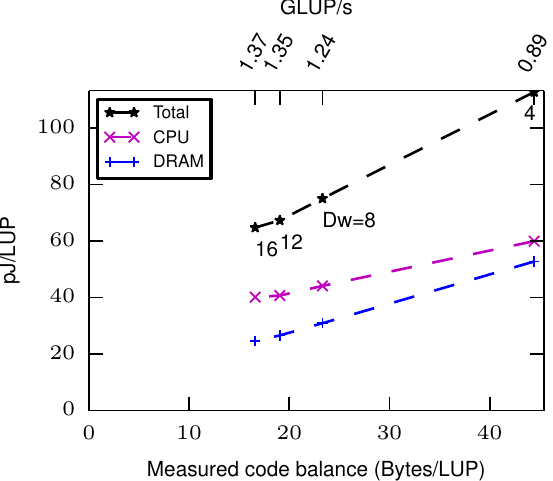}
        \label{fig:7_pt_var_code_balance_energy}
	}
	\caption{Using Intel Ivy Bridge, energy vs.\ code balance for the seven-point stencils at 
          several diamond tile sizes, separately for DRAM and
          \gls{cpu} and as a total sum. The corresponding performance of each experiment is shown on the top $x$-axis. The annotation at each point represents the used diamond width. 5WD is used in the experiments.}
    \label{fig:code_balance_energy}
\end{figure}

This can be seen more clearly in Fig.~\ref{fig:code_balance_energy} where we have measured the energy to solution with respect to the code balance for 5WD (as a consequence of setting different diamond tile sizes) for both 7-point stencils (the diagram for the 25-point stencil would only contain a single data point per set).
In both cases the DRAM energy varies much more strongly with changing code balance than the \gls{cpu} energy.
This was expected from the observations described above, but the \gls{cpu}  energy dependence is far from weak.
Overall there is an almost linear dependence of energy on code balance, making the latter a good indicator of the former.

\subsection{Thread scaling performance}

All measurement results discussed so far were taken on a full CPU socket.
In order to better understand the shortcomings and advantages of the different
temporal blocking approaches we present thread scaling results in this section.
For each stencil we show the scaling behavior of performance, memory bandwidth,
and measured code balance at a fixed problem size on the 18-core Haswell CPU\@.

\subsubsection{7-point constant-coefficient stencil}

\begin{figure}[tbp]
    \centering
    \subfloat[Performance.]{
        \centering
        \includegraphics[width=\sfwidth]{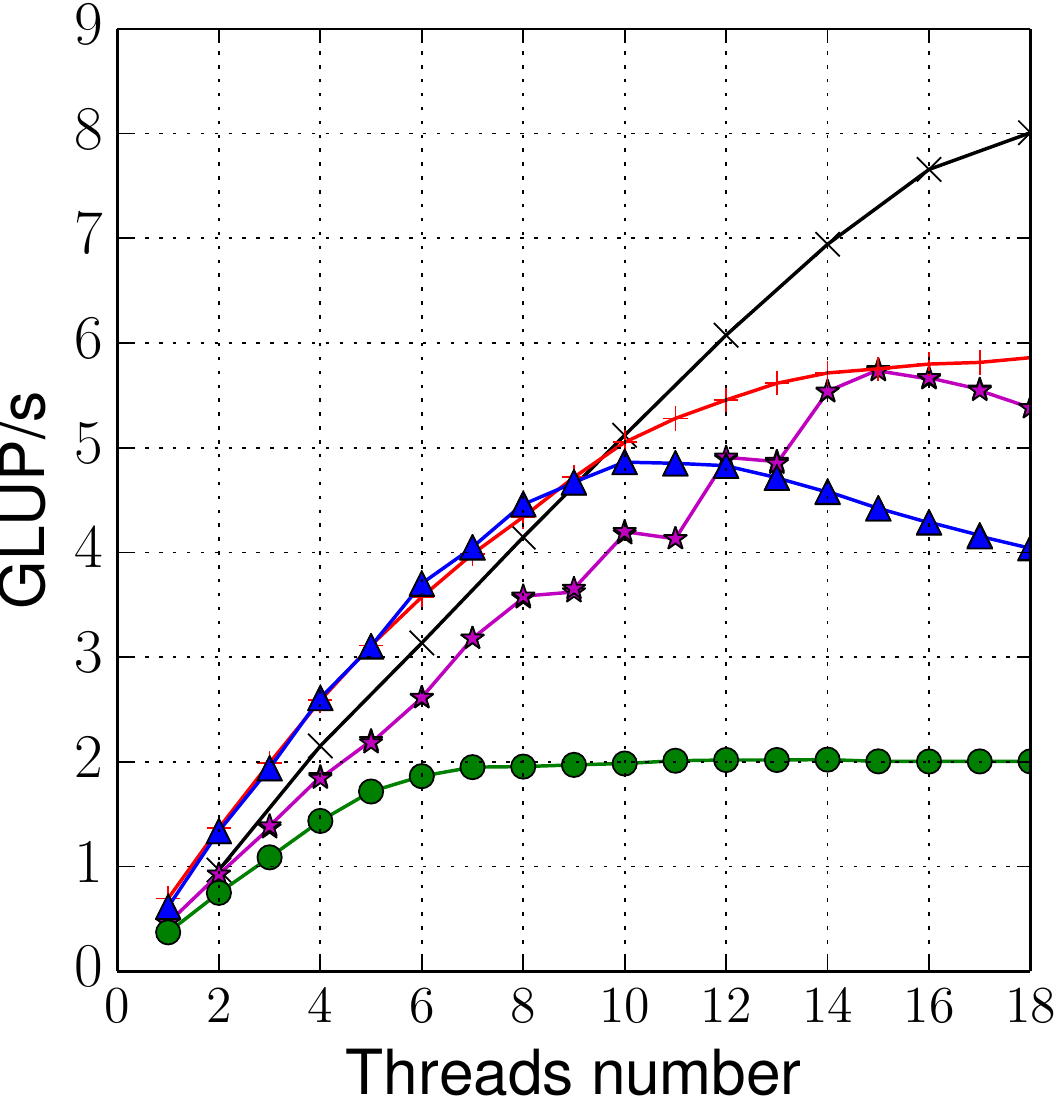}
        \label{fig:hw_7cc_perf_ts} }
    \enskip
    \subfloat[Measured memory bandwidth.]{
        \centering
        \includegraphics[width=\sfwidth]{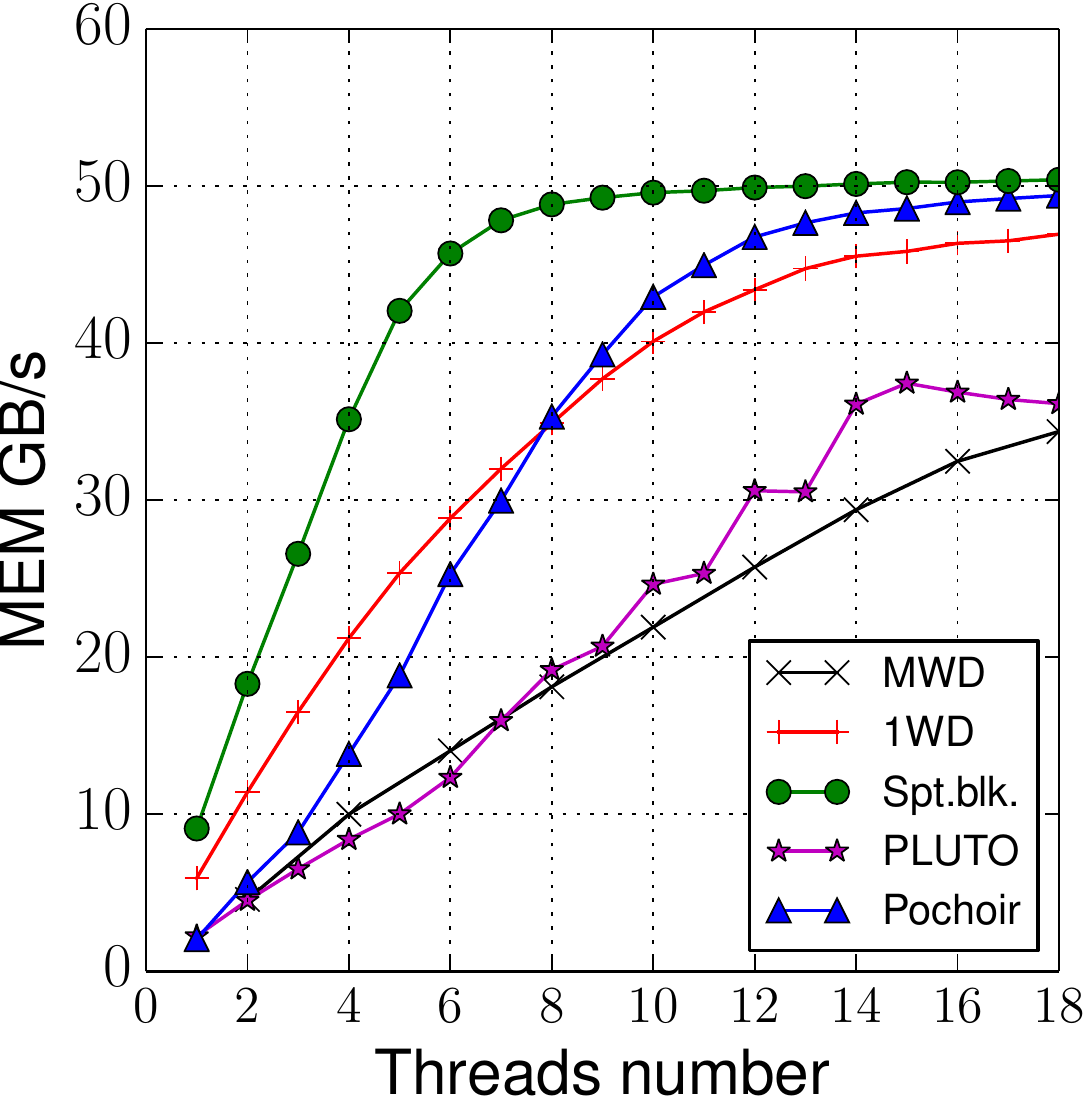}
        \label{fig:hw_7cc_mem_bw_ts} }
    \enskip
    \subfloat[Measured memory transfers per \gls{lup}.]{
        \centering
        \includegraphics[width=\sfwidth]{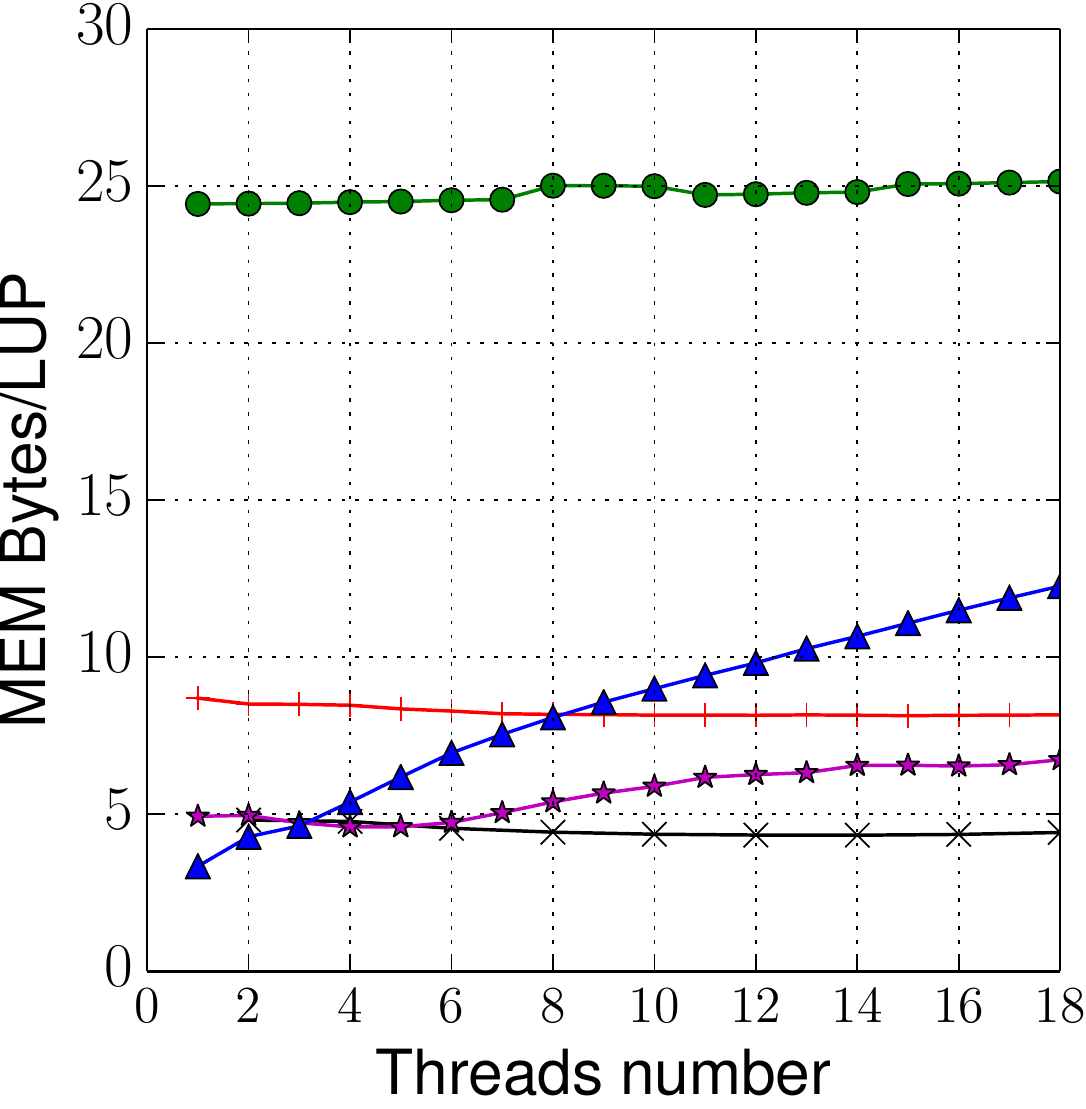}
        \label{fig:hw_7cc_mem_vol_ts} }
    \caption{
    Thread scaling for the 7-point constant-coefficient stencil, showing performance and memory transfer measurements. We compare PLUTO, Pochoir, \gls{swd}, \gls{mwd}, and spatially blocked code variants on the 18-core Haswell socket at a grid size of $896^3$. }
    \label{fig:hw_7cc_ts}
\end{figure}

The thread scaling results for the 7-point constant-coefficient stencil are shown in Figures~\ref{fig:hw_7cc_perf_ts}, \ref{fig:hw_7cc_mem_bw_ts}, and \ref{fig:hw_7cc_mem_vol_ts}. All temporally blocked variants except
Pochoir show a roughly constant code balance with increasing thread count, but only
\gls{mwd} shows good scaling across the whole chip. Pochoir and \gls{swd} clearly
run into the bandwidth bottleneck; the limited scalability of PLUTO is not
caused by traffic issues and will be investigated. \gls{mwd} shows a linearly rising
memory bandwidth utilization, indicating bottleneck-free and balanced execution.

\subsubsection{7-point variable-coefficient stencil}

\begin{figure}[tbp]
    \centering
    \subfloat[Performance.]{
        \centering
        \includegraphics[width=\sfwidth]{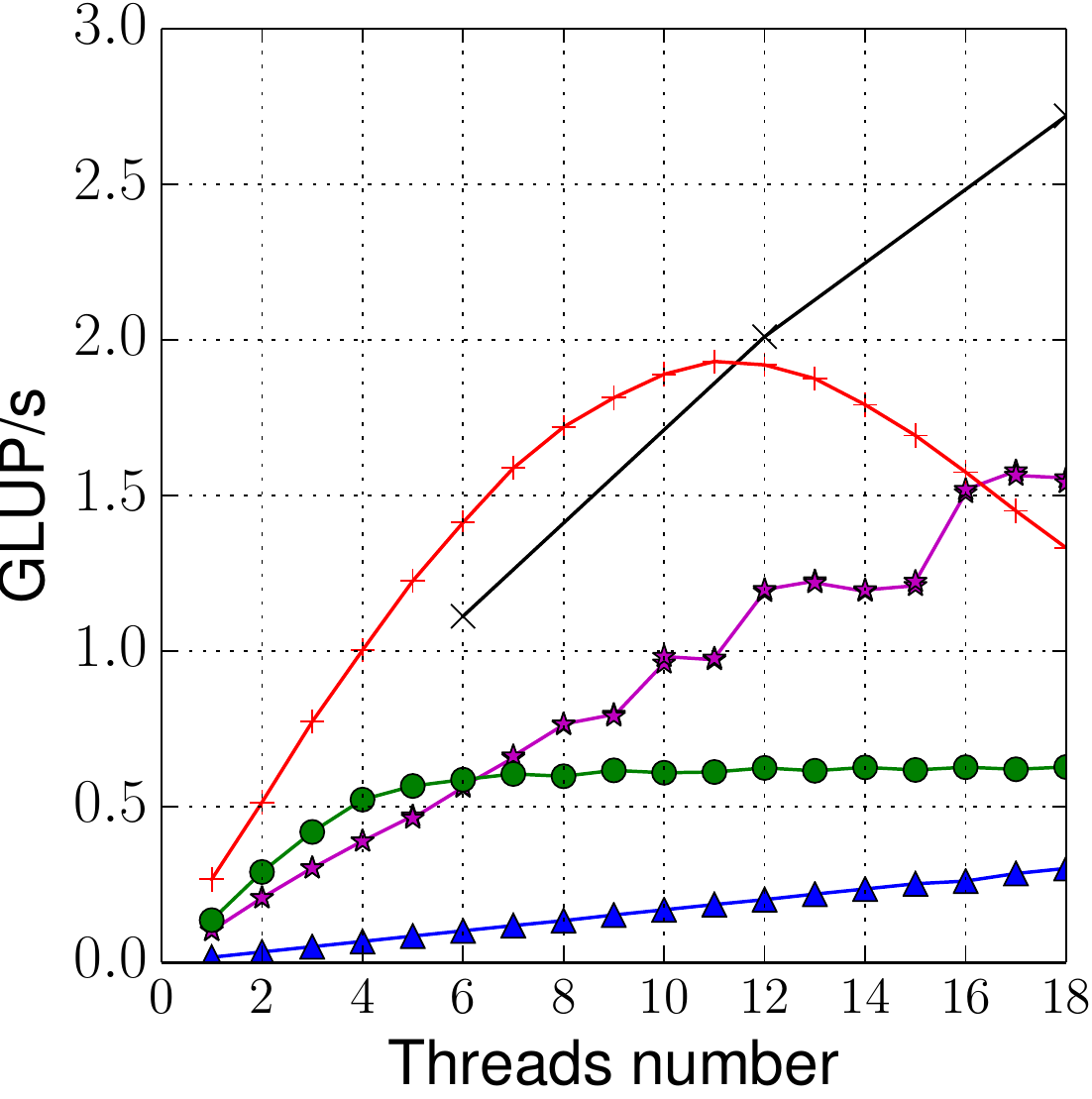}
        \label{fig:hw_7vc_perf_ts} }
    \enskip
    \subfloat[Measured memory bandwidth.]{
        \centering
        \includegraphics[width=\sfwidth]{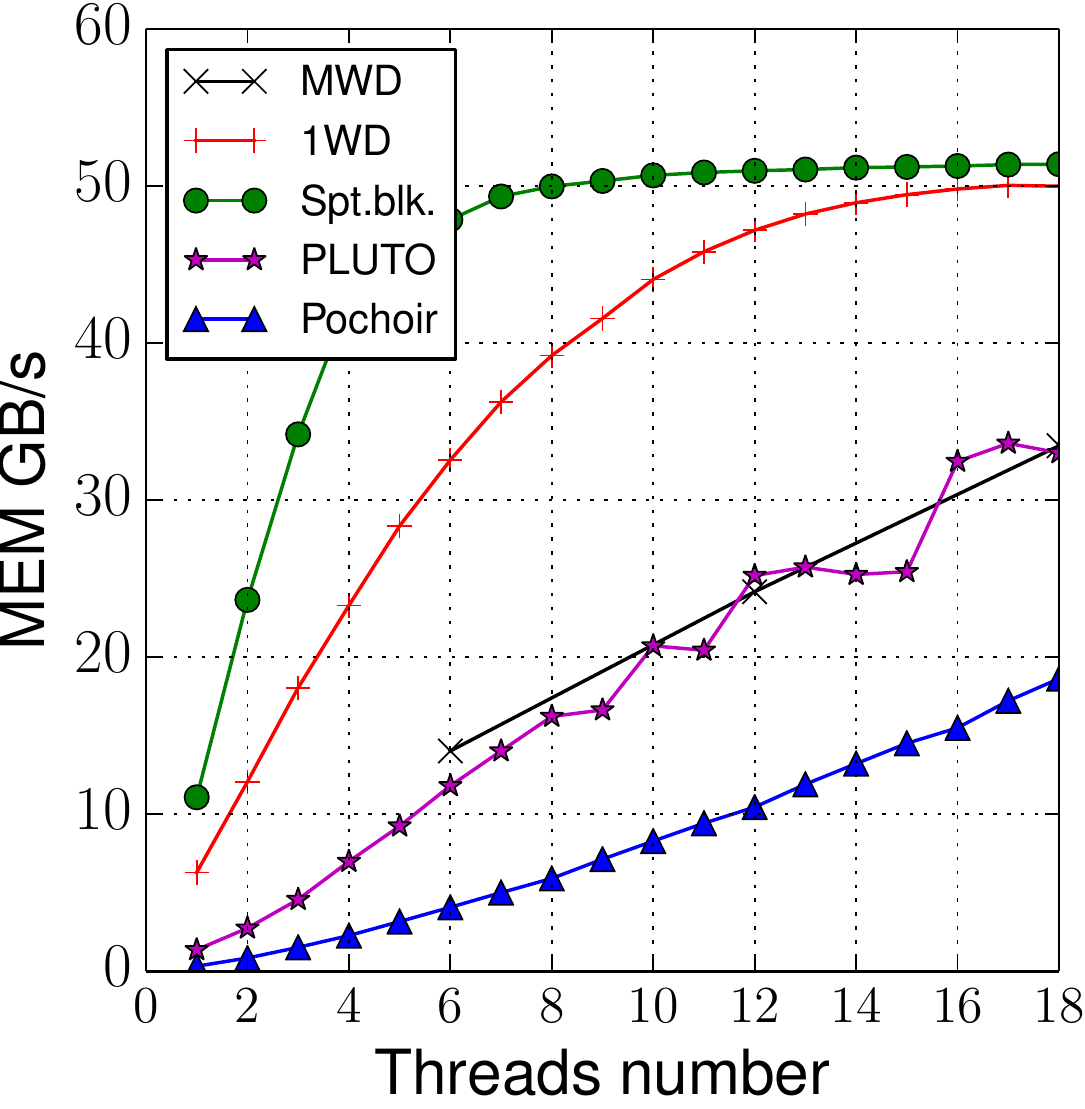}
        \label{fig:hw_7vc_mem_bw_ts} }
    \enskip
    \subfloat[Measured memory transfers per \gls{lup}.]{
        \centering
        \includegraphics[width=\sfwidth]{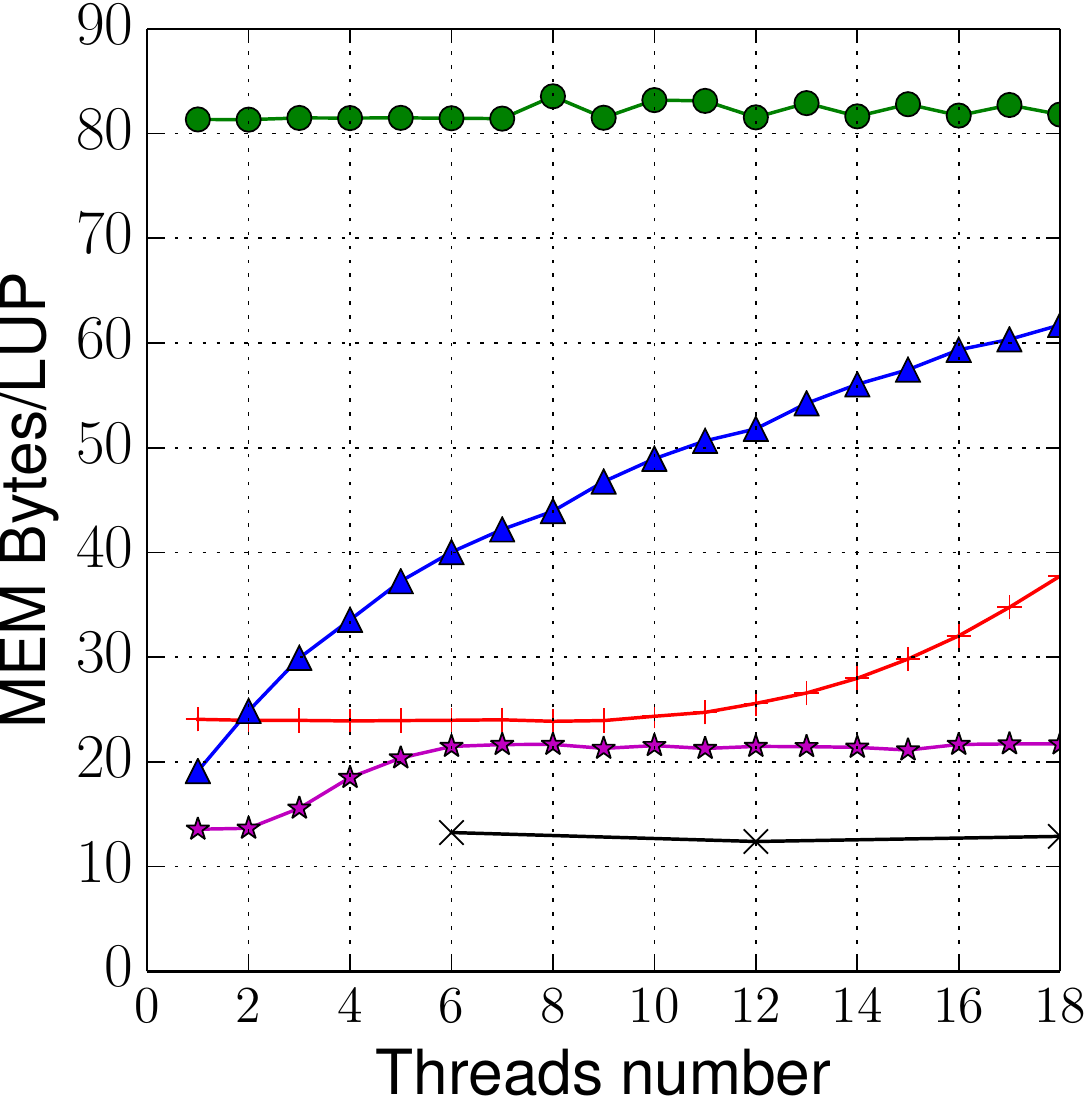}
        \label{fig:hw_7vc_mem_vol_ts} }
    \caption{
     Thread scaling for the 7-point variable-coefficient stencil, showing performance and memory transfer measurements. We compare PLUTO, Pochoir, \gls{swd}, \gls{mwd}, and spatially blocked code variants on the 18-core Haswell socket at a grid size of $768^3$. }
    \label{fig:hw_7vc_ts}
\end{figure}

The thread scaling results for the 7-point variable-coefficient
stencil are shown in Figures~\ref{fig:hw_7vc_perf_ts},
\ref{fig:hw_7vc_mem_bw_ts}, and \ref{fig:hw_7vc_mem_vol_ts}. Again,
\gls{mwd} exhibits a constant, low code balance and good
scaling. Starting at six threads, PLUTO also shows constant code
balance but on a 60\% higher level. Since the memory bandwidth is
roughly the same as with \gls{mwd}, performance also scales at a much
lower level. An interesting pattern can be observed with \gls{swd}: At
rising thread count the shared cache becomes too small to accommodate the
required tiles for maintaining sufficient locality, leading to
a steep increase in code balance beyond ten cores. Since the memory
bandwidth is already almost saturated at this point, performance
starts to break down. This behavior was expected from the
discussion in the earlier sections, but it is evident now that
\gls{swd} would be the best choice on a CPU with only ten cores
but with the same cache size. \gls{swd} is also the only temporal
blocking variant that is not decoupled from the memory bandwidth.
In case of Pochoir the data shows that the decoupling is due to very slow
low-level code, as shown before.

\subsubsection{25-point constant-coefficient stencil}

\begin{figure}[tbp]
    \centering
    \subfloat[Performance.]{
        \centering
        \includegraphics[width=\sfwidth]{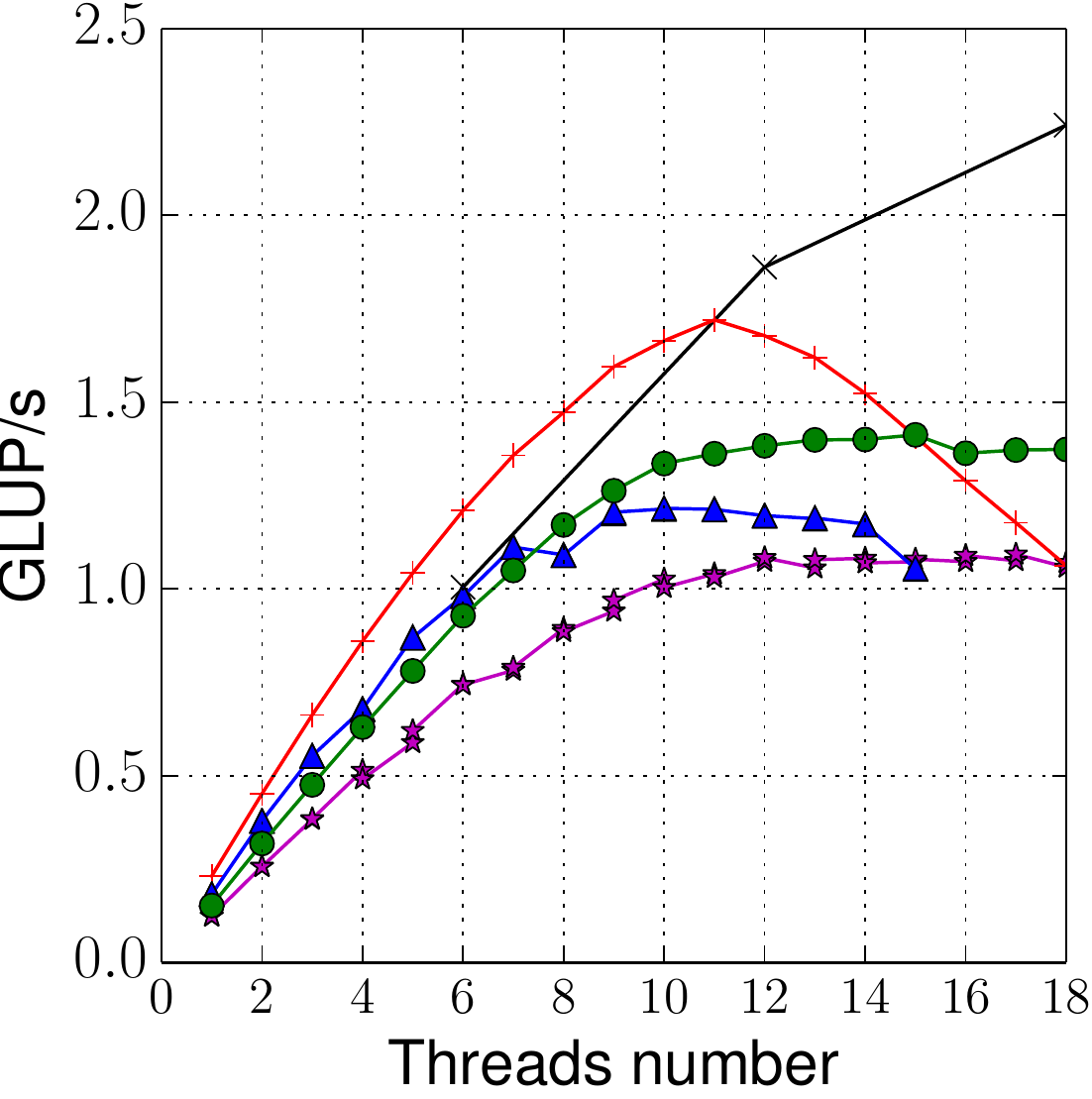}
        \label{fig:hw_25cc_perf_ts} }
    \enskip
    \subfloat[Measured memory bandwidth.]{
        \centering
        \includegraphics[width=\sfwidth]{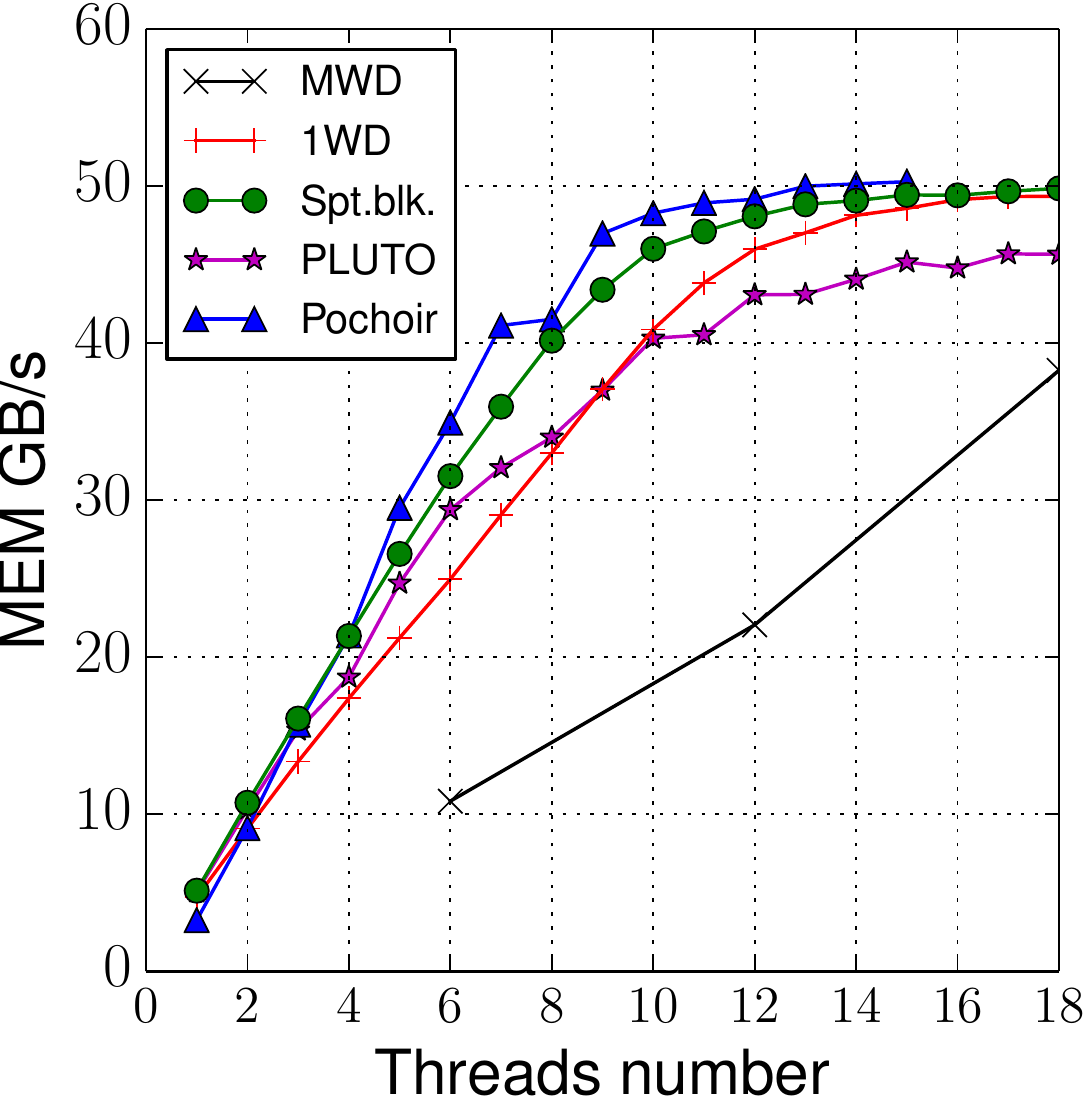}
        \label{fig:hw_25cc_mem_bw_ts} }
    \enskip
    \subfloat[Measured memory transfers per \gls{lup}.]{
        \centering
        \includegraphics[width=\sfwidth]{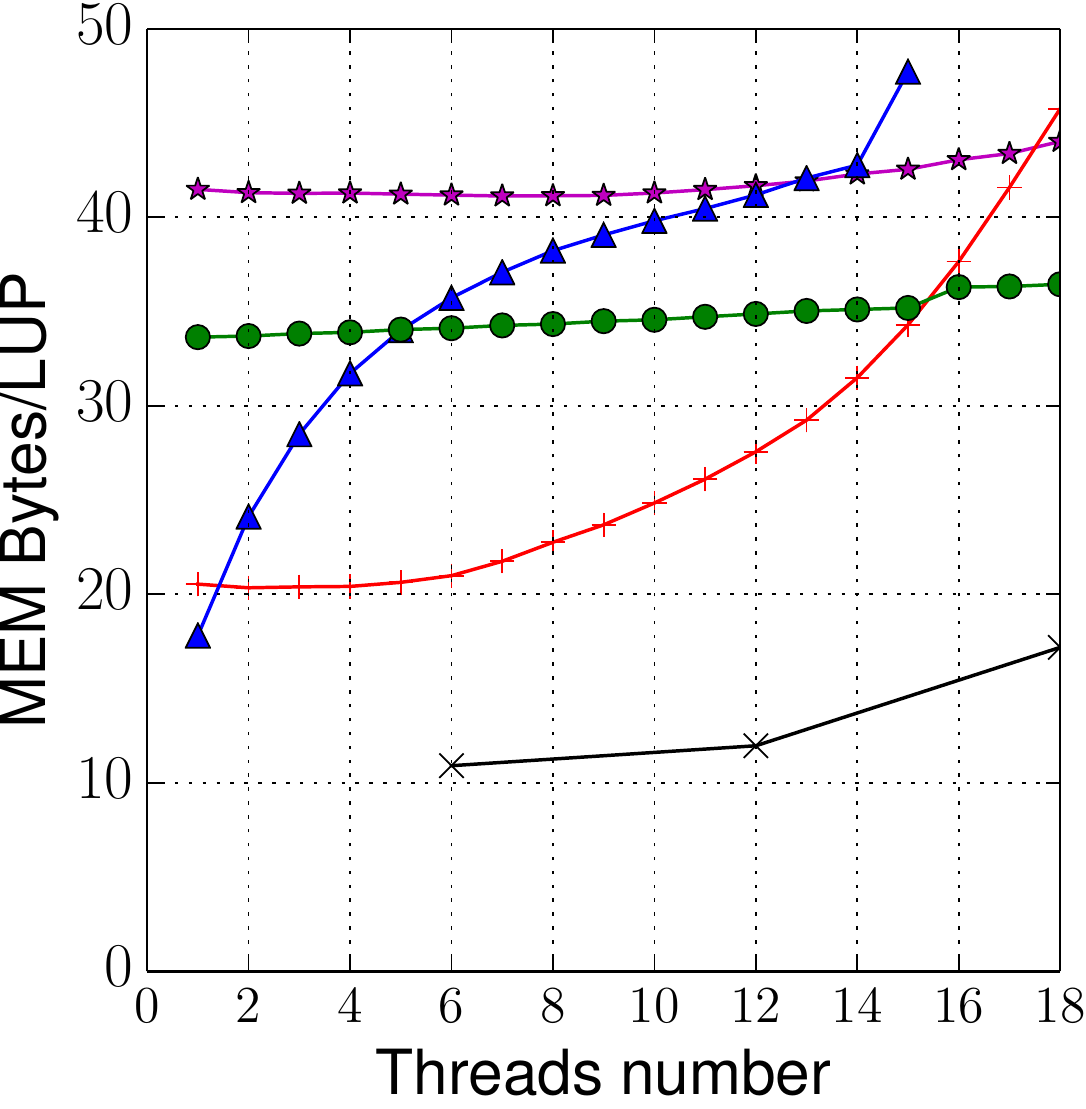}
        \label{fig:hw_25cc_mem_vol_ts} }
    \caption{
    Thread scaling for the 25-point constant-coefficient stencil, showing performance and memory transfer measurements. We compare PLUTO, Pochoir, \gls{swd}, \gls{mwd}, and spatially blocked code variants on the 18-core Haswell socket at a grid size of $896^3$.}
    \label{fig:hw_25cc_ts}
\end{figure}

Thread scaling results for the 25-point constant-coefficient stencil
are shown in Figures~\ref{fig:hw_25cc_perf_ts}, \ref{fig:hw_25cc_mem_bw_ts},
and \ref{fig:hw_25cc_mem_vol_ts}. Due to the massive cache size requirements
of this stencil only \gls{mwd} is still able to decouple from the memory
bandwidth. All other variants show strong saturation, or even a slowdown
in case of \gls{swd} beyond ten threads, which is caused by the same
cache size issues as with the 7-point variable-coefficient stencil. Again,
\gls{swd} would be the method of choice if the chip only had ten cores
but the same shared cache size.

\subsubsection{25-point variable-coefficient stencil}

\begin{figure}[tbp]
    \centering
    \subfloat[Performance.]{
        \centering
        \includegraphics[width=\sfwidth]{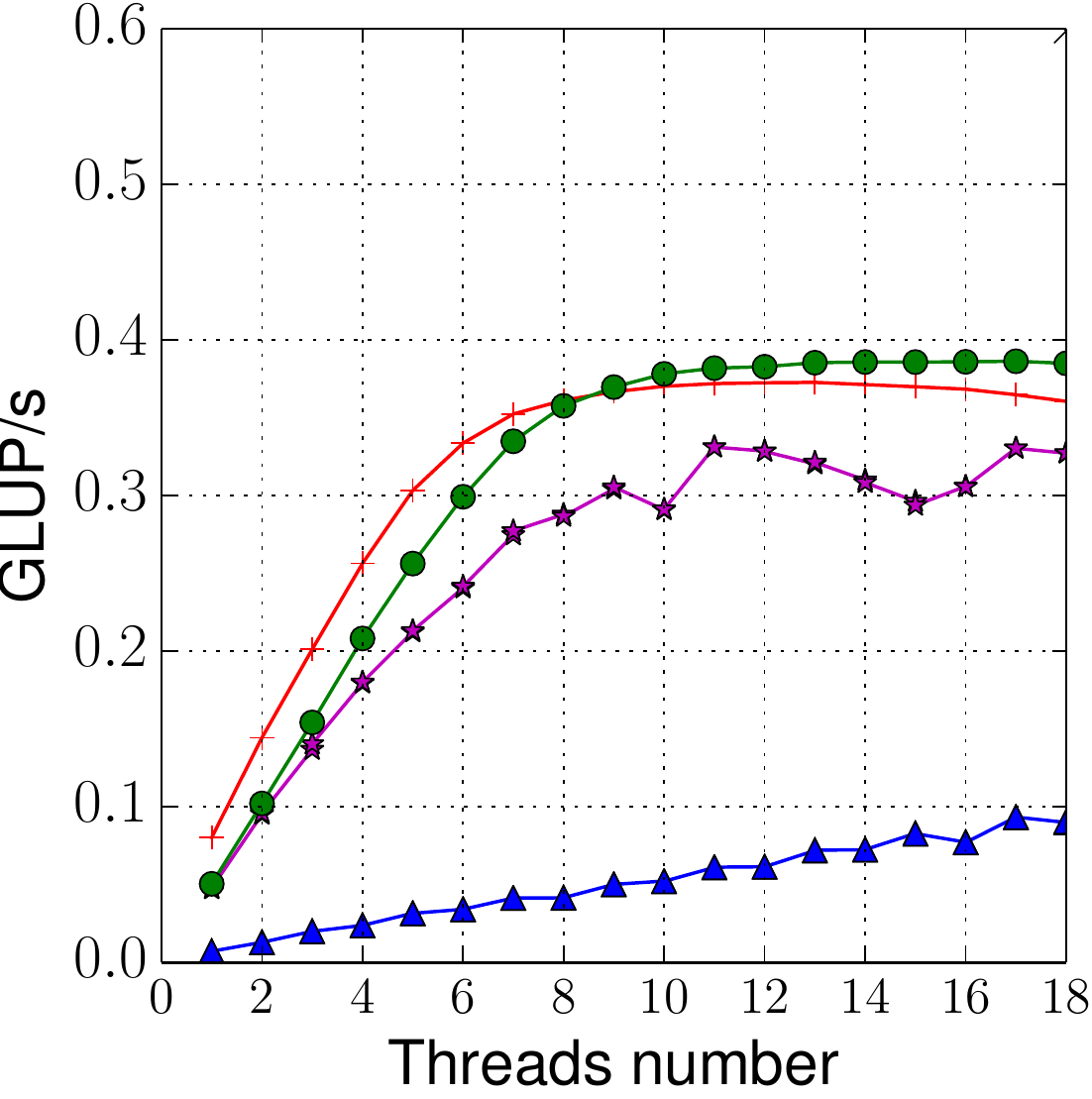}
        \label{fig:hw_25vc_perf_ts} }
    \enskip
    \subfloat[Measured memory bandwidth.]{
        \centering
        \includegraphics[width=\sfwidth]{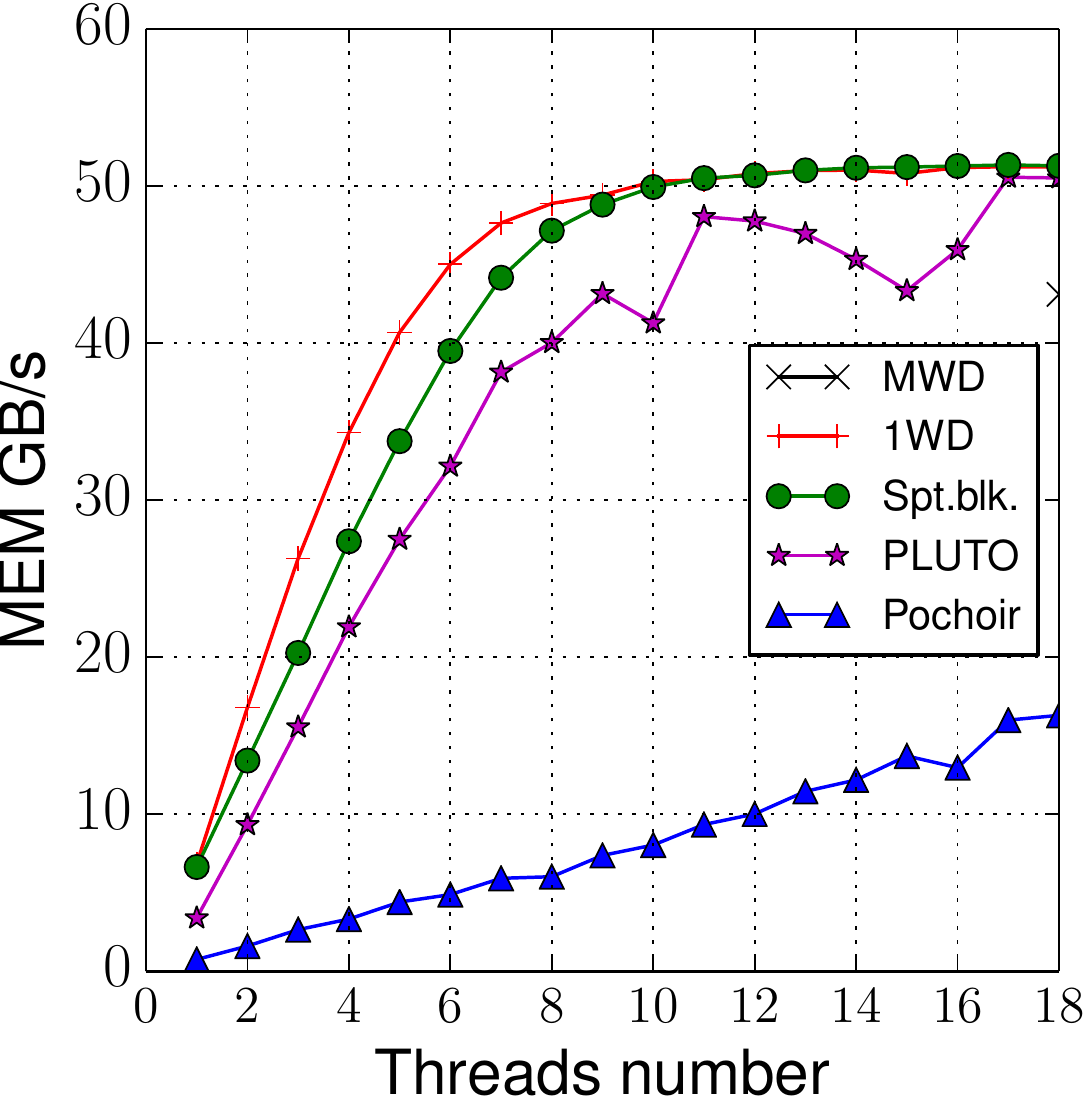}
        \label{fig:hw_25vc_mem_bw_ts} }
    \enskip
    \subfloat[Measured memory transfers per \gls{lup}.]{
        \centering
        \includegraphics[width=\sfwidth]{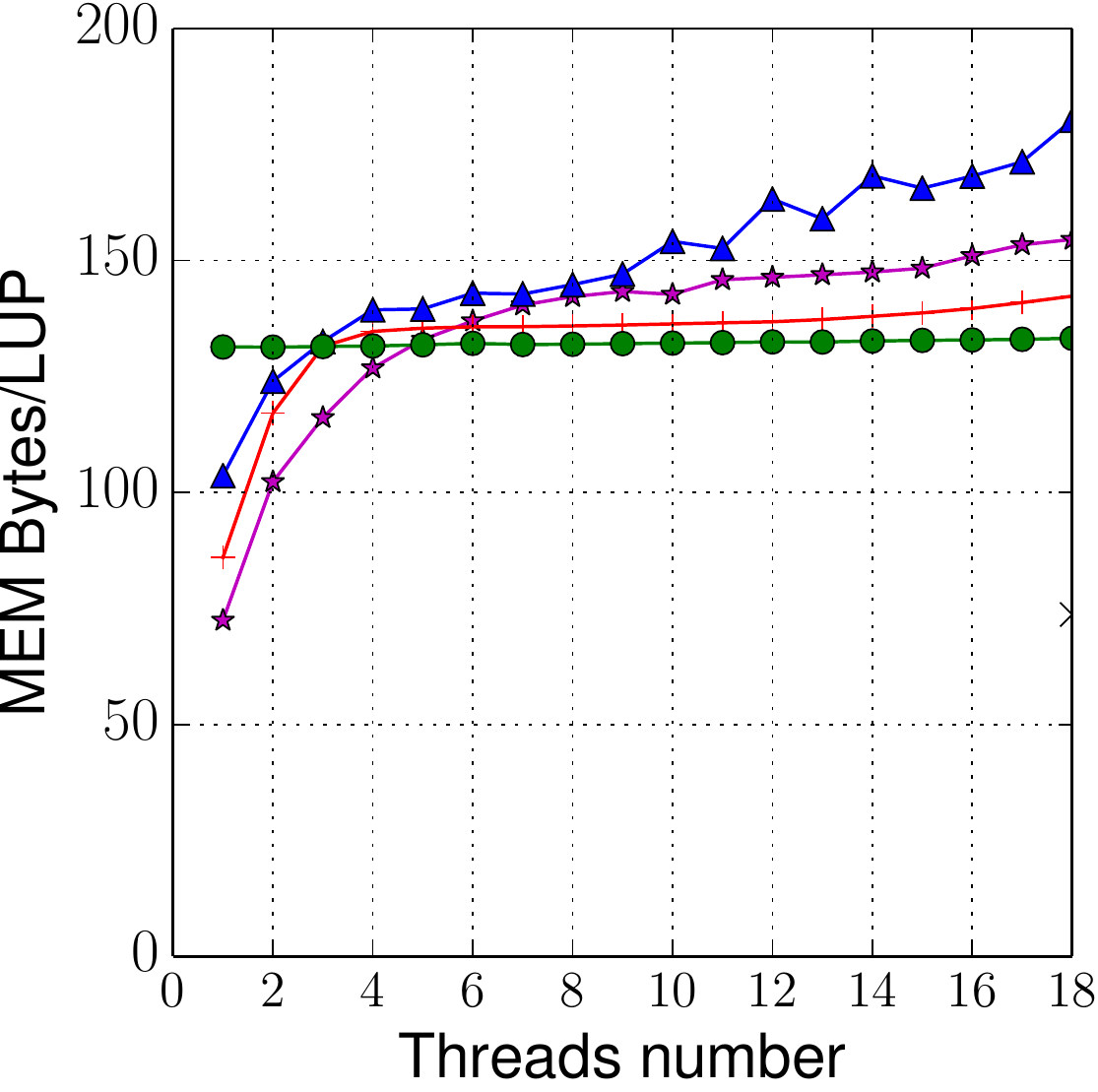}
        \label{fig:hw_25vc_mem_vol_ts} }
    \caption{
    Thread scaling for the 25-point variable-coefficient stencil, showing performance and memory transfer measurements. We compare PLUTO, Pochoir, \gls{swd}, \gls{mwd}, and spatially blocked code variants on the 18-core Haswell socket at a grid size of $768^3$. }
    \label{fig:hw_25vc_ts}
\end{figure}

Thread scaling results for the 25-point variable-coefficient stencil
are shown in Figures~\ref{fig:hw_25vc_perf_ts}, \ref{fig:hw_25vc_mem_bw_ts},
and \ref{fig:hw_25vc_mem_vol_ts}. Note that there is now only
a single data point in each figure for \gls{mwd} (at 18 threads). All
variants except \gls{mwd} exceed even the spatial blocking code balance
beyond five threads and thus show strong performance saturation,
with the exception of Pochoir, which again suffers from code
quality issues. Even if one were able to accelerate the
Pochoir code so that it could saturate the bandwidth, it would
still end up at a lower performance level than all others (about
0.3\,\GLUPS).

\section{Related work} \label{sec:related_work}


The arguably most comprehensive overview on stencil operators and
associated optimization techniques (including spatial and temporal
blocking) on modern processors was given in~\cite{Datta:EECS-2009-177}. Since then, some new ideas about
efficient cache reuse in stencil computations have come up, most of
which are variants of temporal blocking. Spatial blocking alone,
although it may reduce the code balance considerably, is not able to
achieve decoupling from the memory bottleneck in many practically
relevant situations. In the following we restrict ourselves to prior
work in stencil optimizations for standard cache-based microprocessors. 

Several tiling techniques have been proposed in the literature that
can be leveraged for temporal blocking. These include parallelogram
tiling, split tiling, overlapped tiling, diamond tiling, and hexagonal
tiling (see~\cite{orozco2011locality,Zhou2013} for reviews). Out of these, diamond tiling
has emerged as a promising candidate for the stencil schemes and
processor architectures under investigation here, and has
been covered extensively, for example, in~\cite{Strzodka:2011:CAT,Bandishti6468470,grosser2014hybrid,grosser2014ppl}. 

The wavefront technique is a further development of the
hyperplane method as introduced in~\cite{Lamport:1974}. It can be
combined with other tiling approaches as shown in~\cite{Strzodka:2011:CAT,Wonnacott845979,nguyen18993}, and in~\cite{wellein5254211}. The latter work was the
first to leverage shared caches in multicore CPUs for improved cache
reuse in stencil algorithms.

Apart from the tiling technique there is another criterion that can be
used to classify cache optimization approaches: they are either cache
oblivious, i.e., they require no prior knowledge of the hardware to
find optimal parameter combinations
\cite{frigo1999cache,Frigo:2005:COS:1088149.1088197,Tang2011,Strzodka:2010:COP:1810085.1810096},
or cache-aware, i.e., they use either auto-tuning
\cite{Datta:EECS-2009-177} or predictive cache size models
\cite{Strzodka:2011:CAT}.

Producing optimal code by hand is tedious and error-prone, so there
is a strong tendency to develop software frameworks that can handle
this process (semi-) automatically. PLUTO~\cite{Bondhugula2008} is a source-to-source converter using the polyhedral model, CATS~\cite{Strzodka:2011:CAT} is a library, Pochoir~\cite{Tang2011} leverages a domain-specific language together with
a cache-oblivious algorithm,  PATUS~\cite{Christen6012879} employs a DSL and auto-tuning, and a DSL that uses split-tiling is developed in~\cite{henretty2013stencil}.  Gysi \etal~\cite{gysi-modesto} have developed MODESTO, a
stencil framework using models to arrive at predictions about the optimal
transformations on the target system. 
Recently, a review  of stencil optimization tools that use polyhedral model has been published in~\cite{Wonnacott2013}.

The emergence of energy and energy-related metrics as new optimization
targets in HPC has sparked intense research on power issues in recent
years. For instance, the realization that low energy and low time to
solution may sometimes be opposing goals has led to activities in
multi-objective auto-tuning
\cite{bbalaprakash2013anl,Gschwandtner2014_euro_par}. However, 
there is very little work that tries to connect power dissipation
with code execution using simple synthetic models to gain 
insight without 
statistical or machine learning components. Hager \textit{et al.}~\cite{CPE:CPE3180}
have constructed a simple power model that can explain
the main features of power scaling and energy to solution on
standard multicore processors. Choi \textit{et al.}~\cite{Choi:2013:RME:2510661.2511392}
follow a slightly different approach by modeling the energy 
consumption of elementary operations such as floating-point 
operations and cache line transfers. In this work we try to
pick up some of those ideas to establish a connection between
energy consumption on the CPU and in the DRAM with the performance
and, more importantly, with the code balance of a stencil algorithm.

We classify the prior work related to our \gls{mwd} approach in two
categories: using separate cache block per thread and utilizing cache
block sharing.

\subsection{Related work using separate cache blocks per thread}

We describe the temporal blocking issues in using dedicated cache block per thread in Section~\ref{sec:motivation}.
Insufficient data reuse is a consequence in these approaches, given the \gls{llc} size limitation, having to use long strides in the leading dimension, and the increasing number of cores in contemporary processors.

Our work is close to the work in~\cite{nguyen18993}, the diamond tiling extension of~\cite{Bandishti6468470} in PLUTO, and the CATS2 algorithm~\cite{Strzodka:2011:CAT}.
They combine wavefront temporal blocking with diamond tiling and parallelogram tiling in the case of \cite{nguyen18993}.

Nguyen \etal~\cite{nguyen18993} introduce a technique called 3.5D blocking.
They perform wavefront blocking along the $z$-axis and parallelogram tiling along the $y$-axis.
The whole domain is divided among the threads across the $y$-axis, where each iteration is advanced at once using a global barrier.

Our \gls{swd} implementation is very similar to CATS2 and PLUTO's diamond tiling.
While we use CATS2 tiling choices along each dimension, PLUTO performs diamond tiling along the $z$-axis, and parallelogram tiling along the $y$- and $x$-axes.
\gls{swd} is similar to PLUTO because one diamond tile is scheduled to each thread, where a wavefront of short parallelogram tiles is updated in-order along the $y$-axis.
The tiles are usually kept long along the $x$-axis, using the whole domain size in this dimension. Both methods are efficient in terms of minimizing the synchronization
overhead among threads and in terms of data reuse. However, our approach
can improve on this in several respects:

In order to expose sufficient parallelism, PLUTO and CATS2 require a large domain size in
the diamond or parallelogram tiling dimension. If this condition is not fulfilled,
some threads will run idle with PLUTO\@. CATS2 can in this case fall back to CATS1,
which has less stringent requirements. Still both approaches will be hard
to adapt to emerging many-core architectures where the cache size per thread
will most probably decrease. 

Neither PLUTO nor CATS2 assume a shared cache among the threads. As we have shown
in Sect.~\ref{sec:motivation}, memory bandwidth-starved stencil updates may easily
run out of cache space and may thus be unable to decouple from the memory
bottleneck. This issue is even more severe in view of the expected evolution
of processor architectures towards lower memory bandwidth and cache size per
thread, the Intel Xeon Phi coprocessor being a prominent example.


Cache-oblivious stencil computations where introduced in~\textit{et al.}~\cite{Frigo:2005:COS:1088149.1088197}.
Strzodka \textit{et al.}~\cite{Strzodka:2010:COP:1810085.1810096} introduced a parallel cache-oblivious approach that uses parallelograms to have the desired recursive tessellation.
Tang \textit{et al.}~\cite{Tang2011} introduced the Pochoir stencil compiler.
Recursive trapezoidal tiling is used by performing space and time cuts.
The algorithm was improved recently by Tang \textit{et al.} in~\cite{Tang:2015:CWI:2688500.2688514} by reducing the artificial data dependencies produced in previous work.

While these implementations use asymptotically optimal algorithms, the constant factor can be large, more than double compared to other tile shapes in one dimension.
Trapezoidal and parallelogram tiling are provably sub-optimal in maximizing the data reuse of loaded cache blocks compared to diamond tiling, as shown in~\cite{orozco2011locality}, and compared to wavefront temporal blocking, as shown in this work.
The need of using long strides along the leading dimension significantly increases the cache block size requirement, making practical implementations far from the ideal promise in the asymptotically optimal cache-oblivious algorithms.

To the best of our knowledge, utilizing cache block sharing among the threads does not work in cache-oblivious algorithms.
For example, the performance of the selected thread group size is not only affected by the cache subsystem bandwidth and size, but also by the synchronization cost of the threads.
As a result, cache block sharing techniques can achieve better performance as they use larger cache blocks than the optimal ones of dedicated cache blocks per thread, given the constraint of having to use long loops in the leading dimension.

\subsection{Related work utilizing cache block sharing}

The idea of utilizing  cache block sharing by multiple cores was proposed in~\cite{wellein5254211}.
They use parallelogram tiling along the $y$-axis and multi-core wavefront temporal blocking along the $z$-axis.
This work was extended in~\cite{wittmann10,Treibig2011130}.
They use relaxed synchronization in the thread group and assign one thread group per cache domain.
To maintain the intermediate values across parallelogram tile updates, data is copied to temporary storage to keep the intermediate time steps from the boundaries of the tiles.

Their work alleviates the cache capacity limitation that appears when using separate cache blocks per thread.
It has the advantage of reducing the total cache memory requirement by using fewer cache blocks while utilizing all the available threads.
They also introduce the thread group concept, where each thread group shares a tile.
This paper improves on the cache block sharing concept to achieve further cache block saving, more data reuse, more concurrency, flexibility with parameterized tiling, and auto-tuning.

The wavefront data is pipelined across the threads in Wellein \etal\ 's work by assigning one or more time steps for each thread.
A space is imposed between the working set of each thread to achieve concurrency while ensuring correctness.
This requires extending the cache block size in the spatial dimension, without increasing the in-cache data reuse.
For example, the cache block size is doubled to achieve the same data reuse, compared to single thread wavefront variant, when a stencil radius of unity is used and a single time step is assigned per thread.
This cost is significant when using variable-coefficient stencils, which load many bytes per grid cell.
Since each thread updates data from different time steps, an equal amount of work across the time steps of the wavefront is required.
They achieve load balancing by using parallelogram tiling in the other dimension.
This does not allow their work to explore more advanced tiling techniques, such as diamond tiling.

A more recent cache block sharing work, called \emph{jagged tiling}, is proposed in~\cite{shrestha2014jagged}.
The polyhedral model is employed to generate code for intra-tile parallelization using the PLUTO framework.
This work is applied over one- and two-dimensional stencil computations.
They use multi-core wavefront tiles, similar to our work along the $z$-axis, and an optimized runtime system for fine-grained parallelization to schedule the work to threads.
Intra-tile thread synchronization is performed through a dependency mask table of the size of the intra-tile tasks to track the ready-to-update work.
When a thread takes a task, atomics are used to avoid race conditions.
This work was extended in~\cite{Shrestha:2015:LAC:2738600.2738620}.
They combined their introduced jagged tiling approach with the diamond tiling extension of the PLUTO framework to allow concurrent start at the inter- and intra-tile levels.
The new approach is called fine-grain (FG) jagged polygon tiling.
They show results for the three-dimensional 7-point constant-coefficient stencil, running experiments using a grid of fixed size $480^3$, which is friendly to the cache memory.
The major difference in the structure of their tiling approach is that they provide intra-tile concurrency along the diamond tile dimension by stretching the diamond tile along the space dimension, i.e., without increasing the data reuse in time.

Shrestha's work has the advantage of utilizing the polyhedral model in their cache block sharing algorithm.
Their work is implemented in a general source-to-source transformation framework making it more generic and more usable.
Their work is currently restricted to stencil computations, as diamond tiling in PLUTO works only with stencil computations.

However, by providing the diamond tiling and the intra-tile concurrency along the same dimension, they effectively revert to smaller diamond tile sizes and update groups of adjacent diamond tiles together.
The only reuse in the thread group is at the boundary of the sub-tiles.
In contrast, our \gls{mwd} approach allows the thread group to share one large diamond tile, providing more in-cache data reuse.
Figure 5 of their paper \cite{Shrestha:2015:LAC:2738600.2738620} shows an example of their two-level tiling.
The diamond tile is split into nine sub-tile updates for fine-grained parallelization.
Using the same cache block size in space, our \gls{mwd} approach allows for 15 sub-tile updates (i.e., 67\% more data reuse) and our approach is not limited to a thread group size of three, as in their example.
In other words, they compromise tile cache block size for more intra-tile concurrency.
They also allow the intra-tile task to update more than one time step.
This imposes unnecessary data dependencies across sub-tiles in the same row.
Finally our work has the advantage of providing a parameterized thread group size and auto-tuning.

\section{Conclusions}
\label{sec:conclusion}

The performance of stencil computations tend to under-utilize the
compute resources of processors for memory-resident data sets unless
proper temporal blocking optimizations are applied. In this work we
have proposed novel temporal blocking algorithms providing performance
advantages over previously published state of the art techniques. Our
optimizations also show reduced memory bandwidth pressure, which makes
them suitable for future systems that are expected to be more starved
for main memory bandwidth.

To demonstrate the capability of our solution we have performed
extensive benchmark studies using four ``corner case'' stencils on two
modern Intel server CPUs (Ivy Bridge and Haswell). Building on
previous work about diamond tiling and wavefront temporal blocking we
show that separate cache blocks per thread, as used for instance
in cache-oblivious algorithms, are often not sufficient to decouple
from the main memory bandwidth, especially for large problem sizes and
stencils with variable coefficients. Our novel \emph{multi-dimensional
  intra-tile parallelization} approach allows for a more efficient use
of the cache size and exploits high intra-tile concurrency, which is
very advantageous for stencils with high data traffic requirements.
To this end we introduce thread parallelism in the leading dimension
instead of tiling it to reduce the working set in the threads' private
caches. This enables a better utilization of hardware prefetching in
the shared cache level, and it reduces the danger of TLB shortage.  By
making the intra-tile parallelism parallel to the time axis we also
maximize the data reuse in the private caches.

All algorithms were implemented in our open-source testbed framework
\emph{Girih}. Our parameterized implementation provides a controllable
trade-off between fine-grained synchronization and memory bandwidth
pressure. An auto-tuning component determines optimal parameter
combinations for the stencil and architecture at hand.

We have constructed accurate models for the cache size and data
transfer requirements of our optimized implementations. The traffic
model can predict the optimal code balance as a function of the
stencil radius, the tile parameters, and the number of domain-sized
streams (``arrays''), and was shown to be very accurate for the four
stencils under investigation if the predicted cache block size fits
into about half the shared cache size. We use the models to reduce the
effort of the auto-tuner, but they are also of general value since
they predict the expected code balance reduction of temporal blocking
before an actual implementation is done.

Our approach achieves better performance than PLUTO, Pochoir,
1WD/CATS2, and optimal spatial blocking with most grid sizes.
Furthermore it is the only scheme that is consistently faster than
optimal spatial blocking, especially for higher-order (long-range)
stencils. Based on hardware counter measurements we show that
or code makes optimal use of the computational resources as
predicted by a phenomenological performance model. By varying the
thread group size we have explored opportunities for reducing the
memory bandwidth pressure significantly at the cost of a slight
performance degradation. On systems where the energy consumption of
the memory chips is significant this leads to lower energy to
solution at non-optimal performance, providing a striking example
where ``race to halt'' does not apply, i.e., where slower code
is more energy efficient.

Since our temporal blocking solution provides lower code balance
than previous work it is, in our opinion, best suited for future,
bandwidth-starved systems. The concepts behind our work 
may also be applied beyond the specific field of regular stencil
codes, e.g., for unstructured grids.

\section{Future work and outlook} \label{sec:future_work}

We discuss the applicability of our proposed approach in future \gls{hpc} systems, which are expected to have deeper memory hierarchies and long vectorization units.
We also discuss how the \gls{mwd} approach can fit in the architecture of \gls{gpu} accelerators.
Stencil computations have many types and applications.
We show how our method can handle other stencil types and particular applications' requirements, such as adaptive time stepping.
We discuss a special variant of Krylov subspace solvers, which is an interesting application for stencil optimization frameworks.
Our work can be integration with existing stencil frameworks, and it can utilize more optimization techniques, so we discuss these directions further.

\subsection{Integrating \gls{mwd} in future systems}
Our \gls{mwd} approach provides efficient solution to alleviate memory-bandwidth-starved processors, where cache size per thread is not sufficient to hold a cache block that provides the required data reuse.

Future processors are anticipated to have deeper memory hierarchies~\cite{6968745} and long vectorization units.
We discuss the integration of our \gls{mwd} approach with other techniques to address these issues.


\begin{figure}[tbp]
\centering
\includegraphics[width=14cm]{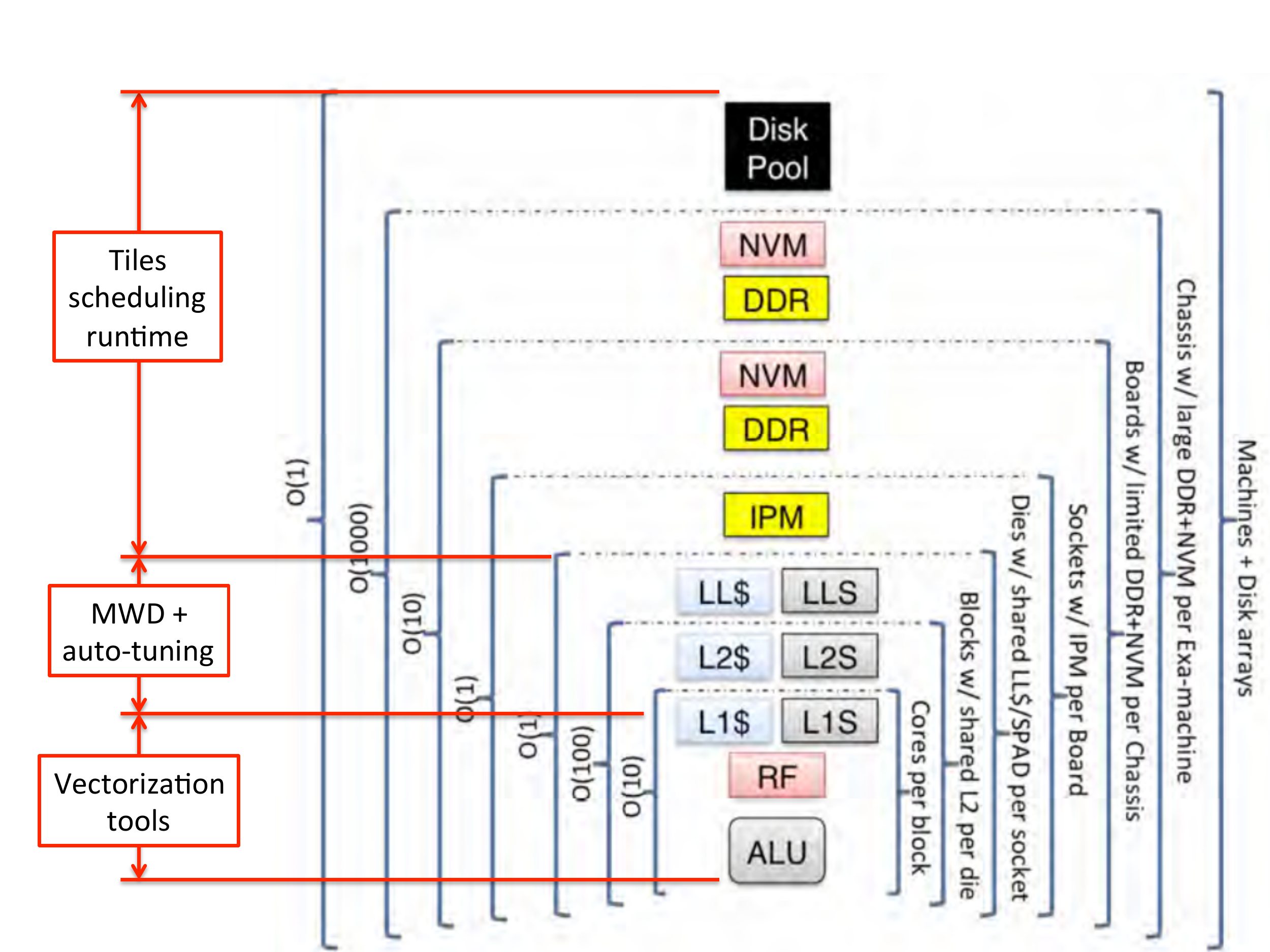}
  \caption{Outlook for integrating MWD with other techniques in future architectures. Figure with courtesy of Pete Beckman, Argonne National Laboratory.}
\label{fig:mwd_memory_outlook}
\end{figure}

Figure \ref{fig:mwd_memory_outlook} shows an example of potential future processors, where our \gls{mwd} approach can be used coherently with two approaches: 
1) Scheduling runtime system of the \gls{mwd} tiles can be used to perform hierarchical blocking of the larger and slower memory levels
2) Vectorization tools can utilize the long unit strides memory accesses provided by \gls{mwd} tiles.

\subsubsection{Handling deeper memory hierarchies with \gls{mwd}}

The runtime system would be invoked infrequently, as each \gls{mwd} tile involves updating millions of \glspl{lup}.
It may serialize (i.e., block) thread groups assignment to \gls{mwd} blocks only for those requesting/completing new tiles.
Our implementation shows negligible impact of this synchronization on the performance.

Cache oblivious techniques \cite{frigo1999cache,Frigo2005,Tang2011} are good candidates for the runtime system. 
They use space-filling-curves to provide automated hierarchical cache blocking for arbitrary memory levels with optimal asymptotic lower bound on the data transfer.
As discussed in the \ref{sec:related_work}, cache-oblivious algorithms face several challenges in utilizing the \gls{cpu} and its nearby memory levels.
On the other hand, utilizing them in the runtime system, at the granularity of our \gls{mwd} tiles, does not affect these resources, as the smallest building block is large with architecture-friendly memory accesses.

\begin{figure}[tbp]
\centering
\includegraphics[width=9cm]{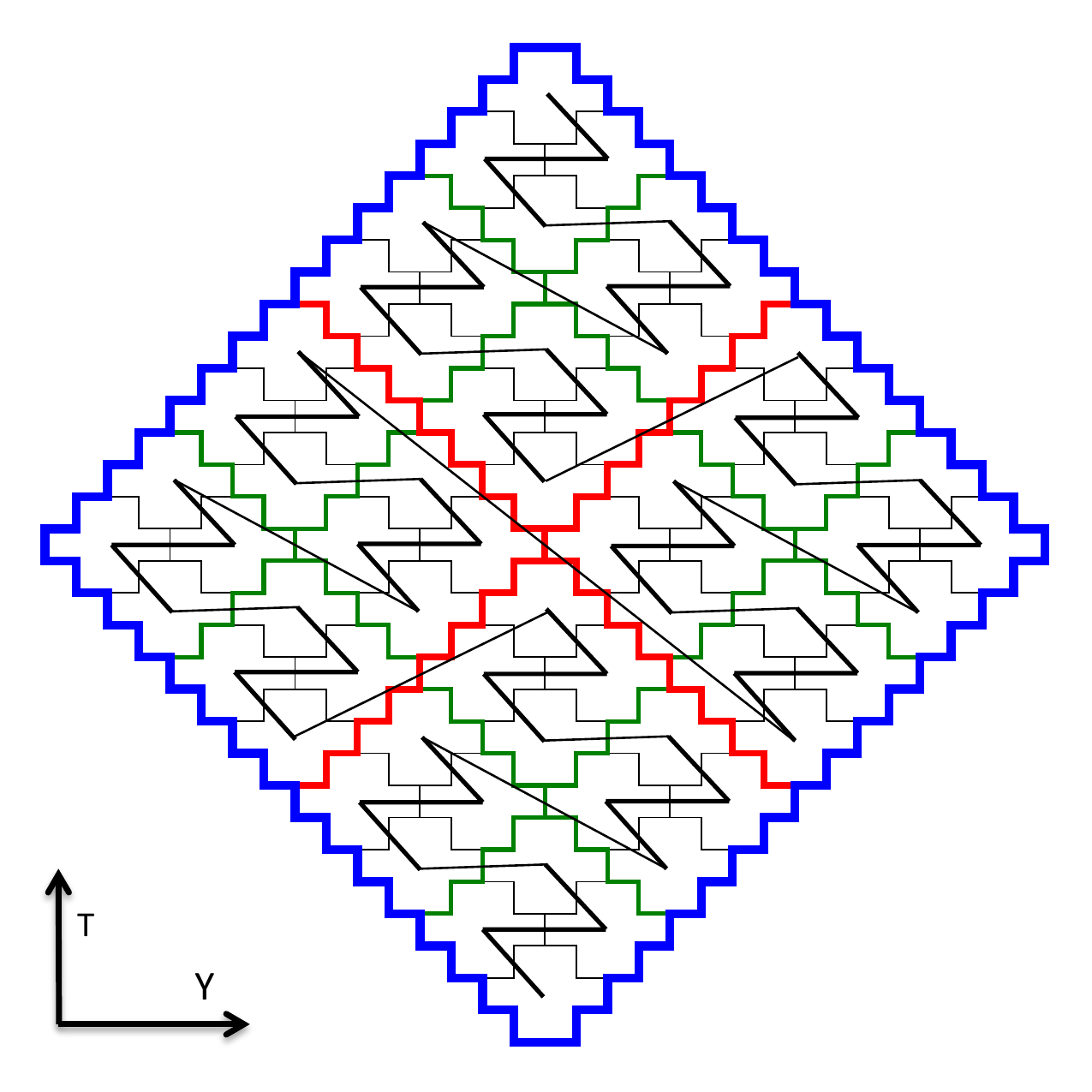}
  \caption{Utilizing Z-ordering space-filling-curves to visit diamond tiles hierarchically. Four tessellation levels are shown using black, green, red, and blue colors for the diamonds' boundaries of three tessellation levels. This recursive tessellation nature of diamond tiles make them good candidates for cache-oblivious algorithms}
\label{fig:cache_oblivious_diamond}
\end{figure}

The recursive tessellation nature of diamond tiles makes them good candidates for space-filling-curve algorithms, as shown in Fig. \ref{fig:cache_oblivious_diamond}.

The extruded \gls{mwd} tile may be split along the $z$-axis using parallelepiped-tiling to control the tile size and provide more concurrency.
This would require using multi-dimensional space-filling-curve similar to~\cite{Tang2011}, with the difference of using diamond tiling along the $y$-axis, instead of split-tiling in all dimensions.

Other approaches may provide better performance with architectural and application considerations using priority queues with certain priority setup criteria.
Combined with other criteria, \gls{lifo} can improve data-locality across tiles updates for the deeper memory hierarchy.

For example, it is shown in~\cite{ernst2014} that better performance can be achieved in Intel \gls{knc} many-core processor when the cores are assigned separate data without read sharing access.
This suggests giving higher priority to non-adjacent extruded diamond tiles to be updated simultaneously by different cores.
Ideally, updating the tiles in checkerboard fashion can guarantee no adjacent tiles updates by the cores.
On the other hand, maximizing adjacent tiles updates is favorable in multi-core processors where uniform memory access is available at the shared level cache, like in~\cite{wellein5254211} work.

\subsubsection{Handling long vectorization units}

In stencil computations, efficient utilization of long vector units usually require updating long unit-strides and performing register manipulation for adjacent stencil updates.
The extruded diamond tiles in our \gls{mwd} approach provides contiguous access of long strides along the $x$-axis.
Moreover, our implementation provides array padding and aligned memory allocation to minimize vector loads across the cache lines from the L1 cache to the \gls{cpu}.
Contiguous and aligned memory accesses make our tiles friendly for efficient vectorization by the compiler and other vectorization techniques, as in \cite{malas_ijhpca_stencil,Caballero:2015:OOM:2751205.2751224}.
The long unit-strides also result in less loop and vectorization prologue and epilogue overhead.

\subsection{Tiles software prefetching}
Our \gls{mwd} approach utilizes the hardware prefetching efficiently by allowing access to long contiguous strides of data.
However, the hardware prefetching unit may not be able to bring all the required data from main memory to the \gls{llc} in the right time, which may be caused by the large number of data streams.

Since our approach successfully reduces the cache size and main memory bandwidth requirements, we can utilize the saving in these resources to perform software prefetching.
The wavefront tiles in the extruded diamond tile are updated successively.
A double-buffering technique can be utilized: while updating a given wavefront tile, use software-prefetching to load the data of the next wavefront tile.
Since the data of successive wavefront tiles largely overlap, the ratio of loaded data to the performed update is usually minimal.

Prefetching distance is usually tuned to efficiently load and use the prefetched data in the right time.
In our case, it might be sufficient to set up the code to prefetch the next block or two.
The prefetching distance would be decided implicitly through the wavefront-diamond tile size in the cache memory, as the auto-tuner would select tile sizes that fill half the usable cache size to save the other half for prefetching.

\subsection{Perspective on integration with accelerators}

\glspl{gpu} have been gaining more importance in \gls{hpc} systems in recent years.
Our generic framework is not only suitable for \gls{gpu} architectures, but also the diamond tile shape allows decomposing the domain among \glspl{cpu} and \glspl{gpu} with relaxed synchronization scheme as we do in the distributed memory setup of the work.
Moreover, the hierarchical shape of the diamonds (i.e., each diamond can be divided to 4 equal diamonds) allows setting different tile sizes for the \glspl{cpu} and the \glspl{gpu} while maintaining the tessellation of the subdomains.
As for the update mechanism of the extruded diamond blocks in the \gls{gpu}, each block can be updated by a \gls{smx} of contemporary Nvidia \glspl{gpu}.
The threads of the \gls{smx} can be kept busy by exploiting the concurrency along the $x$-axis.
In terms of cache block size, contemporary \glspl{gpu} have sufficient cache memory to hold the wavefront data of small diamond tiles.

We present more ideas for using our \gls{mwd} approach in \gls{gpu} accelerators using NVIDIA GK110 Kepler as an example.
The NVIDIA GK110 \gls{gpu} \cite{nvidia_kepler_gk110} uses up to 15 \gls{smx} units, sharing 1536 KiB of L2 cache memory.
Each \gls{smx} has 64 KiB of shared memory that can be configured as L1 cache (hardware-controlled) and shared memory (programmer-controlled).
The hardware allows three L1/shared memory splits: 16k/48k, 32k/32k, and 48k/16k.
Each \gls{smx} contains 192 single-precision CUDA cores, 64 double-precision \glspl{fpu}, 65536 32-bit registers and is configured with 32 threads/warp.

NVIDIA Kepler provides synchronization mechanisms within the \gls{smx},  
using the synchronization features of the \gls{ptx} programming model.
On the other hand, it may be impractical/difficult to perform fine-grain sharing across \gls{smx} units, as the hardware controls the scheduling order of the \gls{smx} units' work at runtime.
This suggests assigning at least one extruded diamond tile per \gls{smx}.
The shared memory of the \gls{smx} can be utilized to perform our intra-tile parallelization approach.
This allows the threads of the \gls{smx} to maximize the data reuse in the shared memory and reduce the pressure on the global memory bandwidth.

Every 32 threads in a warp can be considered as a long \gls{simd} unit, as they perform the same operation over multiple adjacent data elements.
Since our \gls{mwd} approach provides long unit-strides along the $x$-axis, it can efficiently utilize the warp's threads by assigning them contiguous 32-cells strides along $x$-axis.

Since the wavefront tiles in the extruded diamond are updated in-order, double-buffering technique can be used to reduce the memory latency.
However, our relaxed synchronization scheme might be sufficient to cover the global memory latency, as the leading thread fetches new data while other threads update the shared in-cache data.

\subsection{Handling other stencil types}
We considered Jacobi-style updates in the stencil computation work of this paper.
Another important variant is the Gauss-Seidel-style, where the successive time iterations update the same solution array, as opposed to Jacobi-style iterations.

Gauss-Seidel-style schemes should work using our method in straightforward manner, as our method respects the data dependencies across time steps (i.e., iterations). 
However, the updates would not be ordered identically within each time step due to the spatial blocking performed by diamond tiling and the potential out-of-order scheduling of the diamond tiles in each row of diamonds. 
Using color-splitting like red-black Gauss-Seidel requires doubling the slope of the tiles (both the wavefront and the diamond) to respect the data dependency imposed by the red-black ordering in space.
In contrast to regular Gauss-Seidel ordering, spatial blocking can obey the red-black ordering, as each point update has dependency over the direct neighbors in the same time step.

So far, star-shaped stencil operators are handled in our work, where the stencil operator extends along each axis separately.
Other important stencil operators have shapes that extend diagonally, such as the 27-point box stencil operator.
The major difference in the application of these stencils is the diagonal data dependency of the stencil operator.
The star stencil operator has data dependencies only over the faces of the tile and the subdomain, while the box stencil operator has additional dependencies over the corners and the edges.
The current implementation of our approach can handle such stencil types because our tile shapes already account for data dependencies in the tile's corners and edges.

\subsection{Integrating intra-tile parallelization techniques in stencil frameworks}

Our \gls{mwd} can be integrated in cache-aware stencil optimization techniques that perform explicit tiling.
Several well established frameworks, such as PLUTO and Physis, use these techniques.
For example, the tiled stencil codes of PLUTO can incorporate our techniques.
PLUTO has options to perform diamond tiling and parallelepiped tiling along different dimensions.
Concurrent start is achieved by scheduling diamond tiles to threads using OpenMP.
It also has control over the tiles scheduling scheme.
It may be possible to configure PLUTO to perform diamond tiling along the $y$-axis and small parallelepipled tiles along the $z$-axis.
The wavefront-diamond tiling can be achieved by updating consecutive tiles along the $z$-axis. 
These tiles can be then parallelized using our thread group concept through nested OpenMP parallelization.

OpenMP does not provide straightforward mechanism to perform the intra-tile parallelization and synchronization.
To avoid the complexity of using pthread libraries, it is possible to use the ``phasers'' introduced in~\cite{shirako_2011}.
They proposed data structures and interface to provide threads point-to-point and sub-group synchronization that are suitable for parallelizing stencil computations.

\subsection{Handling adaptive time stepping of \gls{pde} solvers with \gls{mwd}}

Adaptive time stepping is used in explicit \gls{pde} solvers when the maximum wave speed in the solution domain is not known in a priori time of the simulation, or may vary significantly during the iterations of the solver.
The \gls{cfl} condition \cite{cfl_orig} (the English translation of the paper \cite{cfl_english}) determines the time step size limit to achieve correct convergence in the \gls{pde} solver.
At any given time step (i.e., iteration), the ratio of the spatial to temporal discretization size $(\delta x / \delta t)$ has to be smaller than the maximum wave speed in the solution domain.
Otherwise, the solver is vulnerable to numerical instability and will not converge to the correct solution.

One of the solutions to this problem is to check the maximum wave speed after each iteration.
If the maximum wave speed violates the \gls{cfl} condition, the solver reverts to the solution domain of the previous iteration and repeat using smaller time step size that obeys the \gls{cfl} condition.

Selecting very small time steps reduces the possibility of repeating iterations, but requires more iterations to arrive to the desired solution.
On the other hand larger time steps increases the probability of violating the \gls{cfl} condition that increases the time to solution.
The ``sweet spot'' is application and data dependent.

Temporal blocking approaches advance the solution domain several iterations at once, which pose challenges to adaptive time stepping approach.
Violating the \gls{cfl} condition results in reverting multiple time steps, wasting more resource and time compared to na\"{\i}ve data update order.
Moreover, regularly checkpointing at correct solution domain iterations is important.
This is challenging in temporal blocking approaches that advance many time steps in some parts of the solution domain, such as the cache-oblivious algorithms.

Our \gls{mwd} can be modified to handle adaptive time stepping.
The most suitable place for checkpointing is the middile of diamond tiles, where the whole solution domain can be stored at one iteration.
This requires modifying the wavefront approach to store the middle time step of certain diamond rows in separate arrays.
The cost of reverting when the \gls{cfl} condition is violated can be minimized.
The runtime tiles scheduler can be modified to increase the priority of tiles at earlier time steps and decrease the priority of tiles at later time steps.
This would decrease the range of the updated time steps, hence decrease the number of reverted time steps when the \gls{cfl} condition is violated.

It is possible to make further reduction in the cost of reverting to correct time iteration.
During the extruded diamond update, the solver tracks the maximum wave speed after the wavefront update.
If the \gls{cfl} condition is violated at any point, all the threads halt their operation and the runtime reverts to the checkpoint using the updated time step size.
This reduces the cost of waiting for complete domain update before reverting.

The same checkpointing for adaptive time stepping can be sued to handle system failure recovery.
This requires storing the data of the checkpoint in a non-volatile storage for recovery.

\subsection{Krylov subspace solvers, a promising applications for \gls{mwd}}
A particular variant of iterative Krylov subspace solvers uses expanded stencil operations in place of a series of individual SpMV operations~\cite{vanroose2013hiding}. 
The motivation for these pipelined methods is synchronization overhead reduction in distributed memory not cache efficiency in shared memory, but the interaction must be exploited for emerging hybrid programming environments.
Krylov solvers use expanded stencil operations through either polynomial pre-conditioners or s-step methods. This approach is promising in the development of extreme scale Krylov solvers. The benefit of our MWD approach can be used to improve the intra-node performance of these approaches.


\begin{acks}
For computer time, this research used the resources of the Extreme Computing Research Center (ECRC) at KAUST. The authors thank the ECRC for supporting T. Malas.

\end{acks}

\bibliographystyle{ACM-Reference-Format-Journals}
\bibliography{references}





\end{document}